\newif\ifcheckpagelimits
\checkpagelimitstrue
\checkpagelimitsfalse

\ifcheckpagelimits
 \documentclass[nofootinbib,pra,aps,twocolumn,showpacs,showkeys,%
 amsmath,amssymb,superscriptaddress,final,reprint,floatfix,longbibliography]{revtex4-1}
 \newcommand{\todo}[1]{}
\else
 \documentclass[pra,aps,twocolumn,showpacs,showkeys,%
 amsmath,amssymb,superscriptaddress,final,reprint,floatfix,longbibliography]{revtex4-1}
 \newcommand{\todo}[1]{{\pdfmargincomment[icon=Note,color=pink]{#1}}}
\fi

\usepackage[table]{xcolor}
\usepackage{helvet} 
\usepackage{lineno}
  \usepackage{mathptmx}
\usepackage{subfigure}
\usepackage{hhline} 
\usepackage{dcolumn}
\usepackage{amsmath,amssymb}
\usepackage{bm}
\usepackage{color}
\usepackage{overpic}
\usepackage{latexsym}
\usepackage{epstopdf}
\usepackage{color}
\usepackage[english]{babel}
\usepackage{latexsym}
\usepackage{stmaryrd}

\usepackage{braket}

\definecolor{mygrey}{gray}{0.94}
\definecolor{myblue}{rgb}{0.2,0.2,0.8}
\definecolor{myzard}{cmyk}{0,0,0.05,0}
\definecolor{mywhite}{rgb}{1,1,1}
\definecolor{myred}{rgb}{1,0.,0.3}

\usepackage[colorlinks=true,citecolor=myblue,linkcolor=myred]{hyperref}
\DeclareMathAlphabet{\mathpzc}{OT1}{pzc}{m}{it}

 \def\ee{\mathord{\rm e}}
 
 \def\ii{\mathord{\rm i}}

\def\half{\textstyle\frac{1}{2}}

\renewcommand{\ii}{{\rm i}}
\renewcommand{\ee}{{\rm e}}
\def\beq{\begin{equation}}
\def\eeq{\end{equation}}

\def\barray{\begin{eqnarray}}
\def\earray{\end{eqnarray}}

\begin{document}

\title{Transversality and lattice surgery: exploring realistic routes towards coupled logical qubits with  trapped-ion quantum processors}


\author{M. Guti\'errez}
\affiliation{Department of Physics, College of Science, Swansea University, Singleton Park, Swansea SA2 8PP, United Kingdom}

\author{M.~M\"uller}
\affiliation{Department of Physics, College of Science, Swansea University, Singleton Park, Swansea SA2 8PP, United Kingdom}

\author{A. Bermudez}
\affiliation{Department of Physics, College of Science, Swansea University, Singleton Park, Swansea SA2 8PP, United Kingdom}
\affiliation{Instituto de F\'isica Fundamental, IFF-CSIC, Madrid E-28006, Spain}

\begin{abstract}
Active quantum error correction  has been identified as a crucial ingredient of future   quantum computers,  motivating the recent experimental efforts to encode   logical quantum bits using small topological codes.   In addition to the demonstration of the beneficial role of the encoding,  a break-even point in the progress towards large-scale    quantum computers will be the implementation of a universal set of gates. This mid-term challenge  will soon be  faced by various quantum technologies, which urges the need of  realistic assessments of their prospects. In this work, we pursue this goal by assessing the capability of current trapped-ion architectures in facing one of the most demanding parts of this quest: the implementation of an entangling CNOT gate between encoded logical qubits. We present a detailed comparative study of two alternative strategies for trapped-ion topological color codes, either a transversal or a lattice-surgery approach, characterized by  a detailed microscopic modeling of both current technological capabilities and experimental sources of noise afflicting the different operations. Our careful fault-tolerant design, together with a low-resource optimization, allows us to determine via exhaustive numerical simulations the experimental regimes where each of the approaches becomes favorable. We hope that our study thereby contributes to guiding the future development of trapped-ion quantum computers. 
\end{abstract}


\maketitle

\makeatletter
\makeatother
\begingroup
\hypersetup{linkcolor=black}
\tableofcontents
\endgroup

\section{\bf Introduction}
\label{sec:intro}

The development of quantum mechanics during the previous century has provided us with a framework to understand the behavior of  microscopic systems in Nature. It is within this framework that scientists have developed various strategies to  control the distinctive features of these  systems according to the laws of quantum mechanics.  Nowadays, the experimental control of these techniques has matured to such a degree that scientists are exploiting them to develop new  technologies with disruptive quantum functionalities. Among these so-called quantum technologies, quantum computers, i.e. devices that exploit  quantum parallelism and entanglement  for quantum information  processing (QIP),  promise to surpass the capabilities of current computers~\cite{nielsen-book,qc_review} and have two facets. On the one hand, the availability of large-scale quantum computers would allow us to solve a variety of problems with a direct impact in society. On the other hand, building such a large-scale device is arguably the biggest quest in the quantum-technology roadmap \cite{qtech_roadmap}. Therefore, it is important to assess the required technological developments that must take place in the future by means of practical and realistic studies.

 \begin{figure*}[t]
 \begin{centering}
  \includegraphics[width=2\columnwidth]{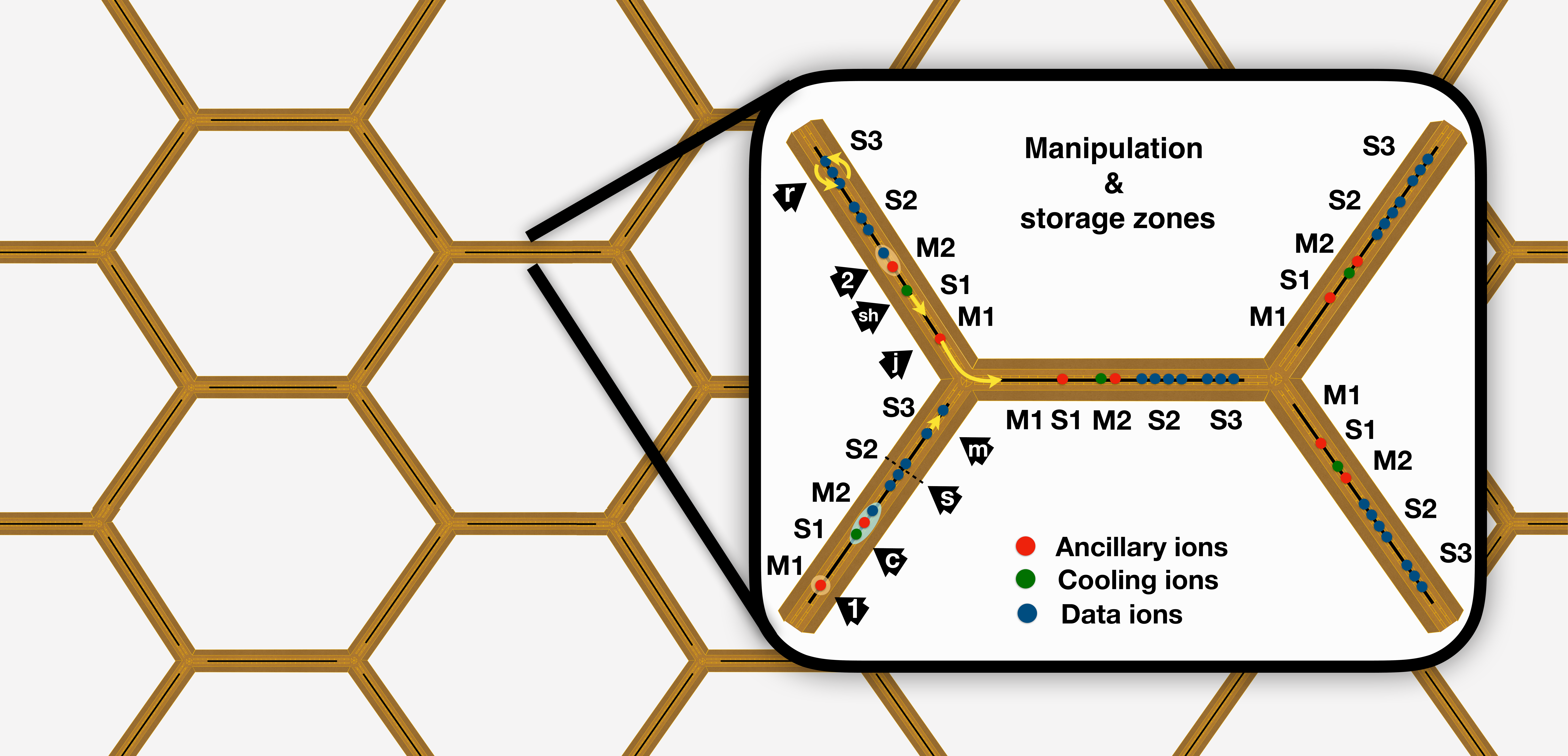}\\
  \caption{\label{Fig:trap_scheme} {\bf Schematics of high-optical access (HOA) segmented ion traps for topological QEC:}  Micro-fabricated segmented  traps connected through $\mathsf{Y}$ junctions forming a honeycomb lattice structure to manipulate mixed-species ion register for quantum information processing. (Inset) Each arm of the trap corresponds to an HOA-2 Sandia trap~\cite{hoa_source}, which consists of a slotted linear section with individually controllable electrodes used for crystal reconfiguration operations among three storage (S1,S2,S3) and two manipulation (M1,M2) regions. These operations are schematically shown by black arrows with the following letters: (r) rotation of an ion crystal, (sh) shuttling of an ion(s), (s) splitting an ion crystal, and (m) merging of sets of ions into a single crystal. The $\mathsf{Y}$ junctions  contain additional electrodes that can be used to (j) shuttle ion(s) between neighboring central regions. Additionally, in the manipulation zones, laser beams that propagate across the surface via the slotted region can be focused onto the ions to manipulate their electronic and motional degrees of freedom: (1) single qubit gates, (2) two-qubit (multi-qubit) entangling gates, and (c) sympathetic laser cooling of the ion crystal. These operations form the toolbox for scalable QEC with trapped atomic ions that is considered in this manuscript. }
\end{centering}
\end{figure*}

Current   technological capabilities have  already enabled the construction of prototype quantum computers with different architectures,  and the execution of small quantum algorithms on them~\cite{qc_review}. However, despite the remarkable level of isolation and exquisite experimental control that has already been achieved, decoherence caused by environmental noise and the accumulation of small errors  have, so far, forbidden to perform large computations of arbitrary complexity proving the supremacy of quantum computers with respect to their classical counterparts. In order to overcome this difficulty, error suppression techniques in combination with active mechanisms that prevent the accumulation of errors must be applied repeatedly during these computations. An important breakthrough in QIP has been the development of strategies ~\cite{qec_shor,calderbank-pra-54-1098,steane-prl-77-793} to  correct such errors without affecting the quantum-mechanical features of the devices: the theory of {\it quantum error correction} (QEC)~\cite{qec_review}. By encoding the  information of a logical quantum bit (i.e.~qubit) redundantly in an enlarged register composed by several entangled qubits, it becomes possible to detect the occurrence of errors  by performing collective measurements, and revert their effect  without compromising the quantum-mechanical features of the computation. An additional challenge arises from the fact that all operations required for these error detection and correction  steps can be faulty themselves. Remarkably, these errors can also be overcome by appropriate designs of the underlying quantum circuits, which prevent the uncontrolled propagation of errors through the hardware. For sufficiently low error rates of all components, the threshold theorem of \textit{fault-tolerant quantum computing} (FTQC) \cite{ft_shor_96,ft_preskill_review} predicts that reliable quantum computations of arbitrary sizes and length will become feasible. In this way, FTQC approaches manage to preserve the reversible unitary evolution in a subspace of a larger quantum register, which is undergoing irreversible non-unitary dynamics due to the external noise and the measurements required for the detection and correction of errors.

Some of the important  challenges in the near- and mid-term horizon of FTQC are $(\mathsf{QEC}$-$\mathsf{I})$ to demonstrate  the beneficial role of a QEC cycle  in an experiment (i.e. the resilience of an encoded logical qubit improves even when  the  required operations are imperfect); $(\mathsf{QEC}$-$\mathsf{II})$ to show  that  the performance of a logical qubit  surpasses that of a bare unencoded qubit; and $(\mathsf{QEC}$-$\mathsf{III})$ to demonstrate that one can perform a universal set of gates with one and two encoded logical qubits. In order to address  various key questions in this respect e.g. the performance of a particular coding strategy, the demonstration of fault-tolerant designs, or the minimal resources and control accuracy required to reach the above break-even points, it is important to perform a detailed study of the specific platform where the QEC code is to be implemented. In particular, it is important to pay special attention to the dominant sources of noise that will afflict the computation in that particular platform, and to take  into account realistic limitations in current and anticipated technological capabilities.

Today, significant efforts  to realize robust QIP based on a plethora of physical platforms are underway, including trapped ions \cite{wineland_review,blatt_review}, neutral atoms \cite{saffman-rmp-82-2313, negretti-neutral-particle-qc}, photons \cite{kok-rmp-photonic-qc} and superconducting circuits \cite{wendin-sc-qubits-review}. Starting with the seminal theoretical proposal~\cite{cz_gate}, laser-cooled crystals of atomic ions stored in radio-frequency Paul traps inside ultra-high vacuum chambers have proved to be a particularly promising  architecture for QIP~\cite{wineland_review,blatt_review}. The possibility of finding an optical cycling transition in certain atomic species, such as the alkaline-earth ions Be$^+$ and Ca$^+$, or the transition-metal ion  Yb$^+$, is crucial for QIP operations such as qubit initialization by optical pumping, or qubit readout by state-selective fluorescence. Two typical qubit choices  select either a pair of hyperfine levels from the ground-state manifold, as occurs for $^9$Be$^+$ or $^{171}$Yb$^+$ (i.e. hyperfine qubits), or a ground state and a meta-stable excited state, as occurs for $^{40}$Ca$^+$ (i.e. optical qubits). In both cases, the closed cycling transitions allow for an extremely accurate readout  with errors that can be as low as $10^{-4}$-$10^{-3}$~\cite{08Myerson,noek13} and comparable initialization accuracies. Single qubit gates with errors in the range of $10^{-6}$-$10^{-4}$ have also been demonstrated for hyperfine qubits~\cite{brown_single_qubit_gates, harty14,16Gaebler}, and errors in the  $10^{-5}$-$10^{-4}$ range have  been reported for optical qubits \cite{philipp_comment}. A distinctive achievement of this QIP platform is the demonstration of high-fidelity multi-qubit entangling gates, which are  mediated by the phonons of the ion crystal and driven by laser-induced  state-dependent dipole forces following the  schemes~\cite{MS_gates,roos_MS_gates,didi_gate}. Starting from the initial high-fidelity entangling gates for optical qubits~\cite{08Benhelm}, recent experiments with hyperfine qubits have reached errors as low as $10^{-3}$~\cite{16Gaebler,16Ballance}. Similar numbers have also been recently achieved for optical qubits~\cite{philipp_comment}. Let us note that experimental efforts are also being devoted into the increase of single- and two-qubit gate speeds~\cite{fast_single_qubit_gates,fast_2_qubit_gates_mar,fast_2_qubit_gates_oxf}, and into the development  of optical addressing of selected pairs of ions from a larger  crystal  to implement an entangling gate~\cite{debnath16}. 
 
 Regarding trapped-ion-based QEC, initial experiments have implemented the 3-qubit repetition code~\cite{3_qubit_code_nist_ions,3_qubit_code_repetitive_qec_ions}, which can detect and correct a single bit-flip classical error. More recently,  a 4-qubit code that can detect any single-qubit quantum error has also been realized~\cite{ft_error_detection_monroe}. More relevant to the present work is the experimental demonstration of the  7-qubit Steane code~\cite{nigg-science-345-302}, a QEC code that detects, but also corrects, an arbitrary single-qubit quantum error. This 7-qubit code can be considered as the smallest  version of the so-called {\it triangular color codes}~\cite{bombin-prl-97-180501}, which are a type of topological stabilizer QEC codes~\cite{stabiliser_qec} with qubits arranged on a two-dimensional (2D) lattice, where errors are detected by  local measurements that only involve groups of neighboring qubits. Similar in spirit to the surface version~\cite{surface_code} of Kitaev's toric code~\cite{toric_code}, the color-code logical qubits are encoded redundantly in a collection of physical qubits showing long-range entanglement. The code's protection against decoherence grows with increasing lattice sizes, based on certain topological aspects of the codes. Due to the locality of the required quantum processing in this class of topological codes, and their remarkable robustness against external noise~\cite{dennis_02,raussendorf07,Katzgraber_09,landahl11},   topological QEC  codes are currently considered as one of the most promising and practical routes towards large-scale quantum computing. Combined with the encouraging progress of trapped-ion QIP and QEC, there is growing interest in the community~\cite{Tomita13,musicq,Tomita14,lekitsch,eQual_1qubit,trout17} in addressing the prospects of achieving the  quantum memory and  processor goals $(\mathsf{QEC}$-$\mathsf{I})$-$(\mathsf{QEC}$-$\mathsf{III})$ using a scalable trapped-ion implementation of a topological QEC code.

One of the practical challenges that must be faced is to integrate all of the  expertise in QIP operations mentioned above, most of which has been  obtained using a particular ion species and a dedicated  apparatus, into a single architecture that can be scaled to larger system sizes. In this respect, there is a clear need~\cite{scaling_tiqc} to go beyond the small  ion chains in linear Paul traps used in the majority of the above experimental demonstrations. Two of the possible routes that are being considered are the so-called quantum charge-coupled devices (QCCD)~\cite{qccd}, and the modular universal ion-trap quantum computing (MUSIQC)~\cite{musicq}. The former is based on micro-fabricated segmented traps where ions can be transported between different storage or manipulation zones, such that the QIP manipulations are only performed in small linear crystals, and thus benefit from the aforementioned  accuracies achieved to date. In addition to these operations, the QCCD requires additional manipulations including a variety of crystal reconfiguration techniques (ion shuttling and crystal splitting/merging~\cite{Rowe02,walther12,Bowler12,fast_ion_splitting}, crystal rotations~\cite{fast_rotation}, and transport across junctions~\cite{blakestad09_xjunctions,blakestad11_xjunctions, moehring11_yjunctions}) in order to achieve  a 2D scalable design. Additionally, due to the heating of the motional degrees of freedom of the ions during these operations, auxiliary ions from a different atomic species/isotope will be required for sympathetic re-cooling of the crystal prior to the entangling gates~\cite{home_09}. The other alternative approach being considered in the community is the MUSIQC scheme, which employs a collection of elementary local units, each of which   might consist of a small ion crystal where QIP operations can be implemented according to the previous schemes. Instead of transporting ions between different trap zones, this scheme  uses photonic interconnects to probabilistically generate entangled pairs between certain communication qubits that belong to  separated elementary units~\cite{moehring_07,Inlek17,hucul15}. Whereas the  QCCD approach offers a scalable method to implement the circuit model of QIP~\cite{nielsen-book}, the  MUSIQC scheme can be considered~\cite{musicq} as a hybrid between the circuit and the cluster-state~\cite{cluster_qc} model  for QIP.

In this article, we  explore the QCCD approach towards fault-tolerant  color-code-based QEC, and focus on an implementation approach based on two-species ion crystals in high-optical access segmented traps~\cite{hoa_source} embedded in a cryogenic environment (see Fig.~\ref{Fig:trap_scheme}). In a recent work~\cite{eQual_1qubit}, a detailed exposition of the capabilities of this setup has been put forth, together with a thorough assessment of the planned experimental progress towards the QEC goals $(\mathsf{QEC}$-$\mathsf{I})$-$(\mathsf{QEC}$-$\mathsf{II})$. Resource-wise, once these goals are achieved, the next logical step  towards the completion of the QEC challenge $(\mathsf{QEC}$-$\mathsf{III})$ is to aim at the experimental implementation  of an entangling CNOT gate between two encoded logical qubits. In this work, we perform a comparative study of two resource-efficient (in terms of required physical qubits) strategies to implement a logical CNOT gate between two logical qubits encoded using the 7-qubit color code: a {\it transversal}~\cite{bombin-prl-97-180501} and a {\it lattice-surgery}~\cite{lattice_surgey_cnot_color_code} CNOT gate. We present detailed fault-tolerant schedules based on an extension of the  QEC trapped-ion toolbox described in~\cite{eQual_1qubit}. In particular, we put a focus on minimizing the required resources to achieve fault tolerance~\cite{eQual_1qubit} in the lattice-surgery protocol by leveraging the recently proposed {\it flag-based readout  schemes}~\cite{flag_based_readout} for fault-tolerant (FT) measurement of multi-qubit stabilizers. We will show that this approach leads to a considerable simplification of the trapped-ion QEC routines, as compared to the required complexity of the FT readout schemes~\cite{FTQC,shor_ft_qec,steane_ft,knill_ft,aliferis_ft_qec} discussed in~\cite{eQual_1qubit}. 
With the specific schedules and a physically-motivated  error model, which includes various quantum channels and multiple parameters  characterizing the error of the different trapped-ion operations, we perform a thorough numerical analysis to compare the performance of the two CNOT strategies based on a judicious application of an importance sampler. We hope that our work, complemented with a resource analysis, where we pay special attention to the complexity of ion-trap junction crossings, provides a useful study to guide future experimental efforts in trapped-ion QIP to achieve the aforementioned goal $(\mathsf{QEC}$-$\mathsf{III})$ in the mid term, and towards a large-scale FT quantum computer in the long term.

This article is organized as follows: In Sec.~\ref{sec:color_codes}, we discuss the $2$D triangular color codes, FT flag-based stabilizer readout procedures, and the two FT strategies to perform a logical CNOT gate that have been studied in our work.  Moreover, we  explain in detail the circuits and procedures needed to perform the  lattice-surgery scheme in a FT fashion.  In Sec.~\ref{sec:trapped_ion_implementations}, we describe the  experimental toolbox of a trapped-ion QCCD architecture, together with a physically-motivated multi-parameter error model, and present the microscopic  schedules for the two alternative CNOT strategies.  In Sec.~\ref{sec:montecarlo_sampling}, we discuss our numerical approach by extending a so-called \textit{subset-based} Monte Carlo sampler  to multi-parameter noise models.  In Sec.~\ref{Sec:schedules}, we present our numerical results for the logical error rates and resources fo the two CNOT strtegies.  Finally,  we conclude and present an outlook in Sec.~\ref{sec:conclusions}.          

\section{\bf Fault-tolerant quantum computation with triangular color codes }
\label{sec:color_codes}

\begin{figure}[t]
 \begin{centering}
  \includegraphics[width=0.95\columnwidth]{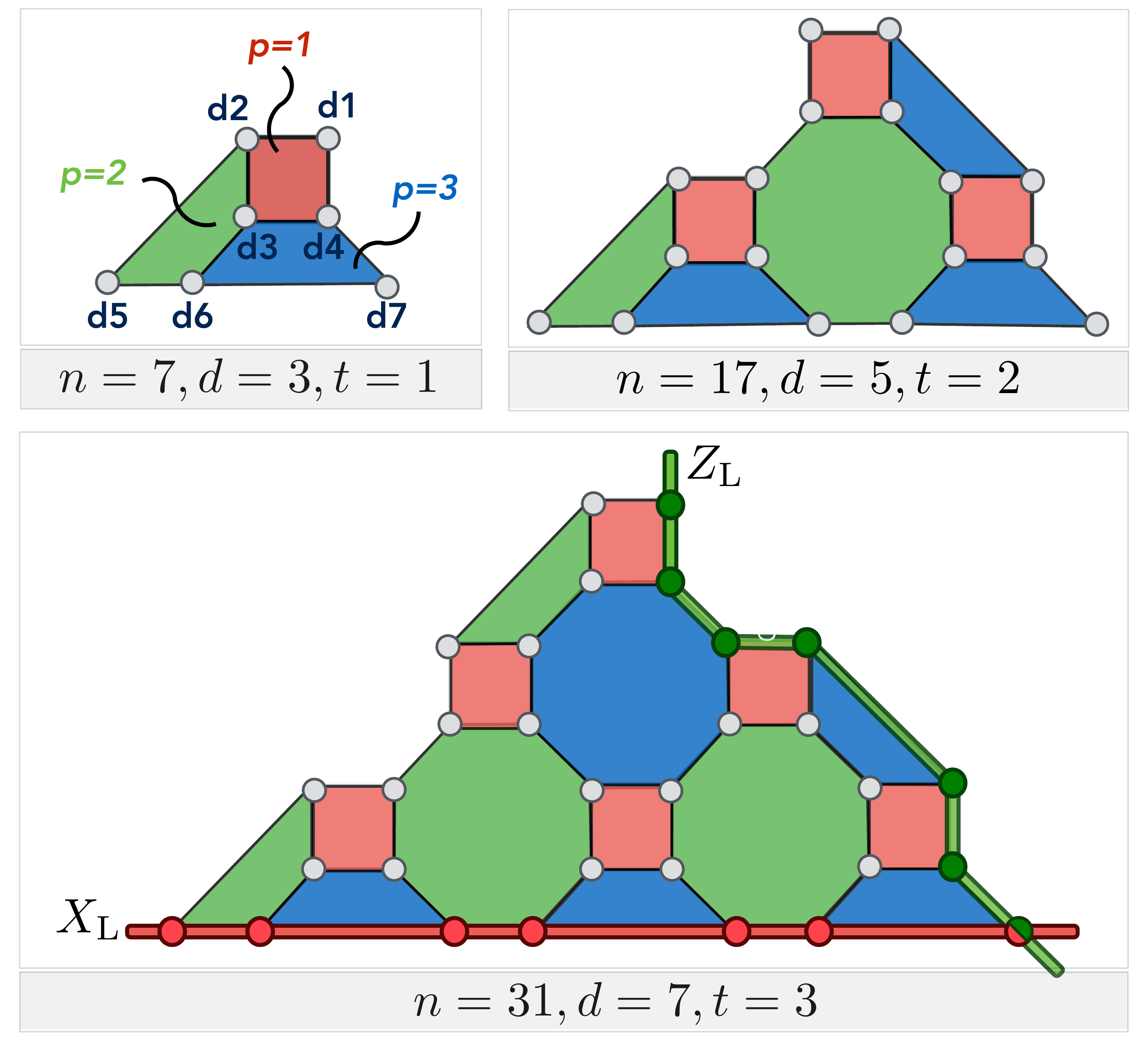}\\
  \caption{\label{Fig:largerCode} {\bf Triangular color codes on the 4.8.8 lattice:} One logical qubit is encoced into several data qubits forming a 2D triangular  code structure. The code space is defined in the usual way via $S_x^{(p)}$ and $S_z^{(p)}$ stabilizer operators~\eqref{eq:satb_larger_code}, each acting on a plaquette  $p$ that involves either 4 or 8 data qubits. We represent instances of distance $d=3,5,7$ triangular  color codes on the 4.8.8. lattice, which allow one to encode a single logical qubit with increasing levels of redundancy and protection. Logical operators such as $Z_L = \bigotimes_{i} Z_i$, and similarly the other logical single-qubit Clifford gate generators $X_L:= \bigotimes_iX_i$, $H_L:= \bigotimes_{i} H_i=\bigotimes_{i} \frac{1}{\sqrt{2}}(X_i+Z_i)$ and $S_L:= \bigotimes_{i} S_i^\dagger= \bigotimes_i\ee^{-\ii\frac{\pi}{4}(1-Z_i)}$,    can be realized transversally, i.e. in a bit-wise manner. In the lower panel we show two examples of how the logical operators can be deformed into strings of Pauli operators along the boundaries of the triangular lattice, as expressed in Eq.~\eqref{eq:log_XZ}.}
\end{centering}
\end{figure}

Let us start this section by reviewing some important properties of the triangular color code~\cite{bombin-prl-97-180501}, which is a 2D stabilizer QEC code with qubits  arranged on the $n$ vertices of a trivalent three-colorable planar lattice with  triangular boundaries.  In particular, we will focus on the so-called 4.8.8 lattice, where each vertex belongs to a square and two octagonal plaquettes with three different colors (see Fig.~\ref{Fig:largerCode}). 
In analogy with the surface code~\cite{surface_code}, triangular color codes only require local quantum processing (i.e.~only neighboring qubits need to be coupled to each other), which is very attractive from an implementation point of view. This locality becomes evident through the definition of the stabilizer generators, which are the following pair of operators   per plaquette $p$, and have a local support on the qubits located at the vertices of the plaquette
\beq
\label{eq:satb_larger_code}
S_x^{(p)}=\bigotimes_{i \in v(p)} X_i, \hspace{2ex} S_z^{(p)}=\bigotimes_{i \in v(p)} Z_i.
\eeq
Here, the product of Pauli matrices $X_i=\sigma_i^x$ and $Z=\sigma_i^z$ involves all qubits that belong to the vertices  of the corresponding  plaquette $p$, which are labelled by the set $v(p)$. This set of commuting operators can be used to define the code space, a subspace within the larger Hilbert space of the $n$ physical qubits $\mathcal{V}_{\rm code}\subset\mathcal{H}$. This subspace  is spanned by all stabilizer eigenstates of eigenvalue $+1$, namely $\mathcal{V}_{\rm code}={\rm span}\{\ket{\psi}\in\mathcal{H}: S_\alpha^{(p)}\ket{\psi}=\ket{\psi},\forall \alpha,p\}$, and can be used to encode $k$ logical qubits redundantly in  the long-range entangled states of the $n$ data qubits. Accordingly, these codes only require local quantum processing to measure these stabilizers~\eqref{eq:satb_larger_code}, which act as parity checks  that allow one to detect when a set of errors has brought the system out of the code space.

 By cutting the lattice along  a triangular boundary with the same odd number of qubits  $d$ per side,  these codes have  $n=\half d^2+d-\half$ data qubits, and  $p\in\{1,\cdots,(n-1)/2\}$ plaquettes (see Fig.~\ref{Fig:largerCode} for the corresponding codes with $d=3,5,7$ and $3,8,15$ plaquettes, respectively). Accordingly, the number of stabilizer generators is $s=2\cdot(n-1)/2=n-1$, such that the triangular codes encode a single  logical qubit $k=n-s=1$~\cite{stabiliser_qec}. Color codes belong to the class of CSS stabilizer codes~\cite{calderbank-pra-54-1098,steane-prl-77-793}. The logical Pauli operators $X_{\rm L},Z_{\rm L}$ for the encoded qubit can be chosen to be transversal, i.e.~they can be obtained from the product of the corresponding Pauli operators over all physical qubits of the lattice, $X_{\rm L}=\bigotimes_{i } X_i,  Z_{\rm L}=\bigotimes_{i } Z_i$. By  multiplying these logical operators by all the $x$-type ($z$-type) stabilizers of the same color, the effect of the logical $X_{\rm L}$ ($Z_{\rm L}$) within the code space is the same as if the operator is deformed to lie exclusively along an edge of the triangular lattice 
 \beq
 \label{eq:log_XZ}
 X_{\rm L}=\bigotimes_{i \in v(e)} X_i,  \hspace{2ex}Z_{\rm L}=\bigotimes_{i \in v(e')} Z_i,
 \eeq
  where the sets $v(e),v(e')$ label the vertices along two arbitrary edges $e,e'$ of the triangle (see the lower panel of Fig.~\ref{Fig:largerCode} for an example). Accordingly, the code distance $d$ coincides with the odd number of qubits per side of the triangle $d$, and thus scales with the size of the lattice. As occurs for other topological codes, and in contrast to the local character of the parity checks~\eqref{eq:satb_larger_code}, the quantum information of the triangular color codes depends on  global features~\eqref{eq:log_XZ}, yielding for low enough error rates an increased  robustness/protection of the encoded logical qubit, when the lattice grows, as more errors $t=(d-1)/2$ are guaranteed to be correctable.

In addition, these triangular codes have the following useful features: among the other possible three-colorable tilings of the 2D plane~\cite{bombin-prl-97-180501}, the 4.8.8. codes {\it (a)} require the minimal number of data qubits $n$ for a given code distance $d$, and {\it (b)} they are the only ones that allow for  a transversal implementation of the full Clifford group. Moreover, as occurs for the other tilings as well, {\it (c)} these codes are CSS codes and thus allow to measure and process the syndrome of phase- ($Z$) and bit-flip ($X$) errors separately. In particular, property {\it (b)} simplifies considerably the achievement of {\it fault-tolerant} (FT) quantum computation.

As mentioned above, the triangular color codes with an odd number of qubits per triangle side $d$  have the capability of correcting up to at least $t=(d-1)/2$ errors. However, the circuits to implement the required QEC cycles or logical operations will contain quantum gates that couple the qubits, or quantum gates that follow a particular measurement outcome. Such circuit elements can copy errors between various data qubits, a situation that is generic for any QEC protocol, and that can reduce, or in the worst case even entirely eliminate, the correcting power of  the code.  The concept of  { fault-tolerant} (FT)  quantum operations,  which use a circuit design that essentially forbids errors from cascading into multiple qubits during the QEC operations, is a crucial concept that underlies one of the most relevant results of QEC: the FT threshold~\cite{FTQC}. This results proves that, provided that the microscopic error rates of a FT circuit are reduced below a certain threshold, the QEC protocol yields and effective error rate for the logical qubits that decreases exponentially with the code distance.  At the expense of a  resource overhead (e.g. more redundancy by code concatenation), arbitrarily-accurate quantum computations are allowed even if one uses faulty operations, provided that these errors are kept below the threshold level. In fact, the practical interest of   topological QEC codes lies  in the high values of the thresholds~\cite{dennis_02,raussendorf07,Katzgraber_09,landahl11} as compared to other QEC strategies such as code concatenation.

To optimally benefit  from the high robustness against errors offered by topological codes, which are defined on increasingly bigger lattices with associated larger logical distances $d$, it is essential to optimize the circuit design for stabilizer readout and, in particular, to work with FT designs whenever possible. Working with non-FT syndrome readout schemes can lead to effectively reduced code distances $d'$, $d' < d$, effectively reducing the protection of logical information. We emphasize that this aspect is critical in particular for the small-scale, low-distance topological codes that will be realized experimentally in the short and mid term - here, the careful consideration of error cascading and the detailed implementation of FT designs will be of paramount importance. 

\begin{figure}[t]
 \begin{centering}
  \includegraphics[width=0.95\columnwidth]{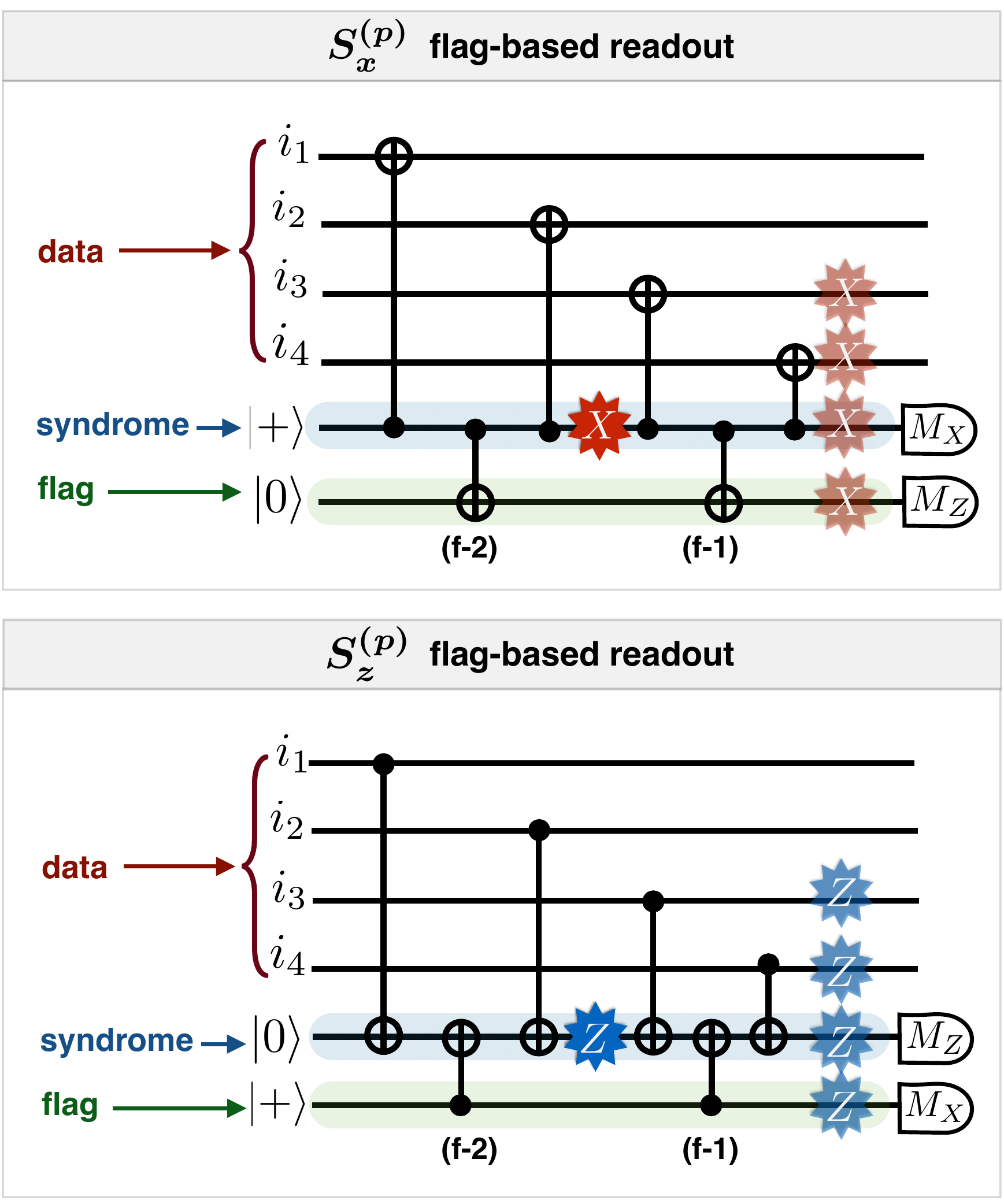}\\
  \caption{\label{Fig:flag_readout} {\bf Flag-based FT readout of weight-4 stabilisers:} (Upper panel)  Scheme for the parity-check measurement of $S_x^{(p)}=X_{i_1}X_{i_2}X_{i_3}X_{i_4}$ by the sequential application of CNOT gates, represented by a solid line with a filled and a crossed circle on the corresponding qubits. This circuit maps the stabiliser information  into the syndrome qubit, which is measured in the $X$ basis.  The data-qubit indexes $\{i_1,i_2,i_3,i_4\}$ belonging to $p$-th plaquette  are sorted in  ascending order according to our choice in Fig.~\ref{Fig:largerCode}, which will be important for the decoding described in the look-up Table~\ref{table_decoding}. A dangerous single bit-flip error is depicted by a red $X$-star at the middle of the circuit. This error would propagate into a pair of bit-flip errors on two data qubits (semi-transparent red $X$-stars), as can be seen by using the CNOT gate conjugation identity $U_{\rm CNOT}^{ c,t} X_{ c} = X_{ c}X_{ t} U_{\rm CNOT}^{ c,t}$ between control $c$ and target $t$ qubits.  To identify this dangerous error, one introduces the {\bf (f-1)} CNOT gate between the syndrome and the extra flag qubit, such that a bit-flip error  also cascades into the flag (semi-transparent red $X$-star), and  can be captured by a $-1$ measurement in the $Z$ basis, signaling that this correlated error  has indeed propagated into the code. The  {\bf (f-2)} CNOT gate is required to map correctly the stabiliser information into the syndrome qubit. (Lower panel) Analogous scheme for the parity-check measurement corresponding to stabiliser $S_z^{(p)}=Z_{i_1}Z_{i_2}Z_{i_3}Z_{i_4}$, where we use the CNOT identity $U_{\rm CNOT}^{ c,t} Z_{ t}=Z_{ c}Z_{ t}U_{\rm CNOT}^{ c,t}$. }
\end{centering}
\end{figure}

\begin{table*}
\begin{center}
\begin{tabular}{|c|c|c|c|c|c|c|c|}
\hline
\hline
\multicolumn{2}{|c|}{No flag triggered $f=+1$} &\multicolumn{2}{c|}{ Flag $S_\alpha^{(1)}$ triggered $f=-1$}& \multicolumn{2}{c|}{Flag $S_\alpha^{(2)}$ triggered $f=-1$}&\multicolumn{2}{c|}{Flag $S_\alpha^{(3)}$ triggered $f=-1$}\\
\hline
Syndrome $\boldsymbol{r}$ & Error & Syndrome $\boldsymbol{r}$ & Error & Syndrome $\boldsymbol{r}$ & Error & Syndrome $\boldsymbol{r}$ & Error \\
\hline
 \cellcolor{mygrey} \hspace{2ex}$(+1,+1,+1)$\hspace{2ex} & \cellcolor{mygrey} no error & $ \cellcolor{mygrey}\hspace{2ex}(+1,+1,+1)\hspace{2ex}$  & \cellcolor{mygrey} $f$  &  \cellcolor{mygrey}$(+1,+1,+1)$ & \cellcolor{mygrey} $f$ &  \cellcolor{mygrey}$(+1,+1,+1)$ & \cellcolor{mygrey} $f$ \\
\hline
 \cellcolor{mygrey}$(+1,+1,-1)$ &\hspace{2ex}   \cellcolor{mygrey}$i=7$\hspace{2ex} &  $(+1,+1,-1)$& $i=7,f$& \cellcolor{mygrey}$\hspace{2ex}(+1,+1,-1)\hspace{2ex}$ & \cellcolor{mygrey}\hspace{2ex}  $i=5,6$  \hspace{2ex} & \cellcolor{mygrey}$\hspace{2ex} (+1,+1,-1) \hspace{2ex}$ &  \cellcolor{mygrey}$i=7\phantom{,f}$  \\
\hline
 \cellcolor{mygrey}$(+1,-1,+1)$ &    \cellcolor{mygrey}$i=5$ &\cellcolor{mygrey}{$(+1,-1,+1)$}&  \cellcolor{mygrey}$i=3,4$ & $(+1,-1,+1)$ &   $i=5,f$    & \cellcolor{mygrey}$(+1,-1,+1)$ & \cellcolor{mygrey}\hspace{2ex}  $i=6,7$ \hspace{2ex} \\
\hline
 \cellcolor{mygrey}$(+1,-1,-1)$ &   \cellcolor{mygrey}$i=6$ &  $(+1,-1,-1)$ &  $i=6,f$  &\cellcolor{mygrey}$(+1,-1,-1)$ &  \cellcolor{mygrey}$i=6\phantom{,f}$ & $(+1,-1,-1)$ &  $i=6,f$    \\
\hline
 \cellcolor{mygrey}$(-1,+1,+1)$ &  \cellcolor{mygrey}$i=1$ & \cellcolor{mygrey}$(-1,+1,+1)$  &  \cellcolor{mygrey}$i=1\phantom{,f}$ & $(-1,+1,+1)$ &  $i=1,f$ & $(-1,+1,+1)$  &  $i=1,{f}$    \\
\hline
 \cellcolor{mygrey}$(-1,+1,-1)$ &   \cellcolor{mygrey}$i=4$ &   \cellcolor{mygrey}$(-1,+1,-1)$ &  \cellcolor{mygrey}$i=4\phantom{,f}$  & $(-1,+1,-1)$ &  $i=4,f$   &  $(-1,+1,-1)$ &  $i=4,{f}$    \\
\hline
 \cellcolor{mygrey}$(-1,-1,+1)$ &  \cellcolor{mygrey}$i=2$ &   $(-1,-1,+1)$ &  $i=2,f$  & \cellcolor{mygrey}$(-1,-1,+1)$ &  \cellcolor{mygrey}$i=2\phantom{,f}$ & $(-1,-1,+1)$ &  $i=2,f$   \\
\hline
 \cellcolor{mygrey}$(-1,-1,-1)$ &  \cellcolor{mygrey}$i=3$ &   $(-1,-1,-1)$ &  $i=3,f$   & $(-1,-1,-1)$ &  $i=3,f$    & \cellcolor{mygrey}$(-1,-1,-1)$ &  \cellcolor{mygrey}$i=3\phantom{,f}$  \\
\hline
\end{tabular}
\end{center}
\caption{{\bf Look-up table for the decoding of the flag-based QEC with the $\boldsymbol{d=3}$ color code:}  If the flagged measurement of  stabiliser  $S_\alpha^{(p)}$ for ${\alpha}=\{x,z\}$ and plaquette $p$     triggers the flag  $f=-1$, the decoding   depends on the subsequent  un-flagged measurements of all three   conjugate stabilisers $\left\{S_{\beta}^{(1)},S_{\beta}^{(2)},S_{\beta}^{(3)}\right\}$, where  $\beta=\{z,x\}$  (i.e. columns 2, 3, and 4). In each of these columns,  the sub-column $\boldsymbol{r}=(r_1,r_2,r_3)$ labels the possible outcomes $r_p=\pm 1$ of these  $\beta$-type stabiliser readouts, whereas the sub-column Error lists the qubit indexes $I$ of the most-like correction $\otimes_{i\in I}\sigma^{\alpha}_i$. On the other hand, if the flag is not triggered  $f=+1$ but a syndrome error  in stabiliser ${\alpha}=\{x,z\}$  is detected, one should use  the three values of the same-type  $\beta=\{x,z\}$ stabilisers to identify the error (i.e.  column 1). The combined values $(f,\boldsymbol{r})$ allow one to identify either the most-likely measurement error (i.e. no error on the data qubits), weight-1 error, or  dangerous weight-2 error of type $\alpha$ that has cascaded into the data qubits, all of which are marked by grey cells. We assume that other possible stabiliser outcomes are due to a weight-2 error stemming from a combination of a measurement error on the flag qubit and a single-qubit error on one of the data qubits. These errors are denoted by a white cell.  }
\label{table_decoding}
\end{table*}

\subsection{Fault-tolerant flag-based schemes for quantum error correction (QEC)}
\label{sec:flag_readout}
One of the crucial operations in QEC is the readout of the parity checks, e.g.~the plaquette operators in Eq.~\eqref{eq:satb_larger_code} for the color code. Since the data qubits cannot be directly measured, the readout  requires mapping the stabilizer information onto a set of ancillary qubits that can be the projectively measured without compromising the logical quantum information stored in the code space. There are well-known strategies for the FT readout of the stabilizers~\cite{shor_ft_qec,steane_ft,knill_ft}, all of which use additional ancillary qubits to map the stabilizer information while avoiding  a non-FT propagation of errors. Considering the particular QCCD trapped-ion implementation where the ion ancillary qubits can be shuttled between different trapping zones and thus re-utilized for various parity-check measurements, one can minimize the required qubit resources~\cite{eQual_1qubit}  by considering  Shor-type FT readouts~\cite{shor_ft_qec,aliferis_ft_qec} where the ancillary qubits are prepared in an entangled GHZ-type state (i.e.~cat-state FT readout). However, the complexity of the QEC schedules for the trapped-ion QCCD implementation is already considerable for the $d=3$ code~\cite{eQual_1qubit}, and increases even further  for a FT readout of higher-distance color codes  involving octogonal plaquettes that require the preparation and validation of cat states of even larger number of qubits (see Fig.~\ref{Fig:largerCode}).

In this subsection, we review a recent  FT readout  for  $d=3$ codes~\cite{flag_based_readout} that minimizes the qubit overhead by avoiding the use of ancillary cat states. Instead, this scheme  adds an extra {\it flag qubit} to the  {\it syndrome qubit} onto which the stabiliser information is mapped. The ancillary flag qubit, in turn, is used to ensure the FT character of the readout.  We  focus from now onwards on the 7-qubit color code, although we emphasize that this  strategy can be generalized to larger-distance color codes~\cite{flag_readout_arb_distance}. The main idea of the flag-based FT readout is to use a single bare flag qubit, and couple it to the syndrome qubit to gather information on whether or not multiple errors have cascaded onto the data qubits which,  if  unnoticed, would compromise fault tolerance of the readout. The flag qubit by itself does not suffice to correct for these correlated errors. However, if combined with subsequent stabilizer measurements using the syndrome qubit, it can be used to unequivocally identify and correct the correlated errors achieving the desired fault tolerance.

To achieve this goal for the  7-qubit code of Fig.~\ref{Fig:largerCode}, one first has to identify the non-FT  propagations where a single  error on any of the ancillary qubits, or two-qubit gates involved, does cascade into two errors in the data qubits. We must thus consider the two types of parity-check measurements of Fig.~\ref{Fig:flag_readout}, where we depict these dangerous  propagations  for  bit-flip and phase-flip errors in the syndrome qubit. To detect these events, the {\bf (f-1)} CNOT gate between the syndrome and flag qubits forces the corresponding $X$($Z$) error to cascade also into the flag qubit, such that it can be  detected by a measurement in the $Z$($X$) basis yielding $-1$ instead of the expected $+1$ (i.e. flag triggering). Note that  the additional {\bf (f-2)} CNOT gate  of Fig.~\ref{Fig:flag_readout} is required to correctly map the stabiliser information into the syndrome qubit, given the action of the  {\bf (f-1)} CNOT gate. In the absence of {\bf (f-2)}, $\ket{+}_{\rm r}\ket{0}_{\rm f}\ket{\psi}_{\rm d}$ would evolve  into $\half\ket{+}_{\rm r}(\ket{0}_{\rm f}+S^{(p)}_x\ket{1}_{\rm f})\ket{\psi}_{\rm d}+\half\ket{-}_{\rm r}(\ket{0}_{\rm f}-S^{(p)}_x\ket{1}_{\rm f})\ket{\psi}_{\rm d}$ under the upper circuit of Fig.~\ref{Fig:flag_readout}. On the other hand, by applying the {\bf (f-2)} CNOT gate, one corrects the effect of {\bf (f-2)}   $\half\ket{+}_{\rm r}\ket{0}_{\rm f}(1+S^{(p)}_x)\ket{\psi}+\half\ket{-}_{\rm r}\ket{0}_{\rm f}(1-S^{(p)}_x)\ket{\psi}_{\rm d}$, and thus obtains the desired mapping of the $\pm 1$   information of $S^{(p)}_x$ by  measuring the readout qubit in the $X$ basis. A similar argument applies to the $Z$ stabilisers of the lower circuit of  Fig.~\ref{Fig:flag_readout}.

\begin{figure*}
 \begin{centering}
  \includegraphics[width=1.7\columnwidth]{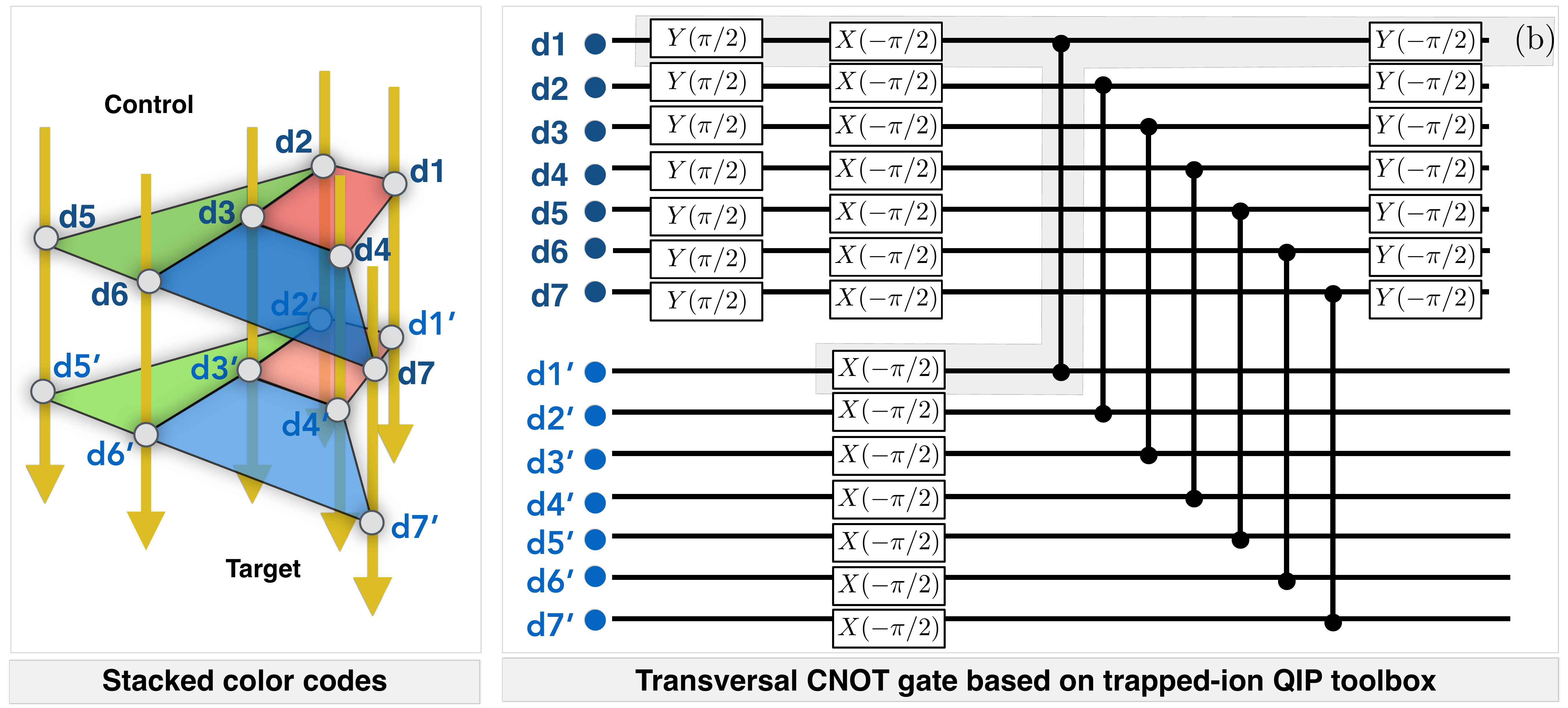}\\
  \caption{\label{Fig:7qubitCode_cnot_transv} {\bf Transversal CNOT gate between two 7-qubit color codes:} (left panel) Two logical qubits are each embedded in 7 data qubits from a couple of  triangular planar codes. The transverse CNOT operator can be visualized by stacking the two logical qubits forming a bilayer, such that each  pair of equivalent data qubits  is coupled by a bare CNOT gate (yellow arrows). (Right panel) Transversal CNOT circuit based on fully-entangling two-qubit MS gates $X^2_{i,j}$ defined below Eq.~\eqref{eq:MS_gate}, and represented by a solid line with two filled circles (not to be confused with a controlled-phase gate). Additionally,  single-qubit rotations along the $x$- and $y$-axis, $X(-\pi/2)$ and $Y(\pm\pi/2)$ defined below Eq.~\eqref{eq:rot_z}, are required to obtain the transversal CNOT. The shaded grey region, labelled by (b), includes a set of operations that will be used repeatedly in the modular microscopic trapped-ion schedules of the following sections.}
\end{centering}
\end{figure*}

 The important point of this scheme is that,  whenever the  flag is triggered,  we can ascertain that  an error must have  occurred. In that case, we can  identify which of the possible one- or two-qubit errors, if any, has indeed propagated into the data qubits   by a subsequent measurement of all the $X$- or $Z$-type stabilisers with un-flagged circuits (i.e. using only the bare readout qubit and switching off the {\bf (f-1)} and {\bf (f-2)} CNOT gates). At FT level  $t=1$,  where one only considers   events with at most one faulty circuit element, Table~\ref{table_decoding} shows how  to correct any of the possible errors, including the dangerous weight-2 errors. On the other hand, if the flag is not triggered but the syndrome measurement signals a $-1$ stabiliser error, we can be certain that an error has occurred without any  two-qubit error  propagating into the data. This entitles us again, at FT-1  level,  to  use the un-flagged circuits to extract the syndrome by measuring all stabilisers, and to correct the possible single-qubit error using the same prescription as in the standard  FT QEC schemes, as  detailed in Table~\ref{table_decoding}. In both cases, the readout finishes after the un-flagged measurements. The last scenario is that neither the flag is  triggered, nor the syndrome qubit of the flagged circuit signals any stabiliser  error. In that case, we can ascertain that no single error has occurred at all, and proceed to the flag-readout of the next stabiliser.

In Sec.~\ref{sec:trapped_ion_implementations}, we will present detailed microscopic schedules for the realization of this readout scheme using a QCCD trapped-ion architecture. To minimize the complexity of the trapped-ion procedures described in Sec.~\ref{sec:schedules_flag_qec}, the ordering in the  measurement of the different stabilisers is $S_x^{(1)}\to S_z^{(1)}\to S_x^{(2)}\to S_z^{(2)}\to S_x^{(3)}
 \to S_z^{(3)}$.
 As it will turn out, the flag-based scheme simplifies considerably the trapped-ion cat-based readout protocols described in~\cite{eQual_1qubit} and is an important improvement for the near-term achievements of the goals $(\mathsf{QEC}$-$\mathsf{I})$-$(\mathsf{QEC}$-$\mathsf{II})$. In the next subsection, we discuss how to merge this scheme with FT implementations of a logical CNOT gate, addressing thus part of the  goal $(\mathsf{QEC}$-$\mathsf{III})$.

\subsection{Fault-tolerant schemes for CNOT gates}
\label{sec:cnot_gates}

As mentioned above, one of the interesting features of triangular 4.8.8 color codes is that  the full Clifford group can be implemented  at the logical level by applying the corresponding  operations in a bit-wise manner, i.e. {\it transversally} \cite{bombin-prl-97-180501}. For single-qubit Clifford gates, it suffices to apply the corresponding unitaries to all of the $n$ data qubits of a single code (see the caption of Fig.~\ref{Fig:largerCode}). The remaining ingredient to implement the full Clifford group is the two-qubit CNOT gate,  and we    review in this section two possible strategies. We first describe its transversal realization by applying CNOT gates between each pair of equivalent data qubits belonging to the control and target blocks.  As shown in Sec.~\ref{Sec:schedules} below, this scheme leads to a trapped-ion schedule of considerable complexity already for the smallest color code. In particular, it has a relatively large overhead of manipulations that shuttle the ions across  junctions of the segmented trap. It is likely that this overhead  impedes its extensibility to larger-distance codes. Therefore, we also discuss below   an alternative strategy based on {\it lattice surgery}~\cite{Horsman-njp-2012} with color codes~\cite{lattice_surgey_cnot_color_code}.

\vspace{1ex}

\subsubsection{Transversal CNOT gate operation}
The CNOT gate  between the $i$-th control  and the $i'$-th target qubits is $U_{\rm CNOT}^{i,i'}=\frac{1}{2}(1+Z_i)1_{i'} +\frac{1}{2}(1-Z_i)X_{i'}$. A CNOT gate between two logical qubits encoded into  two different sets of data qubits $I=\{1,2,\cdots \}$ and $I'=\{1',2',\cdots \}$ with $|I|=|I'|=n$, can be constructed in a bit-wise fashion as follows  $U_{\rm CNOT}=\prod_{i=1}^nU_{\rm CNOT}^{i,i'}$.  Using the transversal logical operators, one can easily verify that the effect of this transversal CNOT at the logical level is ${U_{\rm CNOT}} (X_L\otimes \mathbb{I}){U_{\rm CNOT}}=X_L\otimes X_L$,  ${U_{\rm CNOT}} (\mathbb{I}\otimes X_L){U_{\rm CNOT}}=\mathbb{I}\otimes X_L$,  and ${U_{\rm CNOT}} (\mathbb{I}\otimes Z_L){U_{\rm CNOT}}=Z_L\otimes Z_L$, ${U_{\rm CNOT}} (Z_L\otimes \mathbb{I}){U_{\rm CNOT}}=Z_L\otimes \mathbb{I}$, which realizes the required transformation of basis operators under conjugation by the CNOT gate, and thus proves the validity of the transversal construction. This transversal operation does not take the state out of the code subspace, and by construction enjoys a FT character. Even in the event of a two-qubit error due to a faulty CNOT, or a single-qubit error that cascades into a two-qubit error through the CNOT, these errors correspond to different logical blocks, warranting thus the FT nature of the scheme.

One can visualize this transversal operation by stacking the two logical qubits on top of each other~\cite{dennis_02}, and coupling the respective equivalent data qubits via CNOT gates  (see the left panel of Fig.~\ref{Fig:7qubitCode_cnot_transv}). Note, however,  that  this stacked perspective is a mere visualization for  the envisioned trapped-ion QCCD, where all logical qubits will belong to the same 2D architecture. In order to bring the physically-equivalent qubits close to each other to implement the corresponding physical CNOT gate operations, one must apply a sequence of crystal reconfigurations, the complexity of which   increases dramatically as the distance of the code grows. Therefore, the transversal realization of the CNOT gate compromises one of the central appealing features of topological codes, namely the requirement of \textit{local} quantum processing. 

\begin{figure}[t]
\begin{centering}
  \includegraphics[width=0.9\columnwidth]{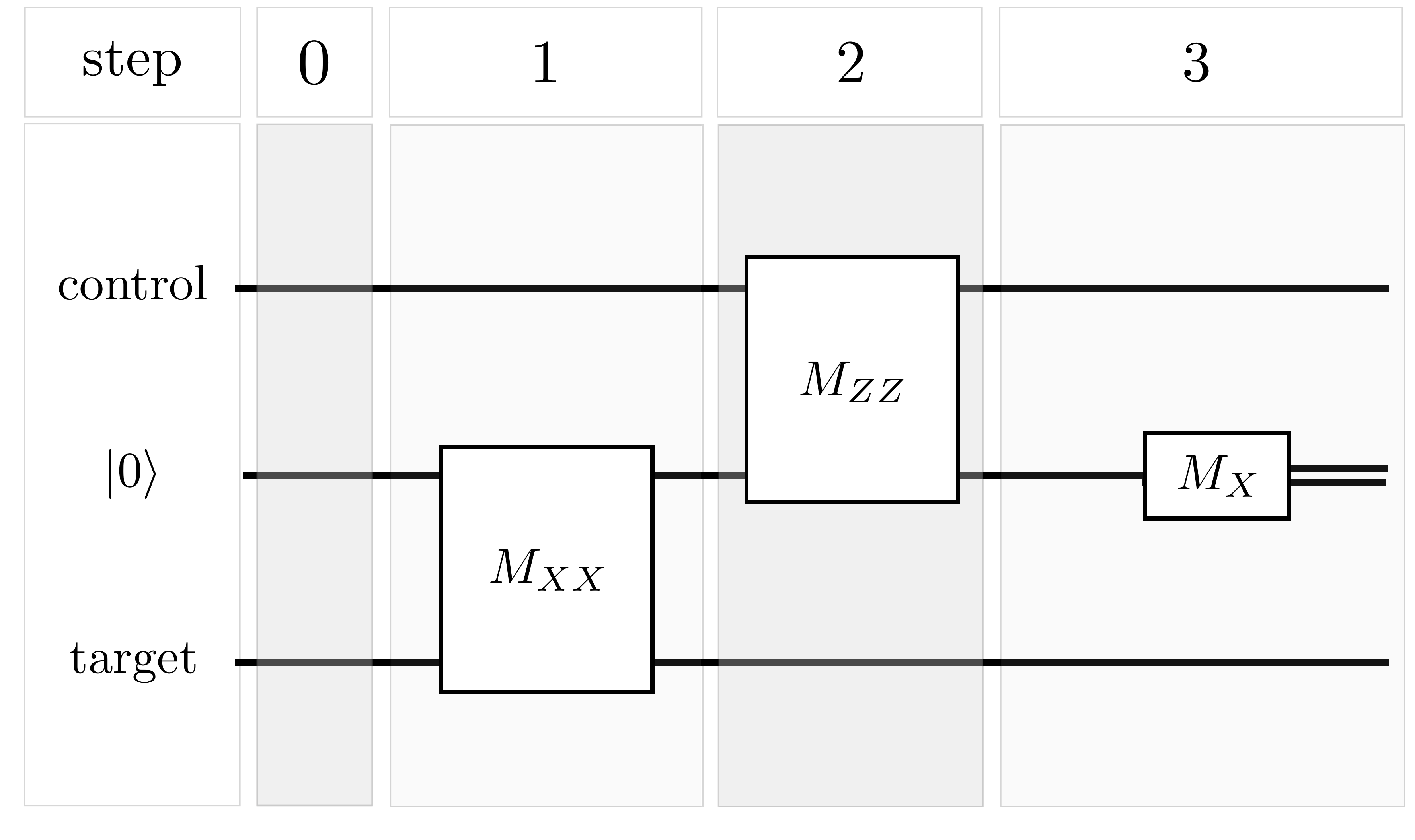}\\
\caption{\label{fig:circ_latt_surg_general} {\bf Teleportation-based circuit for  implementing a CNOT gate:} $M_{XX}$ ($M_{ZZ}$) corresponds to the measurement of the joint operator $X^t X^a$ ($Z^c Z^a$).  Intuitively, the $M_{XX}$ operation pushes a $Z$ ($|0 \rangle$) state from the target to the control while the $M_{ZZ}$ operation pushes a $X$ ($| + \rangle$) state from the control to the target.  The final measurement in the $X$ basis, $M_X$,  is necessary to decouple the ancillary qubit from the other two.  Since the outcome of each measurement operation is random, the final state must be corrected conditionally upon these outcomes: A $Z$ operation must be performed on the control if the total parity of $M_{XX}$ and $M_X$ is odd.  Likewise, an $X$ operation must be performed on the target if the parity of $M_{ZZ}$ is odd.}
\end{centering}
\end{figure}

\vspace{1ex}
\subsubsection{Lattice-surgery CNOT gate operation}

In this section, we describe in detail a protocol to realize a lattice-surgery CNOT gate between two $d=3$ color-code qubits by means of local operations~\cite{lattice_surgey_cnot_color_code}. We focus on developing a careful new FT design for the lattice-surgery approach, as well as on minimizing the required resources in terms of data qubits and  time steps of the protocol, together with  the  optimization of the procedures in view of the trapped-ion QCCD capabilities. 

The generic teleportation-based circuit to perform such a lattice-surgery CNOT is depicted in Fig.~\ref{fig:circ_latt_surg_general}, where the control and the target qubits start in an arbitrary state, while an additional ancillary qubit is initialized  in  state $|0 \rangle$.  This circuit is equivalent to a CNOT gate, up to conditional Pauli corrections based on the outcomes of  measurements of the joint  operators $X^{a}X^t$, $Z^cZ^{a}$, and the single ancillary operator $X^a$, where the superscript denotes the qubit where the operator is acting.  This circuit identity can be verified, as mentioned above for the transversal CNOT, by monitoring the evolution of the canonical operators on the control ($Z^c$ and $X^c$) and  target ($Z^t$ and $X^t$) qubits  (see Table \ref{table:evolution_operators_CNOT}). In the stabilizer formalism, the effect of a CNOT can be visualized by how the canonical operators transform: $Z^c \rightarrow Z^c$, $X^c \rightarrow X^c X^t$, $Z^t \rightarrow Z^c Z^t$, $X^t \rightarrow X^t$.  As seen in Table \ref{table:evolution_operators_CNOT}, measuring the operator $X^t X^a$ has no effect on $X^t$, but collapses $Z^a$ and turns $Z^t$ into $Z^t Z^a$.  The sign of the new stabilizer $X^t X^a$ is completely random ($\epsilon_1$ has equal probabilities of being $+1$ or $-1$.).  In the next step, measuring the operator $Z^c Z^a$ has no effect on $Z^c$, but turns $X^c$ into $X^c X^t X^a$.  Likewise, the sign of the new stabilizer $Z^c Z^a$ is completely random.  Finally, in the third step, measuring the operator $X^a$ decouples the ancillary qubit from the other two.  Just like in the previous steps, the sign of the new stabilizer is completely random.  The total effect of the circuit is therefore: $Z^c \rightarrow Z^c$, $X^c \rightarrow \epsilon_1 \epsilon_3 X^c X^t$, $Z^t \rightarrow \epsilon_2 Z^c Z^t$, $X^t \rightarrow X^t$.

\begin{figure}[t]
 \begin{centering}
  \includegraphics[width=0.9\columnwidth]{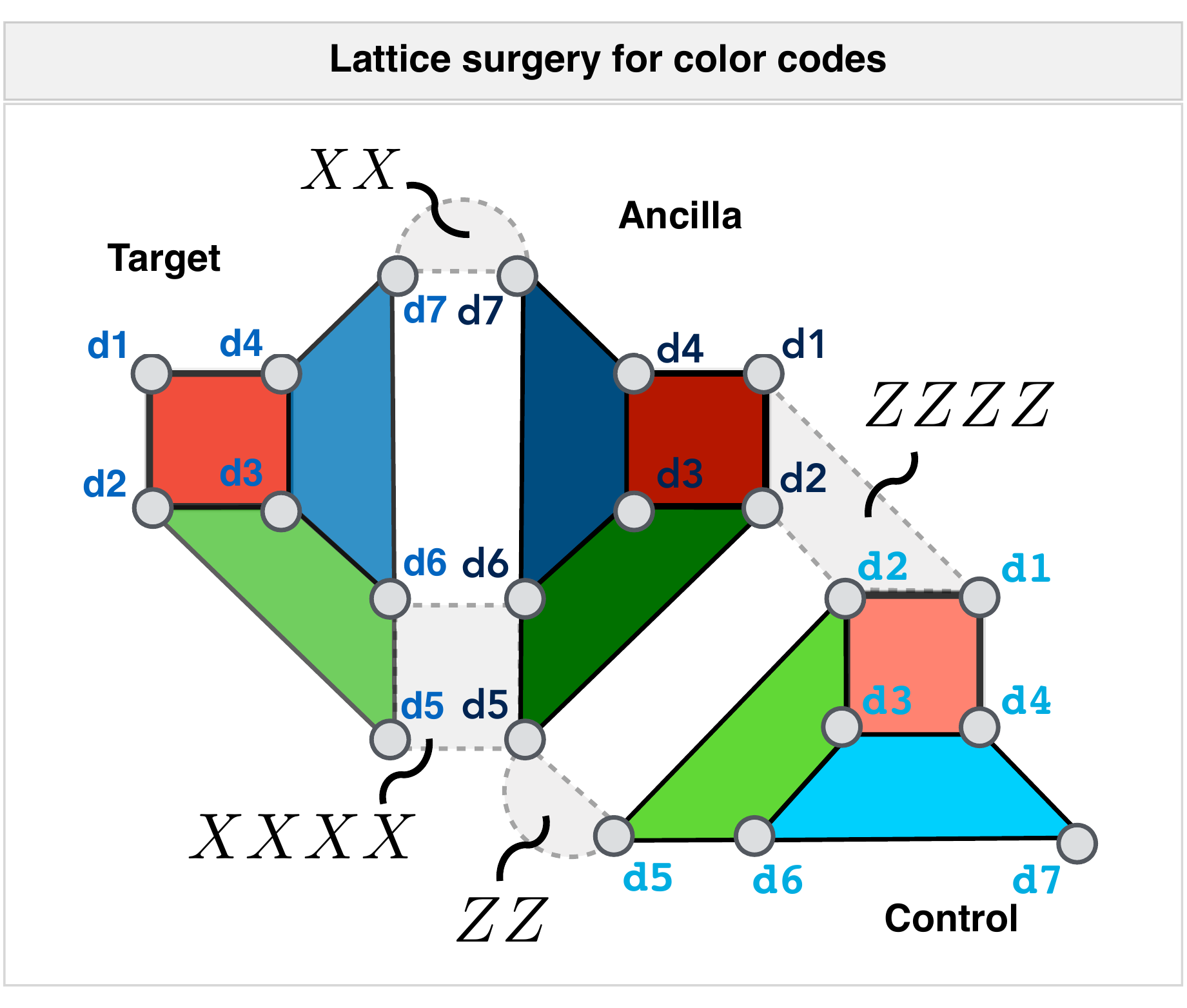}\\
  \caption{\label{Fig:Lattice_Surgery_CC.pdf} \textbf{Lattice-surgery based CNOT gate between two $d=3$ color code logical qubits:} scheme for  the 2D arrangement of the three logical qubits required for the realization of a logical CNOT between the control and target qubit, via intermediate coupling to an ancillary logical qubit. The gray-shaded 2- and 4-qubit operators are measured in order to merge the ancilla qubit with the target and control qubit, respectively. This yields the desired measurements of the joint logical operators $X^{a}_{\rm L}X^t_{\rm L }$, $Z^c_{\rm L}Z^{a}_{\rm L}$.}
\end{centering}
\end{figure}

\begin{table*}
\begin{tabular}{|c|c|c|c|c|}
\hline
  Time step & 0 & 1 (i.e. M$_{\rm XX}$) & 2 (i.e. M$_{\rm ZZ}$) & 3 (i.e. M$_{\rm X}$) \\
  \hline	
   \hline		
 Control & $Z^c$ & \hspace{1ex}$Z^c $\hspace{1ex} & $Z^c$ & \hspace{1ex}$Z^c $\hspace{1ex} \\ \hline
   & $ X^c$ & \hspace{1ex}$ X^c= \varepsilon_1 X^t X^c X^a$\hspace{1ex} & $\varepsilon_1 X^t X^c X^a$ & \hspace{1ex}$ \varepsilon_1 X^t X^c X^a = \varepsilon_1 \varepsilon_3 X^c X^t$\hspace{1ex} \\ \hline
  \hline
Target & \hspace{1ex}$Z^t  = Z^t Z^a$\hspace{1ex} & $Z^t Z^a$ & \hspace{1ex}$Z^t Z^a= \varepsilon_2 Z^c Z^t$\hspace{1ex} & $\varepsilon_2 Z^c Z^t$\\
  \hline 
 & \hspace{1ex}$X^t$ \hspace{1ex} & $ X^t$ & \hspace{1ex} $ X^t$\hspace{1ex} & $ X^t$\\
  \hline  
     \hline
Ancilla &  $Z^a$ & $\varepsilon_1 X^t X^a$ & $\varepsilon_2 Z^c Z^a$ & $\varepsilon_3 X^a$ \\ \hline
\end{tabular}
\caption{\label{table:evolution_operators_CNOT} {{\bf Teleportation-based circuit and stabilisers:} Evolution of the $Z$, $X$ operators of the control and target qubits, and the $Z$ operator of the ancilla qubit throughout the different steps of the teleportation-based  CNOT circuit of Fig.~\ref{fig:circ_latt_surg_general}} (see the discussion in the main text). Here, we have introduced $\epsilon_1,\epsilon_2,\epsilon_3$ denoting the $\pm 1$ results of the three consecutive measurements.}
\end{table*}

\begin{table*}
\begin{tabular}{|c|c|c|c|}
\hline
  Time step & 0 & Merging in M$_{\rm XX}$ & Splitting in M$_{\rm XX}$  \\
  \hline
   \hline
  Target &  $S_{{\rm t},z}^{(3)}=S_{{\rm t},z}^{(3)}S_{{\rm a},z}^{(3)}$ & $S_{{\rm t},z}^{(3)}S_{{\rm a},z}^{(3)}$ & $S_{{\rm t},z}^{(3)}S_{{\rm a},z}^{(3)}=\epsilon_3S_{{\rm c},z}^{(3)}$  \\
    \hline
     & $Z_{{\rm L}}^{\rm t}=Z_{{\rm L}}^{\rm t}Z_{{\rm L}}^{\rm a}$ & $Z_{{\rm L}}^{\rm t}Z_{{\rm L}}^{\rm a}$ & $Z_{{\rm L}}^{\rm t}Z_{{\rm L}}^{\rm a}$  \\
       \hline
      &$X_{{\rm L}}^{\rm t}$ & $X_{{\rm L}}^{\rm t}$ & $X_{{\rm L}}^{\rm t}$  \\
      \hline  
      \hline
       Ancilla &$S_{{\rm a},z}^{(3)}$ & $\epsilon_4X_4$ & $\epsilon_3 S_{{\rm a},z}^{(3)}$  \\
        \hline
      & $Z_{{\rm L}}^{\rm a}$ & $\epsilon_2X_2=\epsilon_2\epsilon_4X_2X_4$ & $\epsilon_2\epsilon_4X_2X_4=\epsilon_2\epsilon_4X_{\rm L}^{\rm t}X_{\rm L}^{\rm a}$  \\      \hline		
\end{tabular}
\caption{\label{table:evolution_operators_CNOT_bis} { {\bf Evolution of the relevant stabilizers during the M$_{\textrm{XX}}$ step}}.  The third $Z$ stabilizers are the only ones that collapse during the merging process, but their product ($S_{t,z}^{(3)} S_{a,z}^{(3)}$) remains well defined.  The splitting process recovers these two stabilizers.  Since they were temporarily undefined, after measuring them again, their eigenvalues are random but equal, at least in an error-free scenario where the eigenvalue of their product remains $+1$.  The new eigenvalues are given by $\epsilon_3$.  We denote the outcomes of the weight-2 and weight-4 merging operators by $\epsilon_2$ and $\epsilon_4$, respectively.}
\end{table*}

The same circuit can be applied at the logical level with encoded qubits, which is particularly convenient for topological codes where the logical $Z$ and $X$ operators can be expressed as transversal chains of   operators, respectively, acting on data qubits that lie at any of the lattice boundaries (see lower panel of Fig.~\ref{Fig:largerCode}). Therefore, by an appropriate choice of the support of these logical operators, the measurement of  the joint  operators $X^{a}_{\rm L}X^t_{\rm L }$, $Z^c_{\rm L}Z^{a}_{\rm L}$ only requires local quantum processing between neighboring data qubits in the 2D layout (see Fig.~\ref{Fig:Lattice_Surgery_CC.pdf}), overcoming the limitations of the   transversal  CNOT gate.  Note, however, that a direct measurement of  the joint logical operators quickly becomes unfeasible as the code distance increases (e.g. for a distance-3 code, one would have a weight-$6$ operator, requiring thus a $6$-qubit cat state to achieve a fault-tolerant readout).  To avoid the associated complexity, as pointed out also in Ref.~\cite{lattice_surgey_cnot_color_code}, it is possible to decompose the joint logical operator measurement into the sequential readout of  lower-weight operators, such that the total combined parity of their outcomes yields the desired FT measurement.  As shown below, an additional important simplification is that these lower-weight operators can be measured fault-tolerantly using a single bare qubit  in the case of the 7-qubit color code, avoiding the overhead of cat-based or flag-based methods.

\begin{figure*}[b]
\begin{centering}
  \includegraphics[width=1.7\columnwidth]{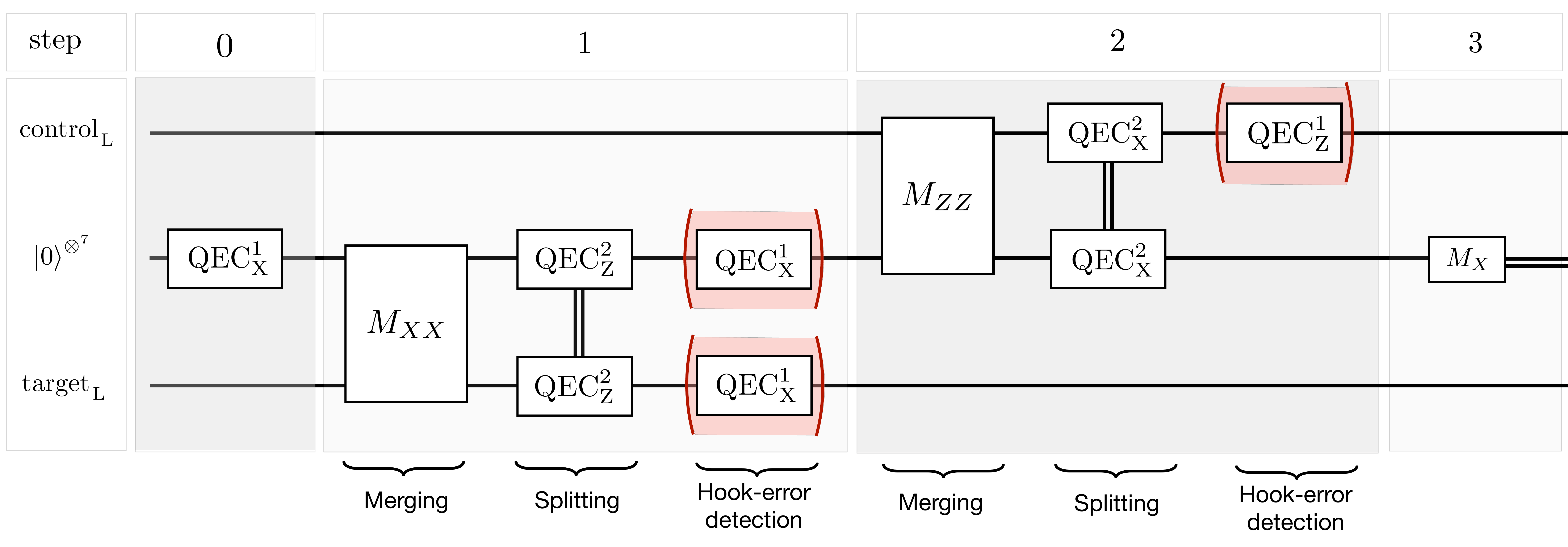}\\
\caption{\label{fig:latt_surg_full} {\bf Full FT circuit for the lattice-surgery-based CNOT:}  The ancillary logical qubit is initialized in the product state $|0 \rangle ^{\otimes 7}$.  A round of $X$-type stabilizer measurement QEC$_{\rm X}^1$ is first performed to project the ancilla state to $|0 \rangle _L$.  It is safe to do this by only a single round of $X$ stabilizer measurements, since single-qubit $X$ errors can be caught at a later stage.  The fault-tolerant implementation of the merging process, $M_{XX}$ ($M_{ZZ}$), is shown in detail in Figure \ref{fig:FT_MXX}.  Notice that during the splitting process the error syndromes for each logical qubit involved in the gate must be shared via classical communication to ensure that the procedure is fault tolerant.  Also, since the stabilizers during the splitting process are measured using flags, it might be necessary to measure the conjugate stabilizers  in the event of a flag being triggered (hook-error detection sub-step).  The red dashed regions  denote  that the operations inside them  are only conditionally applied.  The superscripts denote the number of times the stabilizers need to be measured.  During the hook-error detection stage, the stabilizers can be measured only once and the circuit construction does not require the previous FT constraints.  In the trapped-ion realization, as detailed in sections below, we use a single ancillary qubit and two 5-qubit entangling gates in order to minimize the complexity of the microscopic schedules.  This still ensures that the overall protocol is FT at level $t=1$ since these operations will be performed only if an error has already occurred~\cite{flag_based_readout}.}
\end{centering}
\end{figure*}

\begin{figure*}[t]
\begin{centering}
  \includegraphics[width=1.9\columnwidth]{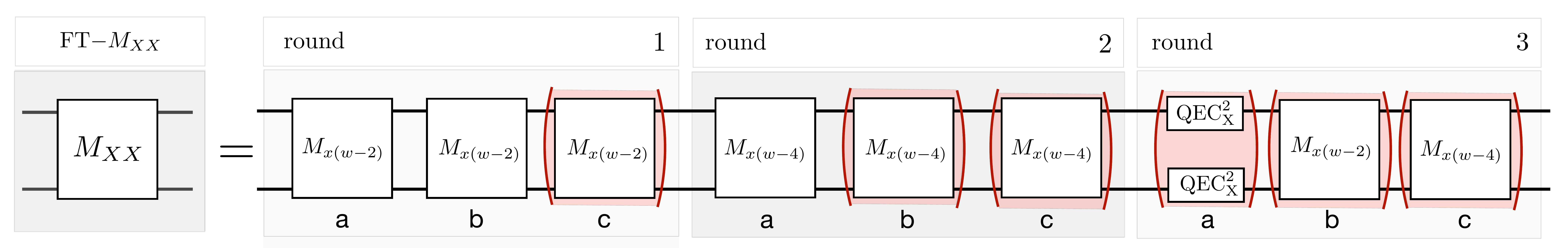}\\
\caption{\label{fig:FT_MXX} {\bf FT circuit to measure the joint logical operator $X_{\rm L} ^t X_{\rm L} ^a$:} $M_{x(w-2)}$ and $M_{x(w-4)}$ refer to the measurement of the merging operators of weight-2 and weight-4 depicted in Fig.~\ref{Fig:Lattice_Surgery_CC.pdf}.  QEC$_X ^2$ denotes two rounds of $X$-type stabilizers, which are necessary to distinguish between $Z$ errors that occurred before and after the measurement of the merging operators, as explained in the main text and illustrated in Fig.~\ref{fig:Lattice_Surgery_CC_error.pdf}.  The merging operators have to be measured at least twice to account for detection errors.  For each operator, if the two first measurement outcomes coincide, we trust them. Otherwise,  if the two outcomes disagree, the operator is measured a third time (the red dashed regions  denote  that the operations inside them  are only conditionally applied). This scheme is tolerant to a single detection fault.  Notice that measuring $X$ stabilizers does not affect the merging action of the $M_{x(w-2)}$ and $M_{x(w-4)}$ operators, since they all commute.  Splitting is only caused by measuring the $Z$ stabilizers. For the $Z$-type joint logical operator $Z_{\rm L} ^a Z_{\rm L} ^c$, one uses an analogous FT circuit with the roles of the $X$ and $Z$ basis interchanged.  Measuring the $X$-type stabilizers must be done in a FT fashion, which requires at least $2$ rounds.  In the first round, the stabilizers are measured with the flag-based readout depicted in Fig.~\ref{Fig:flag_readout}.  If no error is detected during the first round, we stop and trust the result.  Otherwise, we measure the stabilizers a second time to account for detection errors and trust the outcomes of the second round.  Following our low-resource philosophy, since an error has already occurred, during the second round the stabilizers are measured without the previous FT constraints.  If a flag is triggered during the first round of stabilizer measurements, implying that an $X$ error might have propagated from the ancilla, this information is passed on to the splitting sub-step, where the $Z$ stabilizers are measured, in order to correctly interpret the syndrome.}
\end{centering}
\end{figure*}

 \begin{figure}[b]
\begin{centering}
  \includegraphics[width=0.95\columnwidth]{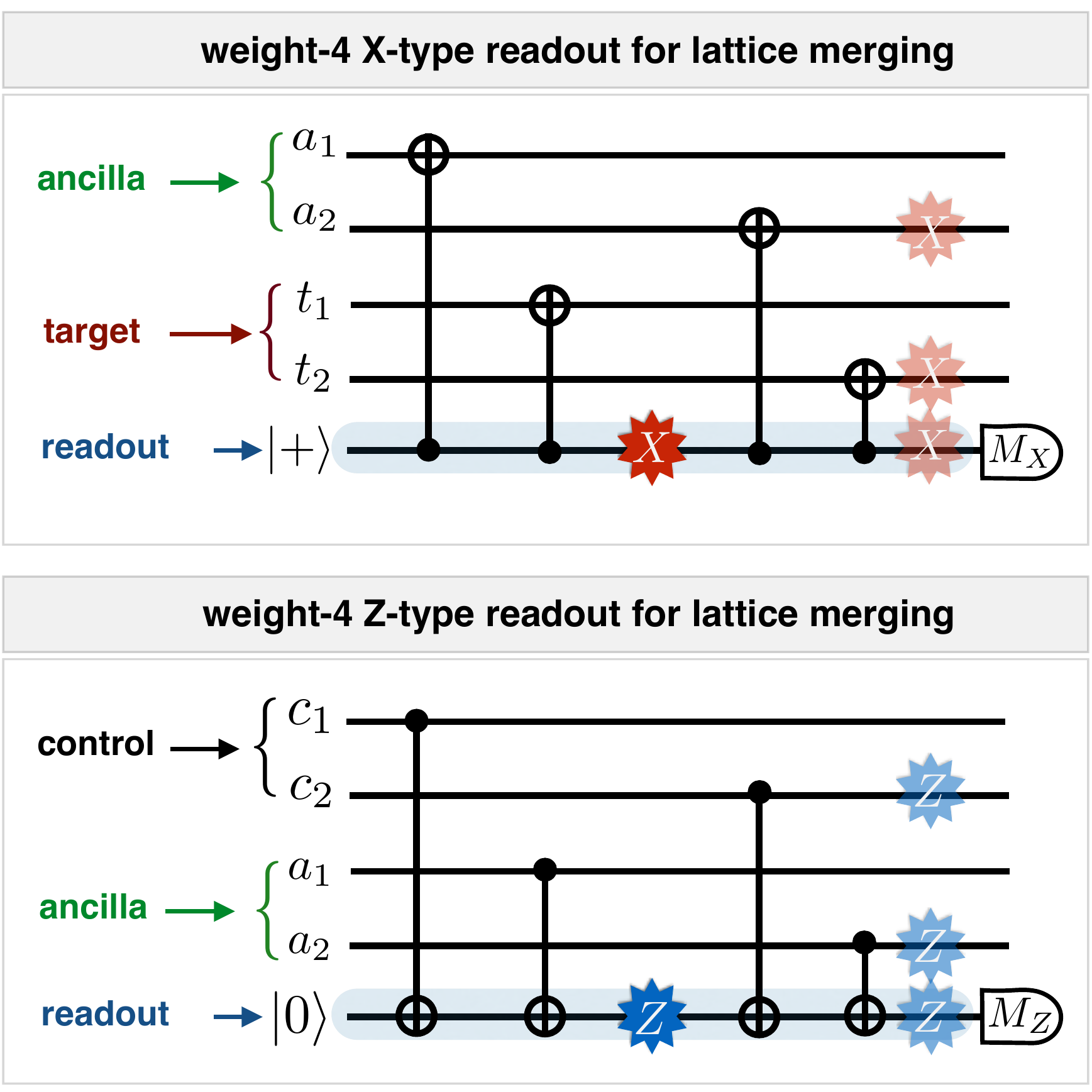}\\
\caption{\label{fig:weight_4_oper} {\bf Circuit for the FT readout of the weight-4 operator:} (Upper panel) By sequencing the CNOT gates for the measurement of $X_{a_1}X_{a_2}X_{t_1}X_{t_2}$,  a single $X$ error on the readout qubit after the second CNOT will propagate into a single bit-flip error on the logical ancillary qubit and a single bit-flip error  on the logical target qubit, both of which are correctable.  All other single-qubit errors propagate to form single-qubit errors, up to the operator $X_{a_1}X_{a_2}X_{t_1}X_{t_2}$ itself. (Lower panel) For the measurement of weight-4 of $Z_{a_1}Z_{a_2}Z_{c_1}Z_{c_2}$ operators, one would use instead a  circuit that inverts the sense of the CNOT gates, and a readout qubit that is initialized in $\ket{0}$ and measured in the $Z$ basis.}
\end{centering}
\end{figure}

\begin{figure}[b]
 \begin{centering}
  \includegraphics[width=0.8\columnwidth]{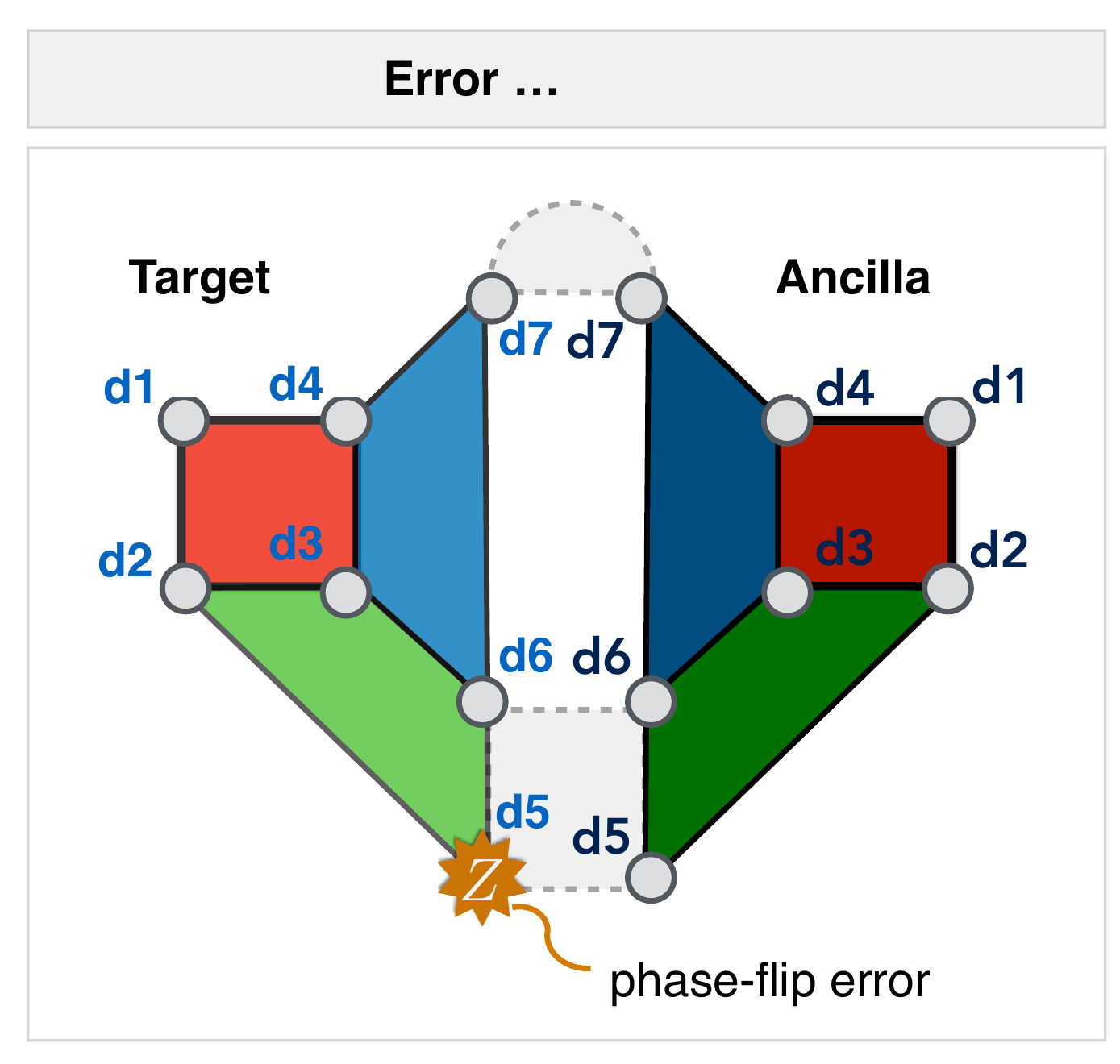}\\
  \caption{\label{fig:Lattice_Surgery_CC_error.pdf} \textbf{Dangerous boundary error in the merging susbetp:} Example of a single error that might result in a logical $Z$ error if not handled properly. If the $Z$ error occurs after the second measurement of the weight-4 boundary operator, applying $Z_5 ^t$ results in the right correction.  On the other hand, if the error occurs before the first measurement of the weight-4 boundary operator, applying a $Z_5 ^t$ will result in a logical $Z$ error.  Instead, one should apply $Z_6 ^t Z_7 ^t$ (see the main text for a more detailed explanation). If the $Z$ error occurs on any of the non-boundary qubits (d$1$-d$4$), the traditional correction works because the error commutes with the joint logical operator on the boundary.}
\end{centering}
\end{figure}

Despite the fact that these lower-weight operators anti-commute with some of the stabilizers of the code, which thus become temporarily undefined, the logical encoded information is not lost. After measuring the low-weight operators depicted as gray plaquettes in Fig.~\ref{Fig:Lattice_Surgery_CC.pdf}, the two logical qubits are temporarily merged into a single logical qubit of an enlarged code.  Notice that for each merging process one stabilizer per logical qubit involved becomes undefined.  For the measurement of the $X_L^t X_L^a$ operator (merging of the target and ancillary logical qubits), the $S_Z^{(3)}$ stabilizers, depicted in blue in Fig.~\ref{Fig:Lattice_Surgery_CC.pdf}, become undefined.  However, their product, which corresponds to an operator of weight-$8$, is still a stabilizer since it commutes with the merging operators.  Subsequently, the original stabilizers of the two logical qubits must be measured to split this code into the original two  codes hosting the logical qubits. It is precisely this merging and splitting process which gives the name of lattice surgery to the  implementation of the teleportation-based circuit of Fig.~\ref{fig:circ_latt_surg_general} at the logical level.  The evolution of the important operators throughout the merging and splitting process is summarized in Table \ref{table:evolution_operators_CNOT_bis}.  

We would like to emphasize that the requirements to convert  this circuit into the desired fault-tolerant procedure are far from trivial, and have not been addressed in sufficient detail previously. We now present a careful exposition of the  steps that are required to achieve fault-tolerance  focusing on optimizing the implementation for  distance-3 color codes, although we emphasize that our construction is readily applicable to larger-distance codes.  

  Fig.~\ref{fig:latt_surg_full} depicts the full FT circuit for the lattice-based logical CNOT between two distance-3 color code qubits.  In step $0$, the ancillary logical qubit is prepared in the state $|0 \rangle_L$ by a flag-based measurement of the $X$ stabilizers 1 time.  The next two steps fo the joint operator measurements of Fig.~\ref{fig:circ_latt_surg_general} must be composed of 3 sub-steps: merging, splitting, and hook-error detection.  The splitting sub-step consists of flagged-based measurements of the conjugate stabilisers.  The third sub-step is conditional, as it is only required whenever a flag is triggered during  the splitting sub-step, such that additional stabiliser measurements are required to detect the weight-2 errors (so-called hook errors).  Finally, the last step  consists of measuring the 7 physical qubits of the ancillary logical qubit in the $X$ basis, and performing classical error correction.  
  
  In order to reduce the resource requirements, our philosophy will be that, as soon as an error is detected at some point in the circuit, the remaining steps can be performed without the same FT constraints, since we only require the whole protocol to be resilient to a single-error event (i.e. FT level $t=1)$.  We note that even if maintaining FT constructions after an error has been detected might lead to lower logical error rates, this is not strictly necessary to achieve a logical error rate whose leading order in the physical noise parameters is quadratic.  Furthermore, in a trapped-ion shuttling-based scheme, switching to non-FT circuits once an error has been detected offers a great advantage with respect to the complexity of the microscopic operations.  We believe that this low-resource philosophy will be useful for the relatively short QEC protocols and FT logical gates that will be experimentally tested in the near future.  It might also prove useful within individual subunits of larger QIP protocols.  However, for sufficiently long protocols, the advantage of the low-resource philosophy disappears.  In this case, maintaining frequent FT QEC steps will certainly be vital to prevent the accumulation of errors \cite{flag_readout_arb_distance}.  Let us now explain the FT requirements for the merging and splitting sub-steps:   

\vspace{1ex}
{\bf (i) Merging.--} The circuit used to measure the joint logical operator $X_L^t X_L^a$ fault-tolerantly is shown in Fig.~\ref{fig:FT_MXX}.  As explained previously, the central point of the merging process is the measurement of weight-2 and a weight-4 operators on the boundary of two logical qubits (cf. Fig.~\ref{Fig:Lattice_Surgery_CC.pdf}).  Following our low-resource philosophy, if an error has been detected previously in the circuit, then we drop the FT requirement.  In this case, it is enough to measure each operator once.  If an error has not been detected previously on the circuit, the full FT machinery is necessary.  In the first place, to account for measurement errors, both the weight-2 and the weight-4 operators must be measured twice.  We start with the weight-2 operator.  If the outcomes of the first and second measurements differ, the operator is measured a third time.  This implies that an error has occurred and it is safe to switch to non-FT circuits from there on, including measuring the weight-4 operator only once.  In contrast, if the two outcomes coincide, the operator is not measured a third time.   The same procedure is performed for the weight-4 operator., which can  be measured fault-tolerantly by using a single bare ancillary qubit.  As shown in Fig.~\ref{fig:weight_4_oper}, this can be achieved by alternating the CNOT gates  between the two logical qubits to prevent a single error cascading into two errors in  the same logical qubit. A similar scheduling underlies the FT readout using bare ancillary qubits in the $d=3$ surface-17 code~\cite{Horsman-njp-2012,Tomita14}.   

After measuring the weight-2 and weight-4 operators, it is crucial to measure the $X$ stabilizers of both logical qubits, in case we are measuring the $X_L^t X_L^a$ operator. When we are measuring the $Z_L^c Z_L^a$ operator, then we must measure the $Z$ stabilizers.  Note that this does not cause a splitting of the two logical qubits, as the $X$ stabilizers do not collapse during this merging process (i.e. splitting occurs only after we measure the $Z$ stabilizers).  Measuring the $X$($Z$) stabilizers is necessary to distinguish between $Z$($X$) errors that occurred before or after the measurement of the joint operator $X_{\rm L} ^t X_{\rm L} ^a$ ($Z_{\rm L} ^c Z_{\rm L}^a$).  Even though a single $Z$($X$) error taking place  before the measurement will eventually be caught, performing the traditional correction step assumes that the error did  occur after the measurement,  such that the correction will result in a logical $Z^c_{\rm L}$($X^t_{\rm L}$) error on the control (target) qubit.  To illustrate this subtlety, consider the error event depicted in Fig~\ref{fig:Lattice_Surgery_CC_error.pdf}.  Imagine we measure the weight-2 operator twice and the outcomes coincide.  We then measure the weight-4 operator twice and the outcomes also coincide.  We then proceed to measure the $X$ stabilizers on each logical qubit and the second stabilizer (light green in Fig~\ref{fig:Lattice_Surgery_CC_error.pdf}) returns a $-1$ eigenvalue, implying that a $Z$ occurred on qubit $5$ of the target logical qubit.  Since the two outcomes of the weight-4 boundary operator coincide, there are only two options: either the $Z$ occurred after the second measurement of the weight-4 operator or before the first measurement.  These two cases result in different eigenvalues of the joint logical operator $X_{\rm L} ^t X_{\rm L} ^a$.  Let us focus on the eigenvalues of the target qubit stabilizer $S_{x,t}^{(2)}$ on the second plaquette, and the joint logical operator $X_{\rm L} ^t X_{\rm L} ^a$:
\begin{itemize}
\item Case 1:   If the $Z$ error occurred after the measurement of the weight-4 boundary operator, the resulting stabilizers would be  $-S_{x,t} ^{(2)}$ and $- \epsilon_2 \epsilon_4 \, X_{\rm L} ^t X_{\rm L} ^a$, where $\epsilon_2$ ($\epsilon_4$) denotes the eigenvalue of the weight-2 (weight-4) boundary operator.  Traditional error correction would imply  applying a $Z$ on target qubit $5$, transforming these stabilizers to $+S_{X2} ^t$ and $+ \epsilon_2 \epsilon_4 \, X_{\rm L} ^t X_{\rm L} ^a$.
\item Case 2:  If, on the other hand, the $Z$ error occurred before the measurement of the weight-4 boundary operator, the resulting stabilizers would be  $-S_{x,t} ^{(2)}$ and $+ \epsilon_2 \epsilon_4 \, X_{\rm L} ^t X_{\rm L} ^a$.  Therefore, the error does not change the sign of the weight-4 operator because \textit{it was not a stabilizer} of the system when the it occurred.  Applying a $Z$ on target qubit $5$ transforms these stabilizers to $+S_{x,t} ^{(2)}$ and $- \epsilon_2 \epsilon_4 \, X_{\rm L} ^t X_{\rm L} ^a$, and causes a logical $Z$ error.  The right correction in this case should have been  to apply the operator $Z_6 ^t Z_7 ^t$, which flips the sign of $S_{x,t} ^{(2)}$ but leaves the joint operator unchanged.     
\end{itemize}

Therefore, when the outcomes of the boundary operators coincide and an error is detected on one of the boundary qubits, it is critical to measure either the weight-2 or the weight-4 boundary operator a third time (i.e.  round $3$ in Fig.~\ref{fig:FT_MXX}).  If the third outcome differs from the previous two, then Case 1 has occurred and we can apply traditional error correction.  If the third outcome coincides with the previous two, then we are dealing with Case 2, we must apply the alternative error correction.  We note that in this final conditional step, it is only necessary to measure one operator: the weight-2 operator is re-measured if a $Z$ error was detected on qubit $7$, whereas the weight-4 operator is re-measured if a $Z$ error was detected on qubits $5$ or $6$. 

\vspace{1ex}
{\bf (ii) Splitting.--}  After measuring the weight-2 and weight-4 $X$($Z$) boundary operators, the next sub-step is the splitting process, where the $Z$($X$) stabilizers are measured to separate the merged lattices (see Fig.~\ref{fig:latt_surg_full}).  Notice that after the merging process only $1$ of the $6$ stabilizers collapses.  After the $ X_{\rm L} ^t X_{\rm L} ^a$ merging, this corresponds to the third $z$-type stabilizer (blue plaquettes in Fig.~\ref{Fig:Lattice_Surgery_CC.pdf}).  After the $ Z_{\rm L} ^c Z_{\rm L} ^a$ merging, this corresponds to the second $x$-type stabilizer (green plaquettes in Fig.~\ref{Fig:Lattice_Surgery_CC.pdf}).  Therefore, in an error-free scenario, it is enough to measure only these stabilizers in order to split the lattices.  In our low-resource philosophy, it is also enough to do this when an error has already happened previously in the protocol.  On the other hand, if an error has not occurred yet,  all $Z$($X$) stabilizers must be measured in a FT fashion in order to both split the lattices and to correct for possible $X$($Z$) errors.  

As explained before, this splitting requires  performing two rounds of stabilizer measurements.  In the first round, we measure the stabilizers in a FT way using the flag-based readout schemes depicted in Fig.~\ref{Fig:flag_readout}.  If no error is detected, we stop and trust the outcomes.  On the other hand, if an error is detected, we measure the stabilizers a second time using un-flagged circuits.  If one of the flags gets triggered during the first round then, after the splitting, we must perform $1$ round of un-flagged  readout of the conjugate   stabilizers to correctly identify the potential hook error.  Notice that it is not necessary to perform the hook-error detection substep on the ancilla logical qubit after the second splitting, as shown in Fig.~\ref{fig:latt_surg_full}, as the  potential $X$-type hook error would not affect the outcome of the subsequent ancillary  measurement in the $X$ basis.

Finally, to correctly perform the error correction after the splitting, the syndromes for each logical qubit must be shared.  Due to the fact that one of the $Z$($X$) stabilizers from each logical qubit collapsed during the merging process, some information about possible $X$($Z$) errors that occurred before the merging is lost.  Therefore, the error syndromes from each logical qubit are shared and if both indicate the presence of errors, a search is performed to determine if the joint syndromes are compatible with a lower weight error that occurred before the merging.


\section{\bf Trapped-ion schedules for a logical CNOT gate with  triangular color codes}
\label{sec:trapped_ion_implementations}

In the introduction, we described the current fidelities for state-of-the-art  QIP operations  that have been achieved in   various trapped-ion laboratories. One of the practical  challenges for the demonstration of  useful QEC is to combine all of these ingredients in a single experimental platform. In this work, we focus on the particular strategy explored in~\cite{eQual_1qubit}, which discusses in detail a trapped-ion QCCD approach that consists on  segmented high-optical-access (HOA) ion traps in a cryogenic environment (see Fig.~\ref{Fig:trap_scheme}), and describes current and expected QIP operations on a mixed-species ion crystal to address the challenges $(\mathsf{QEC}$-$\mathsf{I})$ and $(\mathsf{QEC}$-$\mathsf{II})$. In particular, this approach focuses on $^{40}{\rm Ca}^+$  ions to host the data qubits of an encoded color-code qubit, and use $^{88}$Sr$^+$ ions for sympathetic re-cooling of the ion crystal prior to any phonon-mediated entangling gate between ions residing in the same trap segment. This re-cooling step is an essential step to maintain the high fidelities of the entangling gates between ions  that may have  been separated, shuttled, rotated,  or merged  by using high-speed crystal-reconfiguration protocols.  In Subsec.~\ref{sec:qec_toolbox}, we describe the QEC correction toolbox used in~\cite{eQual_1qubit}, and discuss the additional operations that are required for the implementation of the logical CNOT gate to address the $(\mathsf{QEC}$-$\mathsf{III})$ quantum processor challenge. In Subsec.~\ref{sec:error_model}, we describe a detailed multi-parameter error model that incorporates the different error rates of the various QIP trapped-ion operations.  Together with the microscopic trapped-ion schedules described in Subsecs.~\ref{sec:schedules_flag_qec}-\ref{sec:lattice_surgery_schedules}, this error model can be used to study numerically the actual performance of the QEC schemes presented in the previous section.

\subsection{Extended Toolbox for quantum error correction}
\label{sec:qec_toolbox}
As discussed in~\cite{eQual_1qubit}, the     quantum memory challenges $(\mathsf{QEC}$-$\mathsf{I})$ and $(\mathsf{QEC}$-$\mathsf{II})$ can be addressed using  a single arm of the segmented ion trap in Fig.~\ref{Fig:trap_scheme}. In particular,  cycles of QEC can be implemented using  cat-state FT  schemes for  stabiliser readout~\cite{shor_ft_qec, aliferis_ft_qec}  by re-distributing sets of ions between three storage and    two manipulation  zones, where the   qubits can be manipulated using a trapped-ion  universal set of gates~\cite{schindler-njp-15-123012,Nebendahl-PRA-2009,martinez16}. This set of gates contains the so-called   M{\o}lmer-S{\o}rensen (MS)
entangling operations~\cite{MS_gates,roos_MS_gates}, which are driven by a bi-chromatic laser field that couples the qubits to the center-of-mass mode of the longitudinal  vibrations of an ion string. The MS gate acts globally on the ions of the same crystal that are illuminated by the laser according to  
\beq
\label{eq:MS_gate}
U_{\rm MS,\phi}(\theta)=\ee^{-\ii\frac{\theta}{4} S^2_\phi},\hspace{1ex}  S_\phi=\sum_{i=1}^n(\cos\phi X_i+\sin\phi Y_i)
\eeq
where $\theta$ is controlled by the laser   intensity and pulse duration, whereas  $\phi$ depends on the laser phase. In the following subsections, we will repeatedly use the  $\mathsf{(o1)}$ {\it fully-entangling two-qubit MS gate} for $\theta=\pi/2$, $\phi=0$, which we denote as  $X^2_{ij}(\pi/2)=(I-\ii X_iX_j)/\sqrt{2}$ and  represent as a solid line with two filled circles  touching the corresponding qubits of indices $i$ and $j$ in the circuit. We will also make use of  the $\mathsf{(o2)}$ fully-entangling 5-ion MS gate $U_{\rm MS,0}(\pi/2)$ ($U_{\rm MS,0}(-\pi/2)$), which will be represented by a  solid line with five filled (empty) circles  touching the corresponding  qubits.  

 Additionally, this gate set contains  global rotations around the Bloch sphere 
\beq
\label{eq:rot_xy}
U_{\rm R,\phi}(\theta)=\ee^{-\ii\frac{\theta}{2}S_\phi},
\eeq
controlled by the intensity, phase, and pulse duration of the lasers; as well as  
ac-Stark shifts on  individual qubits
\beq
\label{eq:rot_z}
U_{{\rm R}_j,z}(\theta)=\ee^{-\ii\frac{\theta}{2} Z_j},
\eeq
where $\theta$ is controlled by the intensity of the off-resonant laser beam, its detuning, and the pulse duration. In particular, we will make extensive use of $\mathsf{(o3)}$ {\it single-qubit rotations} around the $x$, $y$, and $z$-axis on a single isolated ion, which we denote  by  $X_{i}(\theta)=\cos(\theta/2) I_i-\ii\sin(\theta/2) X_i$,  $Y_{i}(\theta)=\cos(\theta/2) I_i-\ii\sin(\theta/2) Y_i$, and $Z_{i}(\theta)=\cos(\theta/2) I_i-\ii\sin(\theta/2) Z_i$, which can be obtained from the above gate set using refocusing techniques. As mentioned in the introduction, the closed cycling transition allows also for very accurate $\mathsf{(o4)}$ {\it measurements} of trapped-ion qubits in the $z$-basis $M_Z$ by  state-dependent fluorescence imaging, and $\mathsf{(o5)}$ {\it qubit initialization/reset} into $\ket{0}$ by optical pumping.

In addition to the above tools $\mathsf{(o1)}$-$\mathsf{(o5)}$ to manipulate the internal electronic degrees of freedom of the trapped-ion qubits, we will also exploit  additional techniques to control the external and motional degrees of freedom of the ion crystals. In particular, we consider using  an additional ion species for $\mathsf{(o6)}$ {\it re-cooling} of the ion crystal using sympathetic laser cooling techniques; together with a
a set of crystal reconfiguration techniques that have been already demonstrated experimentally. In particular, we consider fast $\mathsf{(o7)}$ {\it ion  shuttling} of single~\cite{walther12,Bowler12} or small crystals of ions across different segments of a single arm of the trap;  fast  $\mathsf{(o8)}$  {\it ion splitting and merging} of ion crystals~\cite{Bowler12,fast_ion_splitting}; and fast $\mathsf{(o9)}$  {\it ion rotations}  that swap pairs of ions~\cite{fast_rotation} and rotate small ion crystals.  
Finally, the shuttling-based toolbox of~\cite{eQual_1qubit} must be extended to include $\mathsf{(o10)}$  {\it junction crossing} whereby  ions  are transported across junctions~\cite{blakestad09_xjunctions,blakestad11_xjunctions,moehring11_yjunctions} that connect different arms of the segmented trap (see Fig.~\ref{Fig:trap_scheme}). This last operation will be an essential tool for the extensibility of the trapped-ion QCCD approach to QEC in 2D. 

Altogether, the operations $\mathsf{(o1)-(o10)}$ form our trapped-ion toolbox for QEC. In Table~\ref{tab:summary_toolbox}, we summarize the state-of-the-art characteristics for this ion-trap toolbox considering  current experimental results on the above QCCD platform. We also describe the  improvement of each operation that is expected to be achieved experimentally in the near term \cite{eQual_1qubit}.

\begin{table}
  \centering
  \begin{tabular}{|l |c|c|c|c|} \hline\hline
    Operation & Current & Current & Anticipated & Anticipated  \\
      &  dation & infidelity  &  duration &   Infidelity \\\hline
        $\mathsf{(o1)}$  Two-qubit   MS & 40$\mu$s & $1 \cdot 10^{-2}$ & 15$\mu$s & $2 \cdot 10^{-4}$ \\ 
         gate   &  &  &  &  \\ \hline
      $\mathsf{(o2)}$ Five-qubit MS  & 60$\mu$s & $5 \cdot 10^{-2}$ & 15$\mu$s & $1 \cdot 10^{-3}$ \\ 
              gate   &  &  &  &  \\ \hline
      $\mathsf{(o3)}$ One-qubit gate & 5$\mu$s & $5 \cdot 10^{-5}$ & 1$\mu$s & $1 \cdot 10^{-5}$ \\ \hline
     $\mathsf{(o4)}$ Measurement & 400$\mu$s & $1 \cdot 10^{-3}$ & 30$\mu$s & $1 \cdot 10^{-4}$ \\ \hline
      $\mathsf{(o5)}$ Qubit reset & 50$\mu$s & $5 \cdot 10^{-3}$  & 10$\mu$s &$5 \cdot 10^{-3}$ \\ \hline
      $\mathsf{(o6)}$ Re-cooling & 400$\mu$s & $\bar{n} < 0.1$ & 100$\mu$s & $\bar{n} < 0.1$ \\ \hline
      $\mathsf{(o7)}$ Ion shuttling  & 5$\mu$s & $\bar{n} < 0.1$ & 5$\mu$s & $\bar{n} < 0.1$ \\ \hline
       $\mathsf{(o8)}$ Ion  split/merge  & 80$\mu$s & $\bar{n} < 6$ & 30$\mu$s & $\bar{n} < 1$ \\ \hline
        $\mathsf{(o9)}$Ion  rotation  & 42$\mu$s & $\bar{n} < 0.3$ & 20$\mu$s & $\bar{n} < 0.2$ \\ \hline
          $\mathsf{(o10)}$ Junction  & 100$\mu$s & $\bar{n} < 3$ & 200$\mu$s & $-$ \\ 
                 crossing (per ion)  &  & &  &  \\\hline
  \end{tabular}
  \caption{{\bf Trapped-ion QEC toolbox}. Description of current and future trapped-ion capabilities for a QCCD approach to FT QEC. We include the duration and infidelity of  operations dealing with the internal degrees of freedom, and the duration and final mean number of phonons in the longitudinal center-of-mass mode for the operations dealing with the external degrees of freedom. }
  \label{tab:summary_toolbox}
\end{table}

\subsection{Microscopic multi-parameter error model}
\label{sec:error_model}

In the assessment of the QEC capabilities of a particular code in a certain architecture, it is of paramount importance to have a realistic microscopic modeling of the noise that afflicts the different operations. Oversimplified noise models may overestimate the correcting power of a particular approach, such that the current state-of-the-art or envisioned technological improvements might turn out to be insufficient to achieve the aforementioned goals $(\mathsf{QEC}$-$\mathsf{I})$-($\mathsf{QEC}$-$\mathsf{III}$) in the experiments. Therefore, it is crucial to use a realistic microscopic modeling of the relevant sources of noise in our trapped-ion architecture, as emphasized in reference~\cite{eQual_1qubit}. In this work, we employ the same error model  with minor modifications which, contrary to the majority of error models in the literature, uses different noise channels for   different operations that rest on  a detailed microscopic modeling of the trapped-ion hardware.

 The model consists of various stochastic channels of  Pauli errors with $6$ independent noise parameters to account for leading experimental imperfections: 
\begin{enumerate}
    \item \textit{Imperfect qubit initialization and measurement:} modeled as bit-flip ($X$) errors after $|0 \rangle$ state preparation, and measurement in the $Z$ basis, both with  probability $p_m$.
    \item \textit{Single-qubit gate errors:} Pauli (X, Y, or Z) errors after $1$-qubit rotations with the same probability $p_{1q}/3$.
    \item \textit{Two-qubit MS gate errors:} $2$-qubit Pauli errors after $2$-qubit MS gates with the same probability $p_{2q}/15$.
    \item  \textit{Multi-qubit MS gate errors:} $5$-qubit Pauli errors after $5$-qubit MS gates with the same probability $p_{5q}/1023$.
    \item \textit{Dephasing:} Pauli $Z$ errors on all qubits during any crystal-reconfiguration operation, and  on idle qubits not involved in an entangling gate with a probability $p_{\rm idle}$, to account for the collective qubit dephasing.   The parameter is given by the equation $p_{\rm idle} = (1-\exp(-t/T_2))/2$, where $t$ is the duration of the operation and $T_2$ is the standard parameter quantifying the resilience of the qubit's phase coherence.  We employ the anticipated experimental value of $T_2 = 2.2s$ for the trapped-ion optical qubit.  
       \item \textit{Errors during cross-junction shuttling:} Pauli Z errors on all qubits during a crossing through the trap junction with a probability $p_{\rm cross}$.  Although the error during a junction crossing is expected to be of the same form as the error during any reordering operation (i.e. dephasing), we use an independent parameter to be able to scan different junction crossing durations while keeping $p_{\rm idle}$ constant.   
\end{enumerate}

    The duration of the different idle periods, and thus the magnitude of  $p_{\rm idle}$, can have a variety of values depending on the particular trapped-ion schedule. To simplify the numerical simulation,  we fix their ratios to have the same values as the anticipated experimental values, such that the dephasing channel can be applied sequentially without the need of enlarging the multi-parameter set of error rates.  To simplify the simulations, we also set the duration of the $2$- and $5$-qubit MS gates, measurements, and state preparations to be the same as an ion crystal splitting/merging operation.  This is a pessimistic assumption, as can be seen from the durations in Table \ref{tab:summary_toolbox}.  On the other hand, we treat $1$-qubit rotations as instantaneous. As shown below, this is justified because $1$-qubit rotations in our schedules  typically occur concurrently with an ion-crystal reconfiguration step of much longer duration, so that the single-qubit rotations do not add any extra time to the procedure.

A very important noise source during ion reconfigurations and trap-junction crossings is heating of the ion crystals.  To overcome its effects, which would deteriorate the fidelity of subsequent entangling gates, we consider performing sideband cooling (re-cooling in Table \ref{tab:summary_toolbox})   before any entangling gate.  The relatively long duration of these re-cooling steps (more than twice than that of any other operation) causes extra dephasing on the qubits.

\subsection{Schedules for the flag-based QEC}
\label{sec:schedules_flag_qec}

Let us first consider the trapped-ion circuit implementation of the flag-based readout described in Sec.~\ref{sec:flag_readout}, which requires finding  analogues of the CNOT-based circuits of Fig.~\ref{Fig:flag_readout} using the above universal set of gates $\mathsf{(o1)-(o3)}$. 

\begin{figure}[t]
 \begin{centering}
  \includegraphics[width=0.95\columnwidth]{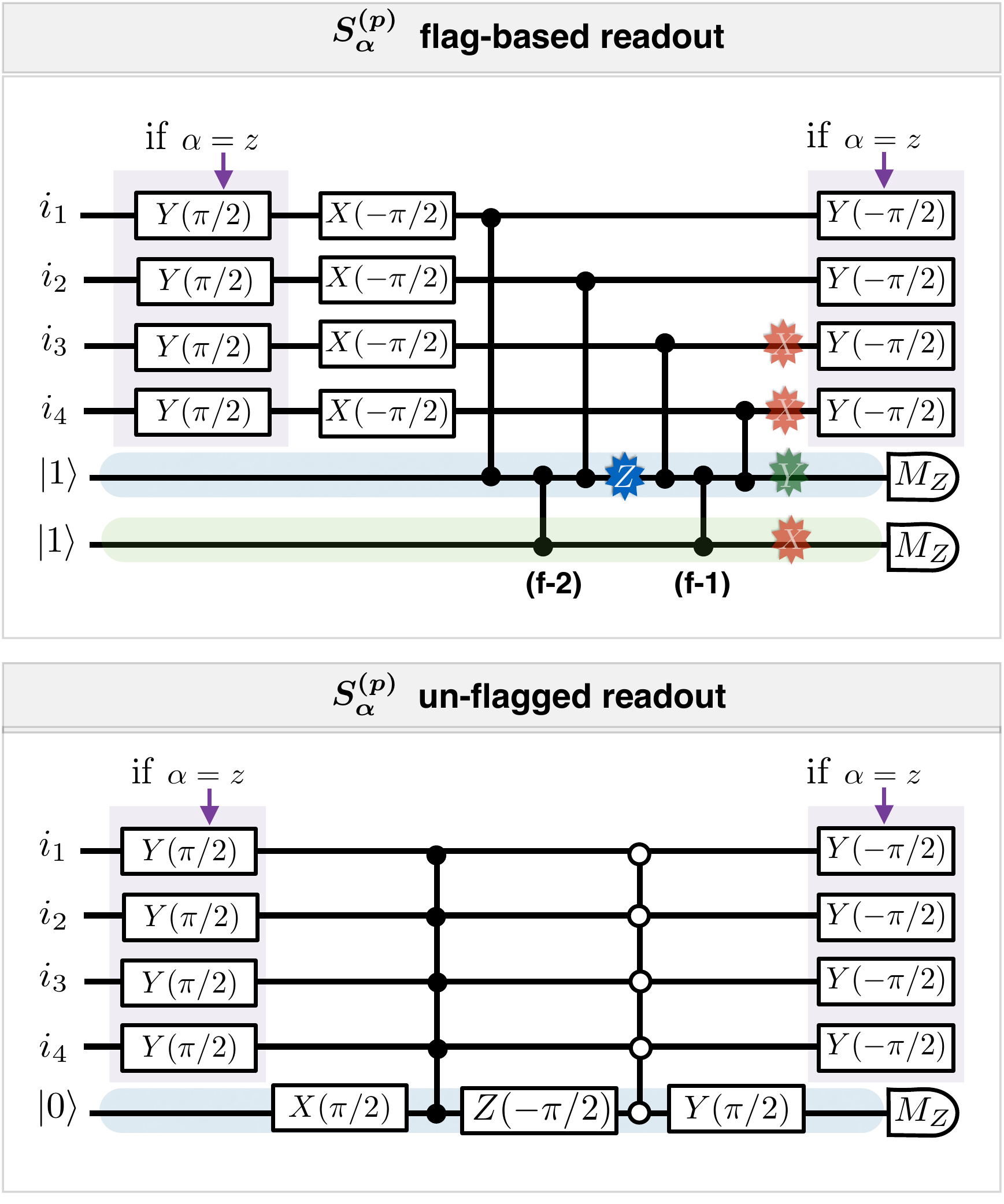}\\
  \caption{\label{Fig:flag_readout_ms} {\bf Trapped-ion flag-based FT readout of weight-4 stabilisers:}  (Upper panel) MS-based circuit for the  parity check measurements based on the flag-readout schemes of Fig.~\ref{Fig:flag_readout}. A dangerous phase-flip error is depicted as a blue $Z$-star at the middle of the circuit, which would propagate into a pair of bit-flip error (red stars) on the data qubits. This can be seen using the MS-gate propagation identities $Z_{i}X^2_{i,j}(\pi/2)\to X^2_{i,j}(\pi/2)Y_{i}X_j$, and $Y_{i}X^2_{i,j}(\pi/2)\to X^2_{i,j}(\pi/2)Z_{i}X_j$. To identify this dangerous error, the {\bf (f-1)} MS gate between the syndrome and an extra flag qubit is introduced, such that a $+1$ measurement in the $Z$ basis signals that this correlated error may have occurred. The  {\bf (f-2)} MS gate is required to reverse the effect of the {\bf (f-1)} MS gate, such that the stabiliser is correctly mapped into the syndrome qubit.  Notice that it is possible to switch off  the {\bf (f-1)} and {\bf (f-2)} MS gates to recover the traditional single-ancilla non-FT stabiliser measurement circuit. (Lower panel) The stabiliser measurement with un-flagged qubits can be implemented using a pair of 5-ion fully-entangling MS gates $U_{\rm MS,0}(\pi/2)$ ($U_{\rm MS,0}(-\pi/2)$), hereby represented by a  solid line with five filled (empty) circles. By sandwiching a single-qubit $Z$-rotation with the two MS gates~\cite{mueller-njp-13-085007}, the stabiliser information can be mapped onto the syndrome qubit directly. Since this un-flagged readout is only performed when an error has already occurred, one preserves the FT nature of the multi-qubit readout to level $t=1$.}
\end{centering}
\end{figure}

In the upper panel of Fig.~\ref{Fig:flag_readout_ms}, we present the MS-based circuits for the flag-based readout of both $X$- and a $Z$-type stabilisers of the 7-qubit color code~\eqref{eq:satb_larger_code}. It is interesting to note that the core structure of the MS-based circuit, as well as the initialization/measurement of the ancillary flag and syndrome qubits,  is the same for both types of stabilisers. The  only difference is  a collection of single-qubit rotations around the $y$-axis that must be applied conditional on the type of   stabiliser being measured. This contrasts the case of the CNOT-based circuits in the upper and lower panels of Fig.~\ref{Fig:flag_readout}. 

\begin{figure}[t]
 \begin{centering}
  \includegraphics[width=0.9\columnwidth]{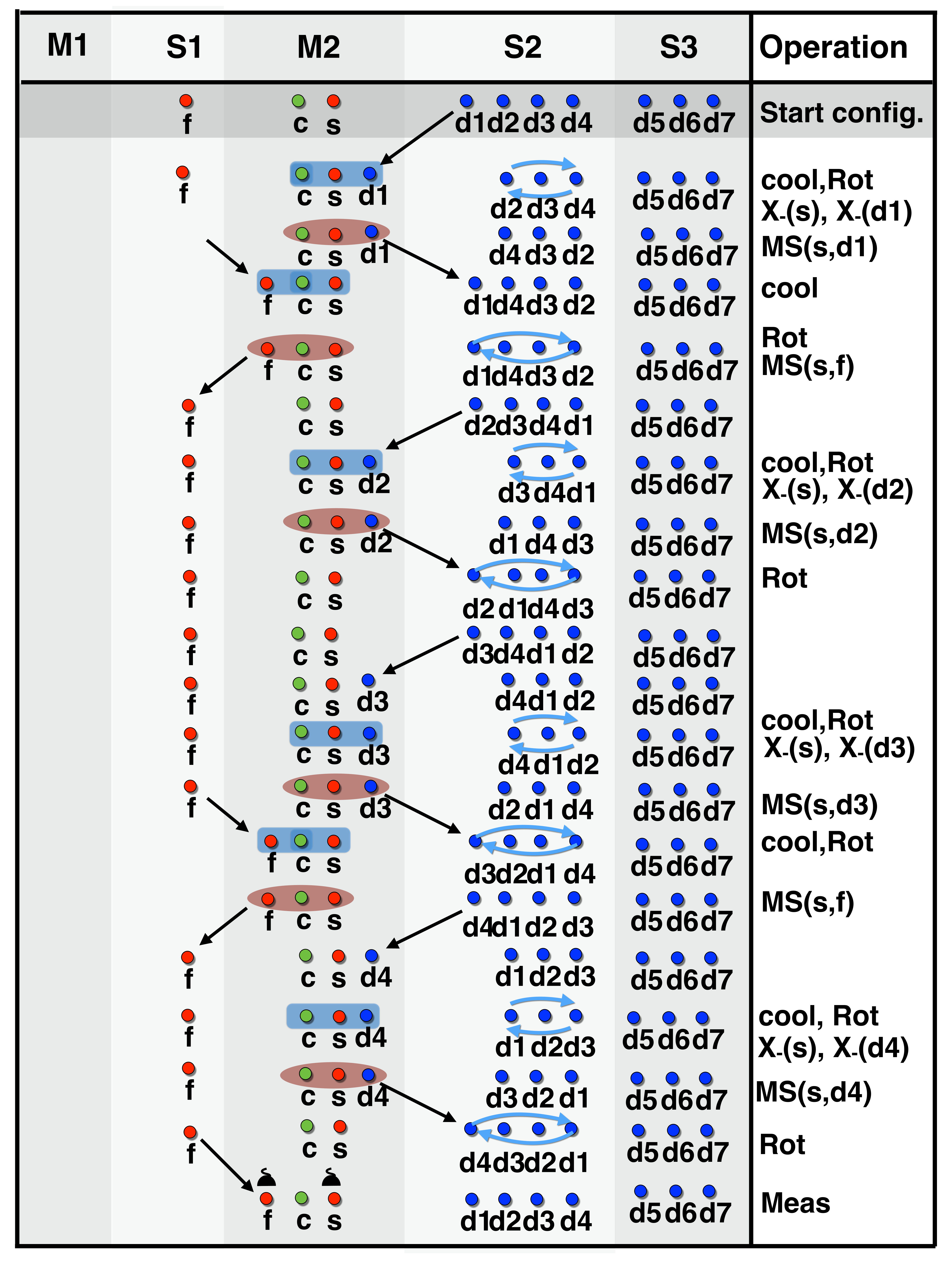}\\
  \caption{\label{Fig:schedule_flag} {\bf Schedule for the flag-based  measurement of $S_x^{(1)}$:} Data qubits $d1,d2,d3,d4,d5,d6,d7$ (blue circles) are distributed among the storage zones $S2$ and $S3$, whereas the ancillary flag $f$ and  syndrome $s$ qubits (red circles) occupy the storage $S1$ and manipulation $M2$ zones, respectively. A cooling ion of a different species is co-trapped in region $M2$. The different lines represent the scheduling of operations required to measure  $S_x^{(1)}=X_1X_2X_3X_4$ using the flag-based circuit of Fig.~\ref{Fig:flag_readout_ms}. Black straight arrows joining two different steps represent the splitting, shuttling and merging of a set of ions that are being transported between two different regions. Blue curved arrows  represent a rotation of the ion crystal. Fully entangling MS gates $X_{ij}^2(\pi/2)$ are represented by a red oval, prior to which a sympathetic re-cooling of the ion crystal must be applied (blue oval). Finally, single qubit gates   $X_{\pm}=X(\pm\pi/2)$, acting on  specific ions,  are listed in the rightmost column. The fluorescence measurements of the flag/syndrome ions are depicted by black detectors and are followed by a reset operation via optical pumping.}
\end{centering}
\end{figure}

Following the philosophy of the flag-based readout, we also depict the occurrence of a dangerous phase-flip error taking place at the middle of the circuit. This error cascades into a correlated pair of bit-flip errors in the data qubits, and has to be detected by our circuit.  As detailed in the caption of Fig.~\ref{Fig:flag_readout_ms}, due to the {\bf (f-1)} MS gate, the error also propagates onto the flag qubit, and  can be detected by a $-1$ measurement in the $z$-basis, instead of the expected $+1$. 
As occurs for the CNOT-based circuits, the role of the  {\bf (f-2)} MS gate between  syndrome and  flag qubits is essential to correctly map the stabiliser information into the syndrome qubit. Essentially, the combination of the two fully-entangling gates leads to a product state for the bipartition between the data-syndrome qubits and the flag qubit. Accordingly, even if we can detect the events where a correlated error may have occurred by inspecting the measurement outcome of  the flag qubit, the  MS gates do not interfere with the  mapping of the stabiliser information into the syndrome qubit.

As described in detail in Sec.~\ref{sec:flag_readout}, whenever the flag is triggered, one can identify which single-qubit or  two-qubit error has propagated into the data qubits by performing a subsequent measurement of all the stabilisers (see Table~\ref{table_decoding}). At FT-1 level, these measurements can be performed using bare ancillas (i.e. un-flagged circuits). This is particularly interesting for the trapped-ion implementation, which  allows for multi-qubit fully-entangling gates $\mathsf{(o2)}$, and allow one to  simplify the un-flagged measurements even further. As depicted in the lower panel of Fig.~\ref{Fig:flag_readout_ms}, the stabiliser can be mapped by a particular combination of two 5-qubit MS gates, instead of the sequence of four 2-qubit MS gates. This will simplify considerably the ion crystal reconfiguration operations that are required to implement the above circuit in the QCCD architecture, while preserving the desired fault tolerance.

Once the MS-based circuits have been presented, let us describe the microscopic schedules that can be followed  to implement the flag-based FT QEC in practice using our toolbox  $\mathsf{(o1)-(o10)}$ for the mixed-species ion QCCD of Fig.~\ref{Fig:trap_scheme}. For the 7-qubit color code, we will require 7 data qubits and 2 ancillary qubits (i.e.flag and  syndrome), both of which belong to the same atomic species. Additionally, we will exploit 1 cooling ion of a different species/isotope for sympathetic re-cooling of the ion crystal. The arrangement of these ions within the central region of the segmented trap is depicted in the inset of Fig.~\ref{Fig:trap_scheme}, where each arm of the trap contains 10 ions and allows for  parallel flag-based QEC  for one logical qubit in each arm of the trap. Accordingly, we focus on a single arm, and describe now in detail the required microscopic schedule.

At the top of Fig.~\ref{Fig:schedule_flag}, we depict schematically the storage and manipulation regions that conform the central region of a single arm of the HOA-2 trap. The 10 ions are initially distributed among the different zones as shown in the first line, where we use the same colors and labels as in Fig.~\ref{Fig:trap_scheme}. Each subsequent line represents a different step of the microscopic schedule, and the columns describe the particular ion occupation of each trap zone during such step. The operations to be performed in that step are listed in the right-most column, which describes which of the tools $\mathsf{(o1)-(o10)}$  must be used. Additionally, we also use straight black arrows to depict crystal splitting, shuttling, and merging; and curved arrows to denote crystal rotations (see the caption for further details).  Fig.~\ref{Fig:schedule_flag} corresponds to the microscopic schedule for the flag-based  measurement of the first stabiliser $S_x^{(1)}=X_1X_2X_3X_4$ of the 7-qubit color code (Fig.~\ref{Fig:largerCode}). We have also obtained similar schedules for the flagged and un-flagged measurements of the remaining stabilisers $S_\alpha^{(p)}$. We remark that the complexity of these building blocks  is considerably smaller than the cat-based approaches explored in Ref.~\cite{eQual_1qubit}. We thus believe that the flag-based approach will be a key element for the envisioned trapped-ion QCCD progress towards fault-tolerant quantum computation with color codes. 

These schedules should be applied according to the following procedure. One starts from the flagged schedule for $S_x^{(1)}$ (Fig.~\ref{Fig:schedule_flag}). If an error is detected either in the flag or syndrome qubits, one should measure all stabilisers  $\{S_\alpha^{(p)}\}$ with the un-flagged circuits using multi-qubit MS gates, and keep the outcomes as the final error syndrome. By applying the decoding table~\ref{table_decoding}, one can detect the most-likely error and correct it actively, or by updating the Pauli frame. On the other hand, if no error is detected, one should proceed  with the  flagged schedule of $S_z^{(1)}$, and apply a similar QEC procedure.  In  case  no error is detected, one should proceed similarly moving to  the next plaquette, and so on in case no errors are ever detected.

\begin{figure}[b]
 \begin{centering}
  \includegraphics[width=0.85\columnwidth]{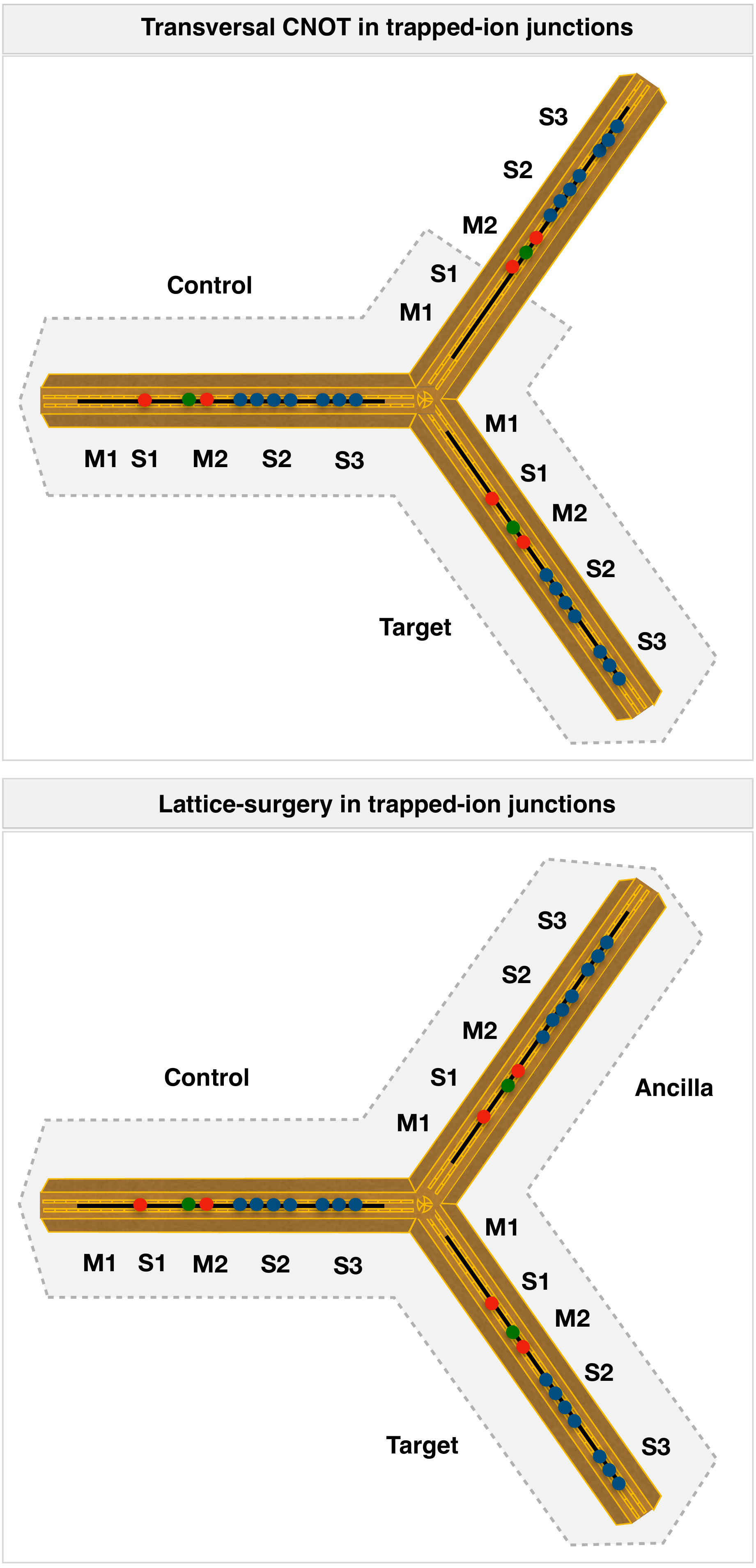}\\
  \caption{\label{Fig:cnot_layout} {\bf Ion layout for the logical CNOT in HOA-2 traps:}   (Upper panel)   The 7-qubit color codes for the control and target logical qubits of a transversal CNOT are stored in the central regions of the  two lower arms using the 10-ion  distribution of Fig.~\ref{Fig:trap_scheme}. The storage and manipulation zones $M1,S1$ of the upper arm are vacated, accommodating the ions of the corresponding logical qubit in the remaining zones, which are mere spectators during the transversal CNOT. Accordingly, the $M1,S1$ sones can be used to simplify the required crystal reconfiguration operations following the schedule described in the main text. (Lower panel)  For a lattice-surgery CNOT, the logical qubit stored in the upper arm is no longer a mere spectator, but acts as the ancillary qubit in the teleportation-based circuit of Fig.~\ref{fig:circ_latt_surg_general}. }
\end{centering}
\end{figure}

\begin{figure*}
 \begin{centering}
  \includegraphics[width=1.6\columnwidth]{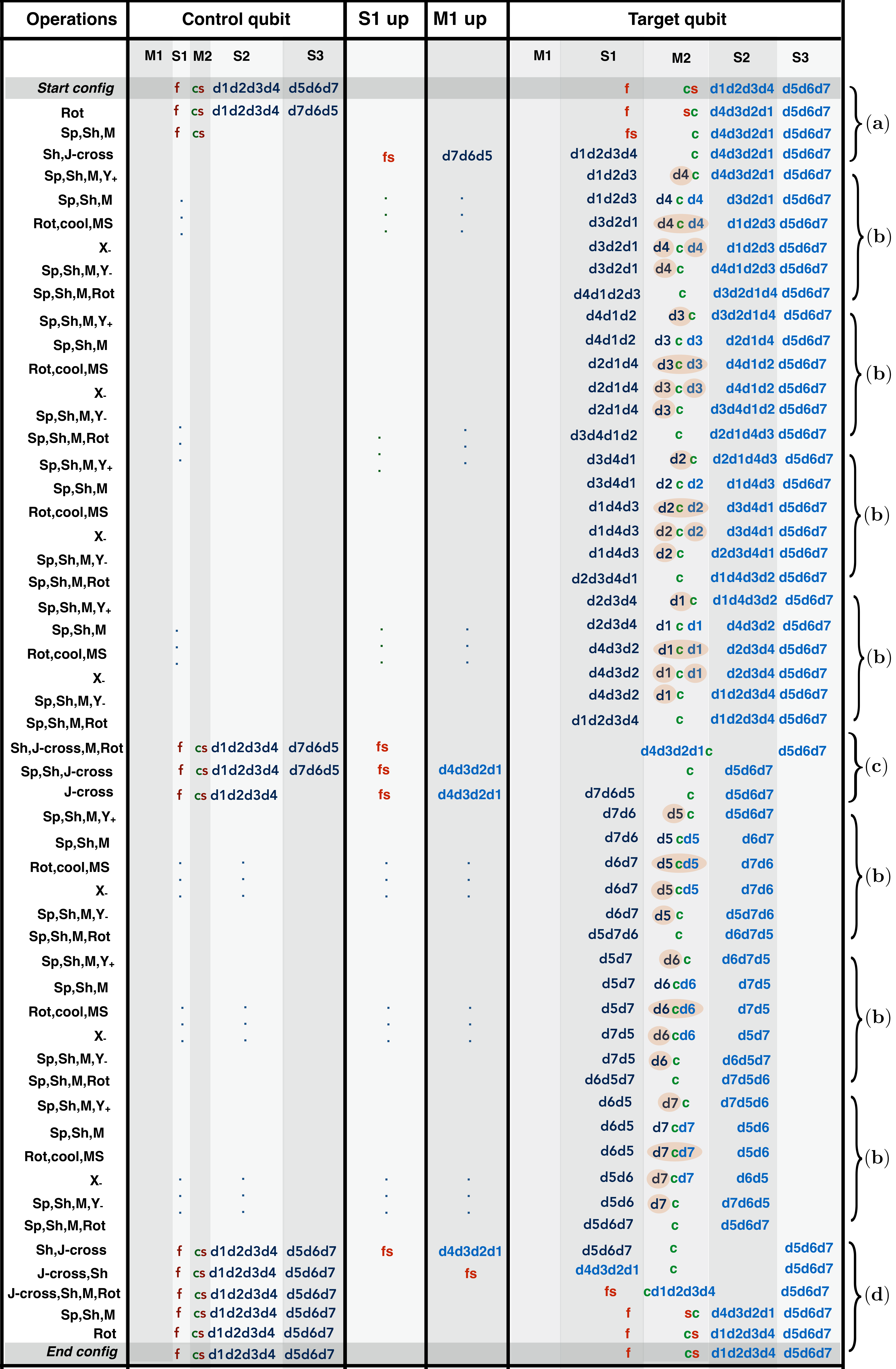}\\
  \caption{\label{Fig:trans_cnot_real_space} {\bf Schedule for the transversal CNOT gate between two 7-qubit color codes:}  A pair of logical qubits, each  encoded into 7 ions,  are distributed in two neighboring arms of the trap together with their corresponding ancillary and cooling ions. We use the same notation as in Fig.~\ref{Fig:schedule_flag}, but eliminate the circles denoting the ions, and simply use their labels (i.e. data $d1$-$d7$, ancillary flag $f$ and syndrome $s$, and cooling $c$) with a different font and coloring to denote the two logical blocks.  We list the operations for the  FT transverse CNOT  routine  in the leftmost column. The remaining columns contain the real-space scheme of the ion-crystal configurations in the different zones  of the  neighboring arms across a $\mathsf{Y}$ junction. On the rightmost column, we group operations into four modules {\bf (a)-(d)}.  }
\end{centering}
\end{figure*}

\begin{figure*}
 \begin{centering}
  \includegraphics[width=1.99\columnwidth]{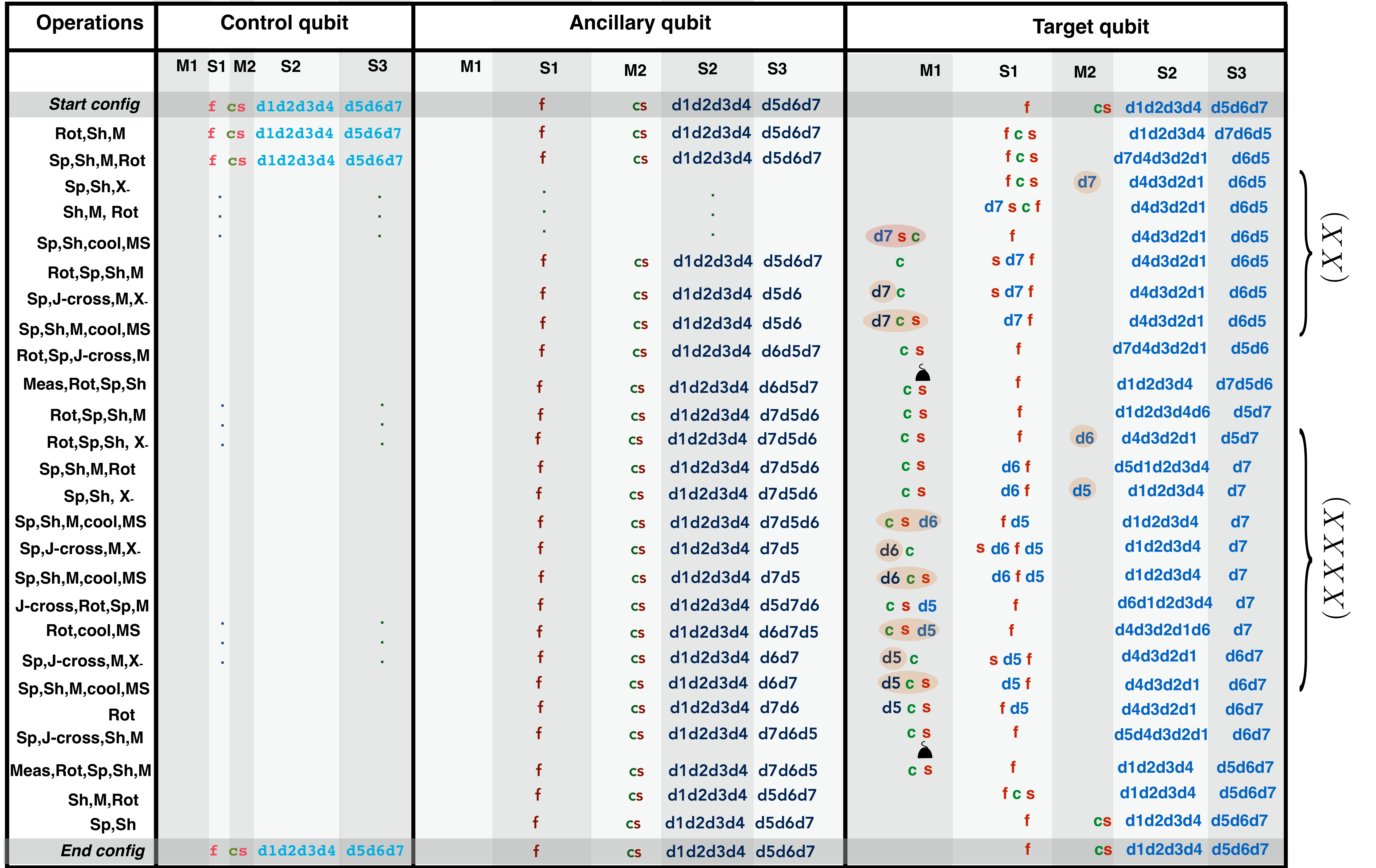}\\
  \caption{\label{Fig:schedule_M_XX_operators} {\bf Schedule for the  lattice-surgery  measurement of low-weight operators  for $M_{XX}$:}  Three logical qubits (i.e. control ancilla and target), each  encoded into 7 ions, are distributed in three neighboring arms of the trap together with their corresponding ancillary and cooling ions. We use the same notation as in Fig.~\ref{Fig:trans_cnot_real_space} with a different font and coloring to denote the three logical blocks.  For the measurement of the $X_aX_t$ logical operator, we need to sequentially measure weight-2 $M_{x(w-2)}$and weight-4 $M_{x(w-4)}$ operators. }
\end{centering}
\end{figure*}

\begin{figure*}
 \begin{centering}
  \includegraphics[width=1.99\columnwidth]{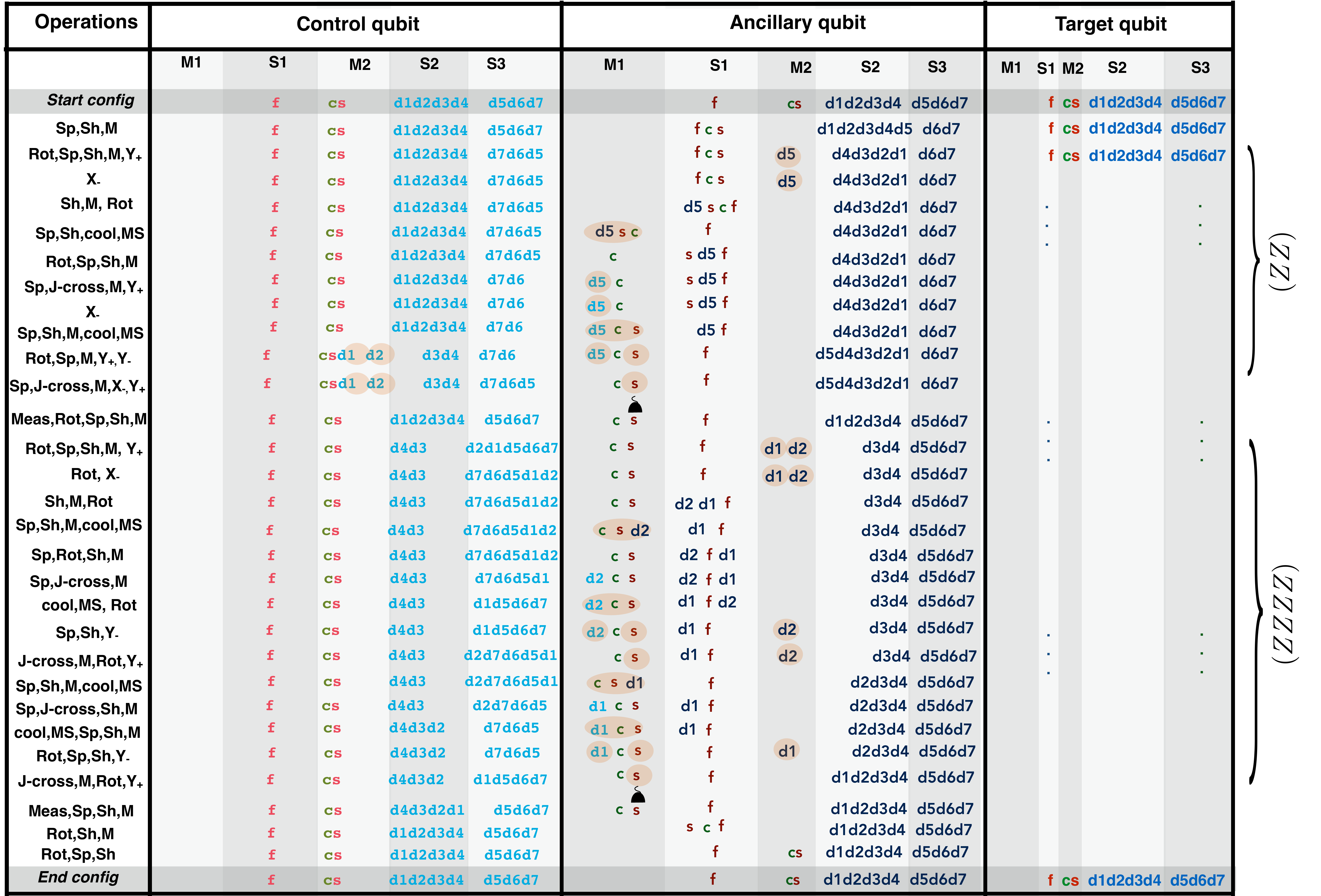}\\
  \caption{\label{Fig:schedule_M_ZZ_operators} {\bf Schedule for the  lattice-surgery  measurement of low-weight operators  for $M_{ZZ}$:} Three logical qubits (i.e. control ancilla and target), each  encoded into 7 ions, are distributed in three neighboring arms of the trap together with their corresponding ancillary and cooling ions. We use the same notation as in Fig.~\ref{Fig:trans_cnot_real_space} with a different font and coloring to denote the three logical blocks.  For the measurement of the $Z_cZ_a$ logical operator, we need to sequentially measure weight-2 $M_{z(w-2)}$and weight-4 $M_{z(w-4)}$ operators. The core module to implement such operations is described by the subsets of operations $(ZZ)$ and $(ZZZZ)$ in the rightmost column, whereas the remaining operations correspond to the reconfiguration steps needed to bring the required ions close to each other, such that these modules can be applied.    }
\end{centering}
\end{figure*}

\subsection{Schedules for a transversal CNOT gate}

Once the schedules for the flag-based QEC in each arm of the HOA-2 trap have been introduced, we can proceed with the  trapped-ion implementation of a  CNOT gate between two logical qubits, which will require shuttling ions across the $\mathsf{Y}$ junctions. We start by considering the transversal realization of the CNOT gate, which can be implemented with the  universal set of gates $\mathsf{(o1)-(o3)}$ according to the scheme shown in the right panel of Fig.~\ref{Fig:7qubitCode_cnot_transv}. This circuit is obtained by using the equivalence of a CNOT gate, up to a global phase, with the sequence of single-qubit  and fully-entangling MS gates $U_{\rm CNOT}^{c,t}=\ee^{-\ii \pi/4}Y_c(-\pi/2)X_c(-\pi/2)X_t(\pi/2)X_{c,t}^2(\pi/2)Y_c(\pi/2)$, which corresponds to the shaded region ${\rm (b)}$ of Fig.~\ref{Fig:7qubitCode_cnot_transv}.

We consider, for simplicity, the transversal CNOT gate between two  logical qubits  hosted in neighboring arms of the segmented  HOA-2 trap, and thus connected through a $\mathsf{Y}$ junction (see the upper panel of Fig.~\ref{Fig:cnot_layout}). Here, we present the initial  configuration  with a 7+2+1 ion crystal (7 data, 2 ancillary, and 1 cooling ions) held in each arm of the trap. As detailed in the caption, the zones $M1,S1$ in the upper arm are vacated to simplify the required crystal reconfigurations.  The microscopic trapped-ion schedule with our toolbox $\mathsf{(o1)-(o10)}$ is depicted in Fig.~\ref{Fig:trans_cnot_real_space}, where the columns  only show  the zones  relevant for the protocol. In the leftmost column, we describe the operations that must be applied at each time step,  leading to the  specific crystal configurations in the next rows. In addition to the operations already introduced for the flag-based QEC, see e.g. Fig.~\ref{Fig:flag_readout_ms}, we also include {\it J-cross}  to indicate when the ions must cross the $\mathsf{Y}$ junction. Besides, we have compressed notation  by suppressing the ion circles, and by grouping sequential operations in a single row. The ions where the sympathetic cooling is applied, or where  other crystal reconfiguration (e.g. rotations) take place, can also be recovered from the current and subsequent ion configurations.

As shown in Fig.~\ref{Fig:trans_cnot_real_space}, the whole routine can be divided into four modules {\bf (a)-(d)}.  Modules {\bf (a)}, {(\bf c)} and {\bf (d)} describe re-orderings of the ions that bring closer certain subsets of physically-equivalent qubits that belong to the control and target logical blocks. These modules contain all the overhead in $\mathsf{Y}$-junction crossings, while  module {\bf (b)} describes operations on  pairs of physically-equivalent ions that can be implemented within a single arm of the trap. This module is the core of the transversal CNOT gate in the shaded region ${\rm (b)}$ of Fig.~\ref{Fig:7qubitCode_cnot_transv}.  Since all physically equivalent ions are to be coupled to each other, we remark that the complexity of the re-orderings and the amount of  $\mathsf{Y}$-junction crossings   will increase considerably as the  distance of the code grows. The transversal approach loses one of the appealing features of topological codes, namely the local nature of quantum processing. In the following subsection, we therefore describe microscopic schedules for an alternative CNOT strategy that maintains this character, and alleviates the increase in complexity for larger-distance codes.

\subsection{Schedules for a lattice-surgery CNOT gate}
\label{sec:lattice_surgery_schedules}

In this subsection, we introduce the microscopic schedules for the lattice-surgery CNOT approach described in Sec.~\ref{sec:cnot_gates}. In particular, we describe  the trapped-ion  $\mathsf{(o1)-(o10)}$ operations that allow one to implement the building blocks of the circuit in Fig.~\ref{fig:latt_surg_full}. The interspersed QEC cycles of this figure can be implemented following our description of Sec.~\ref{sec:schedules_flag_qec}. The remaining parts correspond to the measurement of the joint $M_{XX}$, $M_{ZZ}$ and single $M_X$ logical operators.  The measurement of the ancillary logical $X_{\rm L}^a$ operator can be achieved by measuring all data qubits in the $X$ basis.

According to Fig.~\ref{fig:FT_MXX}, the FT measurement of the joint logical operations requires a sequential measurement of weight-2 and weight-4 operators that involve data qubits belonging to the neighboring boundaries of the logical blocks (see Fig.~\ref{Fig:Lattice_Surgery_CC.pdf}).  In Fig.~\ref{Fig:schedule_M_XX_operators}, we represent the trapped-ion schedule  for the sequence of operations $\mathsf{(o1)-(o10)}$ that are required to measure sequentially the weight-2 and weight-4 $X$-type operators. The control, ancillary, and target qubits, each corresponding to 7 data ions with their corresponding flag, syndrome and cooling ions, are distributed as depicted in Fig.~\ref{Fig:cnot_layout} (lower panel). For the $M_{XX}$ measurement, only the central regions that trap the ancilla and target qubits need to be used, as detailed in Fig.~\ref{Fig:schedule_M_XX_operators}.  The core modules to implement such operations are described by the subsets of operations $(XX)$ and $(XXXX)$  of the rightmost column, whereas the remaining operations correspond to the reconfiguration steps required to bring the required ions close to each other, such that these modules can be applied. A similar scheduling must be used for the $M_{ZZ}$ measurement, albeit focusing now on the control and ancillary blocks (see Fig.~\ref{Fig:schedule_M_ZZ_operators}). A direct comparison with the transversal approach shows that the overhead in crystal reconfigurations, especially $\mathsf{Y}$-junction crossings, is considerably reduced when the ordering of the operations follows Fig.~\ref{fig:FT_MXX}. We conclude that by exploiting the local quantum processing of the lattice-surgery approach, where only neighboring boundary qubits must be coupled to each other, and by a judicious design of the microscopic schedules, it is possible to minimize the overhead in trapped-ion  junctions crossings, which may turn out in an improvement of lattice-surgery methods with respect to transversal approaches already for small-distance codes.

\section{\bf Numerical Monte Carlo stabilizer toolbox and importance  sampling }
\label{sec:montecarlo_sampling}

In this section, we introduce the numerical Monte Carlo  toolbox  used to assess the performance of the previously developed QEC protocols.   As our noise model only involves Pauli operators, and the circuits consist of Pauli preparations/measurements and Clifford unitaries, it is possible to perform stabilizer simulations efficiently. In contrast to the exponential scaling of full-wavefunction numerical simulations \cite{exactsim2015,exactsim2016}, using the stabilizer formalism for a numerical simulation yields   a polynomial scaling of the simulation time with the number of qubits.  In this scenario, the bottleneck is no longer the system size (given by the number of physical qubits and depth of the quantum QEC circuits simulated), but rather the sampling of error configurations.  To optimize this point, we extend a recently developed \textit{subset-based} sampling scheme~\cite{importance_sampling} to a multi-parameter error model, which is of ultimate relevance for the trapped-ion implementation.

\subsection{Standard sampling and single-parameter subset sampler}

 The \textit{traditional} sampler generates an error configuration by traversing the whole circuit and deciding, after each gate, whether or not to insert an error based on a physical error rate.  If we assume that the physical error rate is characterized by a single parameter $p$, then, for each gate, a uniformly distributed random number $r$ between $0$ and $1$ is generated. If $r < p$, an error is inserted after the gate.  
 
 This sampler is appropriate for relatively high error rates and low-distance codes.  However, as the error rate decreases, the probability of actually not inserting any errors on the circuit increases.  Furthermore, for a fault-tolerant procedure on a distance-$3$ code, any single fault is correctable, so error configurations of weight-$1$ will never result in a logical error.  The probability of inserting $2$ errors on the circuit scales as $p^2$, which means that the minimal number of samples to obtain reliable statistics has to be on the order of $1/p^2$.  For a code of distance-$d$, this scaling becomes $1/p^{(d+1)/2}$, such that the sampling becomes very slow in the low-error regime. As an example, for the fault-tolerant lattice-surgery based CNOT circuit, the time required to run $10^5$ error configurations is about $24$ hours on a desktop computer using $4$ cores (Intel Core i7-6700 CPU @ 3.40GHz).  Obtaining a logical error rate for $p = 10^{-4}$ would take around $1000$ days.

\begin{figure*}
\begin{centering}
  \includegraphics[width=1.8\columnwidth]{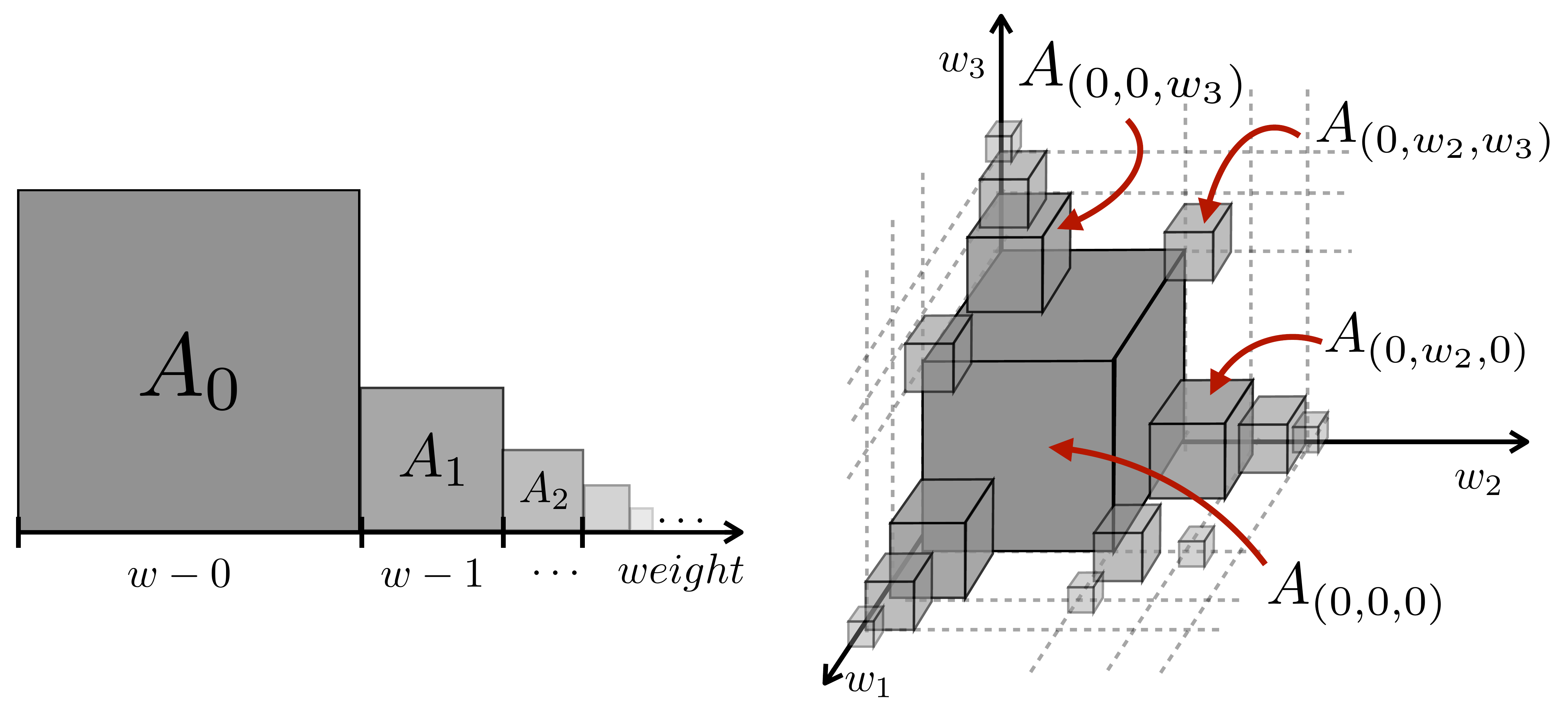}\\
\caption{\label{fig:Sampler_Box} {\bf Schematics of all subsets of microscopic error configurations:} (left panel) The individual boxes labeled as $w-n$ correspond to the set of error configurations of weight-$n$ errors that may occur during the execution of one complete circuit. The area $A_n$ of the subsets is meant to indicate (not to scale) the relative probability of occurrence of each subset. In the limit of small physical error rates, the $w-0$ subset dominates and all other subsets become vanishingly small. In this limit, the traditional sampler becomes inefficient, as it predominantly samples from the zero-error subset. In contrast, the subset-based sampling strategy explores in a targeted way the specific subsets chosen by the user, typically corresponding to the lowest-weight, non-correctable error subsets. (right panel)  Multi-parameter generalization of the error subsets.  Every noise parameter adds an extra dimension to the subset structure, which can be depicted as an hyper-cube (e.g. cube for a three-parameter error model as shown in the picture).  As in the single-parameter case, the hyper-volume of each subset indicates (not to scale) its relative probability of occurrence, which depends on the error rates vector $\vec{p}$.}
\end{centering}
\end{figure*}

In contrast, the \textit{subset-based} sampler relies on dividing all possible error configurations into subsets according to the weight of the error configuration, as illustrated in Fig.~\ref{fig:Sampler_Box}. Instead of performing a direct Monte Carlo sampling on the whole set, this sampler focuses specifically on individual subsets at a time.  If the noise acting on the quantum circuit is described by a single parameter $p$ which quantifies the probability of an error after each gate, then the probability of having $w$ errors is given by
\begin{equation}
A_w = {n \choose w} \, p^w (1 - p)^{n-w},
\end{equation} 
where $n$ is the total number of gates in the circuit. The logical error rate $p_L$ of the circuit (or in general the failure probability) is then 
\begin{equation}
p_L = \sum_{w=0}^{n}  A_w  p_L^{(w)},  
\end{equation}  
where $p_L^{(w)}$ is the logical error rate of the weight-$w$ subset.  The logical error rate for a given subset is multiplied by its probability of occurrence (the combined probability of occurrence of all the error configurations in the subset), which is straightforward to calculate analytically. 

The subset-based sampler procedure then consists of selecting a small number of the most relevant subsets, sampling each of them individually, and computing the contribution of the subset to the total logical error rate. This becomes very efficient if the physical error rate is small, such that the probability of occurrence $A_w$ of weight-$w$ errors on the circuit falls off quickly beyond a critical $w_{max}$, and the truncation to a small number of relevant subset is accurate. 

A convenient feature of the sampler is that it provides upper and lower bounds to the exact logical error rate, based on the total contribution of all subsets that were \textit{not} considered.  A lower bound is obtained assuming (optimistically) that all the subsets that are not sampled would result in a logical error rate of $0$  while an upper bound is obtained assuming (pessimistically) the same error rate would be $1$. Accordingly, we can bound the logical error rate by 
\begin{equation}
\label{eq:bounds}
\sum_{w=0}^{w_{max}}  A_w  p_L^{(w)}  \medspace \leq \medspace p_L \medspace \leq \medspace  \sum_{w=0}^{w_{max}}  A_w  p_L^{(w)} +  \sum_{w=w_{max}+1}^{n}  A_w 
\end{equation}
Notice that $p_L^{(0)} = 0$ since the weight-$0$ subset just corresponds to the error-free circuit.  For FT operations on distance-$3$ codes, $p_L^{(1)} = 0$ as well, and it suffices to start the sampling at $w=2$.  For all of  our circuits, we first sample the weight-$1$ subset exhaustively to check that  every error in this subset is correctable, and we have implemented the FT circuit correctly.  

As anticipated above, the subset-based sampler is appropriate for low physical error rates, when the probability of occurrence of large-weight subsets is vanishingly small.  In this regime, there is no practical difference between the two bounds~\eqref{eq:bounds}, and one can approximate the exact error with great confidence.  For higher error rates, it is always possible to bound the uncertainty on our approximation to the  exact logical error rate by a constant $\delta$.  For a given circuit, noise model, physical error rate $p$, and user-defined tolerance $\delta$, the sampling tools can identify all the subsets that need to be sampled such that
\begin{equation}
\sum_{w=0}^{w_{max}} A_{w}  <  1 - \delta
\end{equation} 
In practice, however, for high enough physical error rates (typically $p > (10^{-3} - 10^{-2})$), $w_{max}$ becomes prohibitively large, and the efficiency advantage of the \textit{subset-based} over the \textit{traditional} sampler disappears.   However, we note that in view of the current trapped-ion fidelities introduced in previous sections, the subset-based sampler is ideally suited to assess QEC protocols in trapped-ion architectures.

\subsection{Importance sampler for multi-parameter Pauli noise}

Let us now generalize the subset-based sampler to a multi-parameter noise model.  Let $m$ be the number of independent noise parameters.  Now, each subset is labeled by a vector of integers $\vec{w} = (w_1, w_2, ..., w_m)$, where each integer refers to the number of errors associated with a particular noise parameter.  The probability of occurrence of subset $\vec{w}$ is now given by
\begin{equation}
A_{\vec{w}} = \prod_{i=1}^{m}  {n_i \choose w_i}   p_i^{w_i} \, (1 - p_i)^{n_i - w_i},
\end{equation}
where $n_i$ is the number of gates (or operations) associated with the noise parameter $i\in\{1,\cdots,m\}$.  In our simulations, since some of the noise sources depend on the duration of the operation, we insert identity gates to account for waiting times on idle qubits.  Accordingly, every ion-crystal reconfiguration operation is represented by a sequence of identity gates whose number depend on the duration of the operation. 

Under a single-parameter noise model, we calculate the logical error rates in an error interval defined by $p_{min}$ and $p_{max}$, where $p_{min}$ is typically $0$.  In a multi-parameter noise model, we now have an error interval $[ p_{min,i}, p_{max,i}]$ for each parameter $i$.  These error intervals together form a hypercube in parameter space.  Just like for the single-parameter sampler, it is always possible to bound the uncertainty of the logical error rate by a constant $\delta$.  For a given circuit, noise model, error vector $\vec{p}$, and user-defined tolerance $\delta$, the sampling tools identify all the subsets $\vec{w}$ that need to be sampled to guarantee 
\begin{equation}
\sum_{\vec{w}=(0,...,0)}^{\vec{w}_{max}} A_{\vec{w}}  <  1 - \delta.
\end{equation} 
However, it is not necessary to calculate this for every error vector $\vec{p}$ in the hypercube of choice.  The hypercube vertex corresponding to the largest error vector $\vec{p}_{max}$ will determine the subsets that must be  sampled in order to guarantee that the difference between the upper and the lower bounds is less than the tolerance $\delta$.  For all error vectors $\vec{p}$ in the hypercube ($|\vec{p}| \leq |\vec{p}_{max}|$) the probability of occurrence of the subsets not considered will be even smaller than for $\vec{p}_{max}$, and so will the difference between the bounds on the logical error rate be.

A very useful feature of the \textit{subset based} sampler is that there is no need to perform Monte Carlo simulations for every individual $\vec{p}$.  To understand this, notice that the logical error rate $p_L^{(\vec{w})}$ of a subset $\vec{w}$ does not depend on the physical error rates given by $\vec{p}$.  The only terms that depend on $\vec{p}$ are the probabilities of occurrence of the subsets $A_{\vec{w}}$ (represented by the size of the boxes in Fig~\ref{fig:Sampler_Box}), which are easily computed analytically.  Therefore, each required subset is sampled once and for all.  Then,  the analytical nature of $A_{\vec{w}}$ allows us to construct smooth functions for the dependence of the logical error rate with respect to various microscopic error rates.  Furthermore, it also allows us to easily and exactly calculate the scalings of the logical error rate with respect to the different error combinations.  For other kinds of simulations, including Monte Carlo simulations with the \textit{traditional} sampler and exact wave-function simulations, extracting scalings involves computing logical error rates for a sufficiently dense set of points in the error vector hypercube, and then fitting the points along different directions to polynomials.  Since the volume (and hence number of points) of the hypercube scales exponentially with the dimension (number of noise parameters), this procedure quickly becomes unfeasible without the \textit{subset based} sampler.

\section{\bf Comparison of trapped-ion CNOT strategies}
\label{Sec:schedules}
 
We have now introduced and gathered all required ingredients to perform a realistic comparative study of the performance of transversal and lattice-surgery CNOT gates between two color-code qubits in a trapped-ion QCCD architecture. To identify the performance of the CNOT protocol separately,  we assume that the initial state is always perfectly prepared in the logical product state $|+ \rangle _L |0 \rangle _L$ in our simulations, which is ideally mapped under the CNOT onto the logical, maximally entangled Bell state $\frac{1}{\sqrt{2}} (|0 \rangle _L |0 \rangle _L + |1 \rangle _L |1 \rangle _L)$. We  then evaluate numerically the performance of the faulty CNOT gate to obtain the logical Bell pair.  For the lattice-surgery approach, we additionally assume that the ancilla logical qubit starts in a perfectly prepared $|0 \rangle _L$ state.  In the following subsections, we present our numerical results for the performance of both schemes, and introduce a detailed resource analysis that helps us in their qualitative understanding, and illustrates the technical complexity that would be required in an experimental realization.

\subsection{Performance of the transverse and lattice-surgery strategies} 

As described in the previous section, we use the multi-parameter subset approach to perform the numerical Monte Carlo sampling of the error configurations, and finally assess the performance of the two logical CNOT strategies.  For the transversal circuit, we have sampled all subsets up to weight $w=7$. For the lattice-surgery circuit, given its higher number of gates and ion reordering operations, we have  sampled all subsets up to weight $w=5$. This will be noticeable in the subsequent plots in the form of the divergence between the upper and lower bounds for the logical error rates setting in earlier (i.e.~already for lower physical error rates) for the lattice-surgery circuit as compared to the transversal circuit.  After the faulty CNOT circuit, we perform one round of perfect QEC to project the final state back to the code space and thereby account for the logical failure rate only for uncorrectable errors. Once this is done, we can compute both the probability for occurrence of a logical $X$ and a logical $Z$ error independently.

\begin{figure*}
\begin{centering}
\includegraphics[scale=0.4]{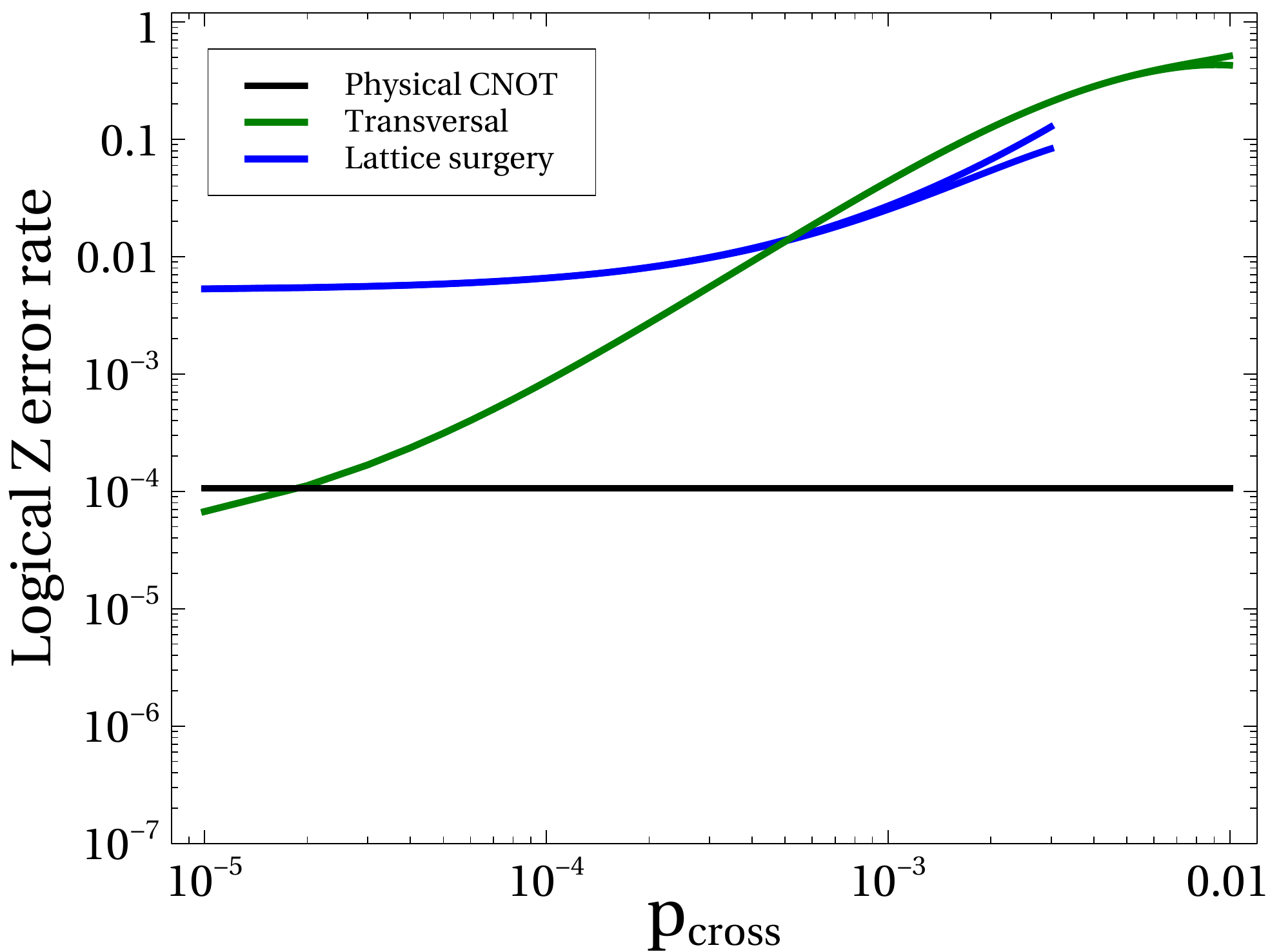}
\includegraphics[scale=0.4]{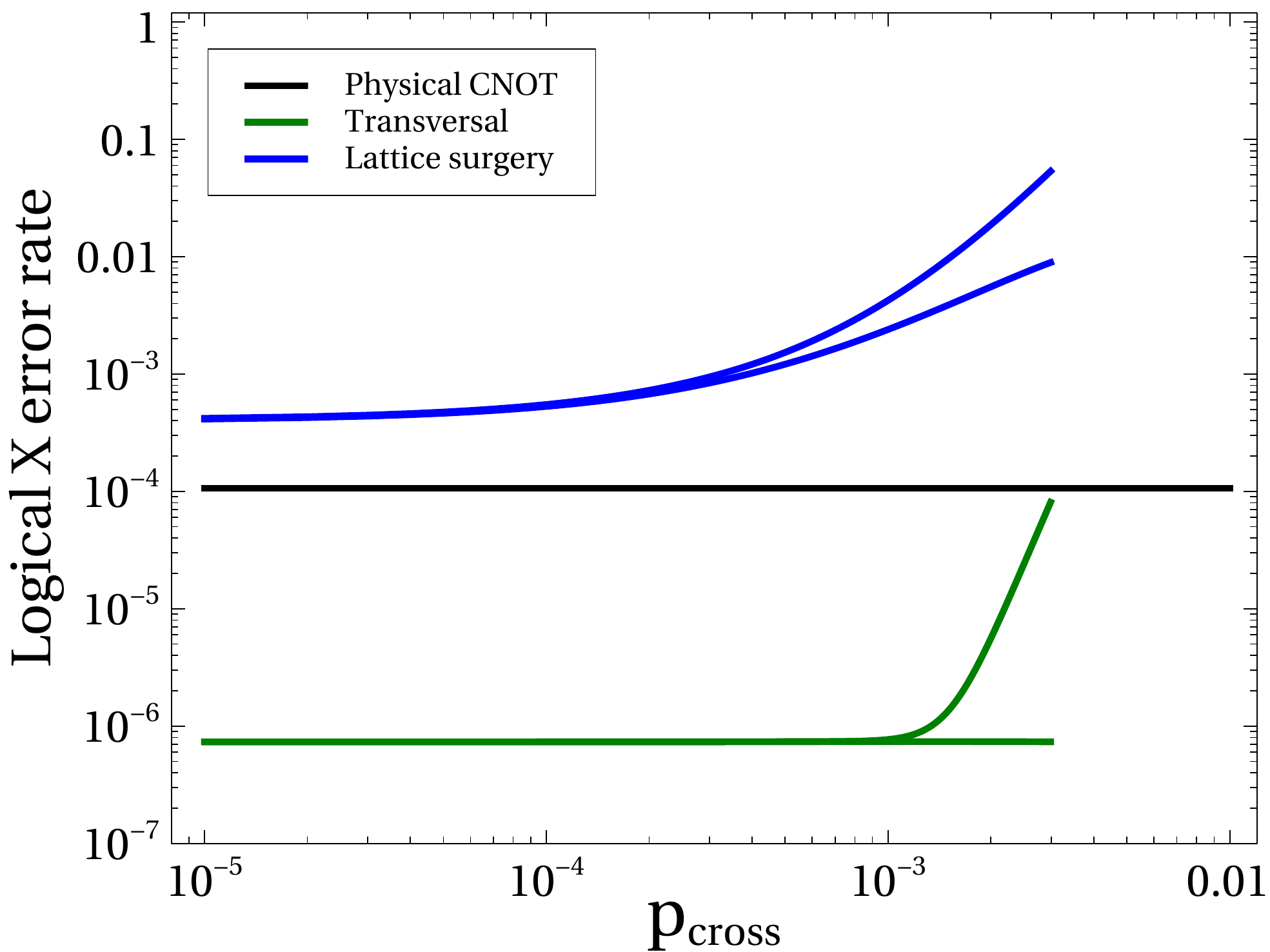}
\end{centering}

\caption{ {\bf Comparison of the logical CNOT strategies as a function of the junction-crossing errors:} Probability of a logical Z (left) and logical X (right) error after the preparation of a logical Bell pair using two alternative CNOT strategies as a function of the error strength associated with a trap-junction crossing.  All the other error strengths and reordering durations are set to the anticipated experimental values.  The horizontal black curve corresponds to the probability of causing a physical Z or X error during the preparation of a physical Bell pair assuming the same error model employed in the simulations.  This error rate is $8 \, p_{2q}/15$ because of the possible 15 Pauli errors after a 2-qubit gate, $8$ cause a $Z$ error on a Bell pair ($IY, IZ, XY, XZ, YI, YX, ZI, ZX$) and a different set of 8 causes an X error ($IX, IY, XI, XZ, YI, YZ, ZX, ZY$).  For each logical error rate, the subset sampler returns a lower and an upper bound, which essentially coincide tightly at low physical error rates and start to diverge at increasingly larger physical error rates, where the importance of the error subsets, which have not been included in the sampling, becomes important.}
\label{fig:logical_error_pcross}
\end{figure*}

Figure~\ref{fig:logical_error_pcross} shows the logical $Z$ and $X$ error rates for the two alternative CNOT approaches as a function of the error strength of the junction crossing (i.e. the dephasing associated to the crossing time). Notice that for both logical CNOT strategies, the logical $Z$ error rate is considerably higher than the logical $X$ error rate, which agrees with the fact that during the reordering and crossing operations, the qubits only experience $Z$-type errors according to the error model outlined above. We remark that, while all other  operations in Table~\ref{tab:summary_toolbox} have been optimized considerably over the last years, the transport of ions across junctions is the most demanding reconfiguration, and still requires more detailed experimental investigation in the current scalable trap designs. It is therefore important to explore how critically the performance and resources of two CNOT schemes will behave depending on the quality with which such junction crossings can be performed. We observe from the simulations, for instance, that the transversal logical CNOT outperforms even the physical CNOT for sufficiently low values of  $p_{\rm cross}$ (see Fig.~\ref{fig:logical_error_pcross}), i.e.~in the regime where the junction crossings can be done relatively fast and therefore associated with low qubit dephasing rates. On the other hand, even with the anticipated values of experimental parameters, the lattice-surgery logical CNOT cannot outperform the physical one. This result is certainly not surprising given the extremely high fidelities already achieved by trapped-ion  gates, and the overhead in the number of microscopic operations that is required to implement the lattice surgery method (see the following subsection on the resource analysis). Instead, this result  reflects the expected fact that the additional advantage of the QEC codes for logical lattice-surgery CNOT gates with respect to the bare CNOT gates will only become appreciable for larger-distance codes.  What it is quite remarkable is that, already for these low-distance codes, the transversal approach could beat the bare CNOT provided that the junction crossing is sufficiently efficient and the other experimental ingredients reach the expected level of accuracy.

The point that we would like  to stress in this manuscript is that, in the event that the junction crossing in the new trap designs cannot reach these high-quality levels, it might be favorable to adopt the lattice-surgery scheme in detriment of the transversal one despite of its larger overhead in terms of the required operations. This becomes clear from the crossing of the blue and green lines in the left panel of Fig.~\ref{fig:logical_error_pcross}, and one can generally expect that will be more appreciable as the distance of the code increases, since the transversal approach will require a higher number of faulty slow crossings. In particular, considering the logical $Z$ errors, when the error associated with the junction crossing is above a certain value, ($p_{\rm cross} > 5.2 \times 10^{-4}$), the more local character of the lattice-surgery CNOT approach, requiring only the coupling of qubits on the lattice boundary of pairs of logical qubits, pays off and allows it to outperform the transversal approach. With the anticipated dephasing time $T_2= 2.2\,$s, this $p_{\rm cross}$ value would correspond to a crossing time of $t_{\rm cross}=$2 ms per ion. A round-trip shuttling of a single $^{40}$Ca$^+$ ion through a junction on a surface trap has been reported to take between $0.96\,$ ms and $3.6\,$ms \cite{Junction_GTRI}. This implies that a one-way shuttling step could take approximately between $0.48\,$ms and $1.8\,$ms, slightly less than our threshold value of $2\,$ms. Faster shuttling has also been reported for $^9$Be$^+$ ions through an $X$ junction \cite{blakestad09_xjunctions,blakestad11_xjunctions}. In this regime, it would be advisable to perform a logical CNOT between two distance-3 color code qubits in a transversal fashion.  Whether the future QCCD experiments with the HOA-2 trap are in one regime or the other will likely  determine the strategy that must be followed to achieve the  $(\mathsf{QEC}$-$\mathsf{III})$ goal.

\begin{figure*}
\begin{centering}
\includegraphics[scale=0.4]{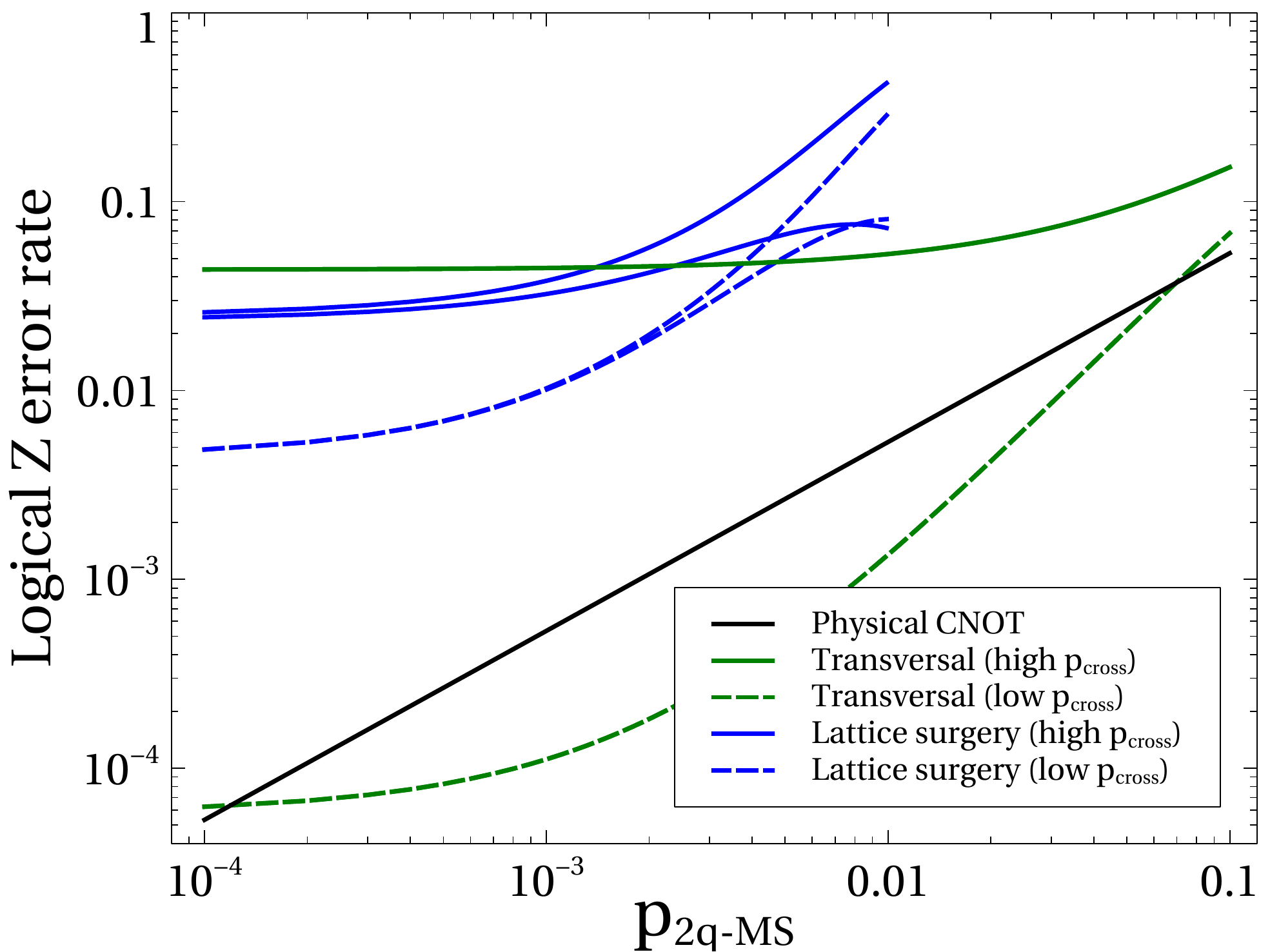}
\includegraphics[scale=0.4]{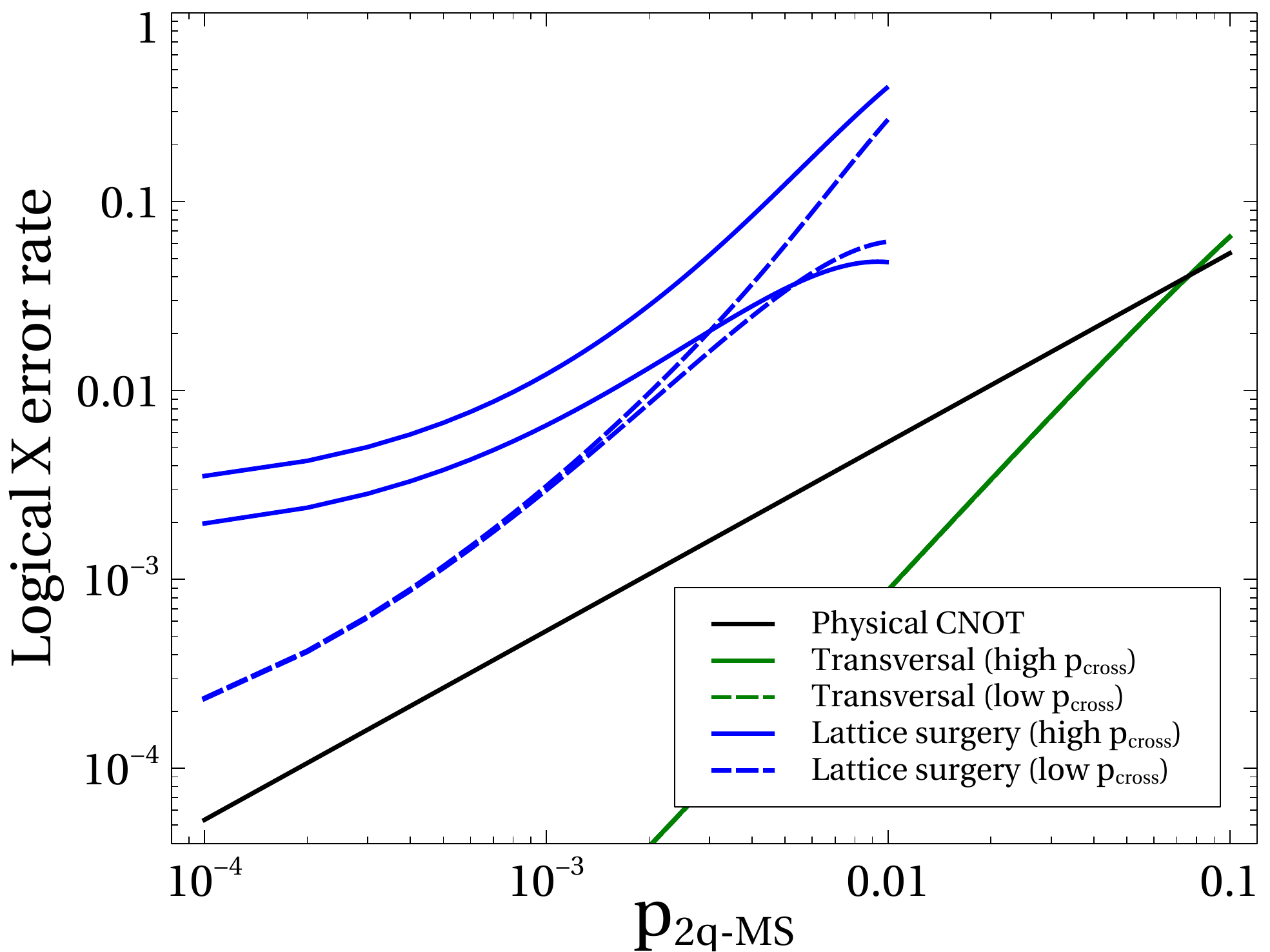}
\end{centering}
\caption{{\bf Comparison of the logical CNOT strategies as a function of the entangling MS gate errors:} Probability of a logical Z (left) and a logical X (right) error after the preparation of a logical Bell pair using two alternative CNOT strategies as a function of the 2-qubit MS gate error strength.  All the other error strengths and reordering durations are set to the anticipated experimental values.  The error strength of the 5-qubit MS gate is set to be $p_{5q} = 5 \, p_{2q}$.  As before, the black curve corresponds to the probability of causing a physical Z error (left) or physical X error (right) during the preparation of a physical Bell pair, which is given by $8 \, p_{2q}/15$, as explained in Fig~\ref{fig:logical_error_pcross}.}
\label{fig:logical_error_MS}
\end{figure*}

Figure \ref{fig:logical_error_MS} presents the logical Z and X error rates for the two alternative CNOT approaches as a function  of the error rate of the 2-qubit MS gate, which is a critical key parameter in the experimental implementation and the most demanding operation on the internal degrees of freedom of the ions. We have performed simulations for two regimes of junction-crossing, a low-overhead regime ($p_{\rm cross} = 10^{-5}$, denoted as \textit{low $p_{\rm cross}$} in Figs.~\ref{fig:resources_gates} and \ref{fig:total_duration_CNOTs}), and a high-overhead regime ($p_{\rm cross} = 10^{-3}$, denoted as \textit{high $p_{\rm cross}$}). In the low crossing overhead regime, the transversal CNOT outperforms the physical CNOT for a very wide range of MS error rates, including the anticipated experimental value. This implies that, if in fact the overhead associated with the junction crossing is low, it would be feasible to perform a logical entangling gate with a lower error rate than its physical bare counterpart, a major experimental breakthrough. The advantage is lost for $p_{2q} > 0.1$, which can be considered the level-1 pseudo-threshold for the transversal CNOT.  

For $p_{\rm cross} < 10^{-4}$, the other noise sources (phase flips during ion re-ordering operations, state preparation, measurement, and 1-qubit gates errors) begin to dominate over the 2-qubit error rate and the advantage of the transversal logical CNOT over its physical counterpart is lost again. For the lattice-surgery logical CNOT, the large number of gates and reordering operations cause the logical error rate to stabilize around $5 \times 10^{-3}$ in the low $p_{2q}$ limit, thus never outperforming the physical CNOT.  This means that even with the anticipated experimental values, we would be above the level-1 pseudo-threshold of the lattice-surgery logical CNOT.  On the other hand, in the high overhead crossing regime, while both logical strategies are above the pseudo-threshold, the relative advantage of the transversal scheme is lost and it becomes better to perform the lattice-surgery scheme for low enough values of $p_{2q}$.  For $p_{2q} > 1.2\times 10^{-3}-2.6\times 10^{-3}$, both strategies result in similar Z logical error rates.  The anticipated experimental value for the 2-qubit MS fidelity lies within this interval. As before, the X logical error rate is lower than the Z logical error rate.  The transversal protocol is particularly resilient to logical X errors, given its lower number of MS gates.

\subsection{Resource analysis for the transverse and lattice-surgery CNOT strategies}

\begin{figure*}
\begin{centering}
\includegraphics[scale=0.35]{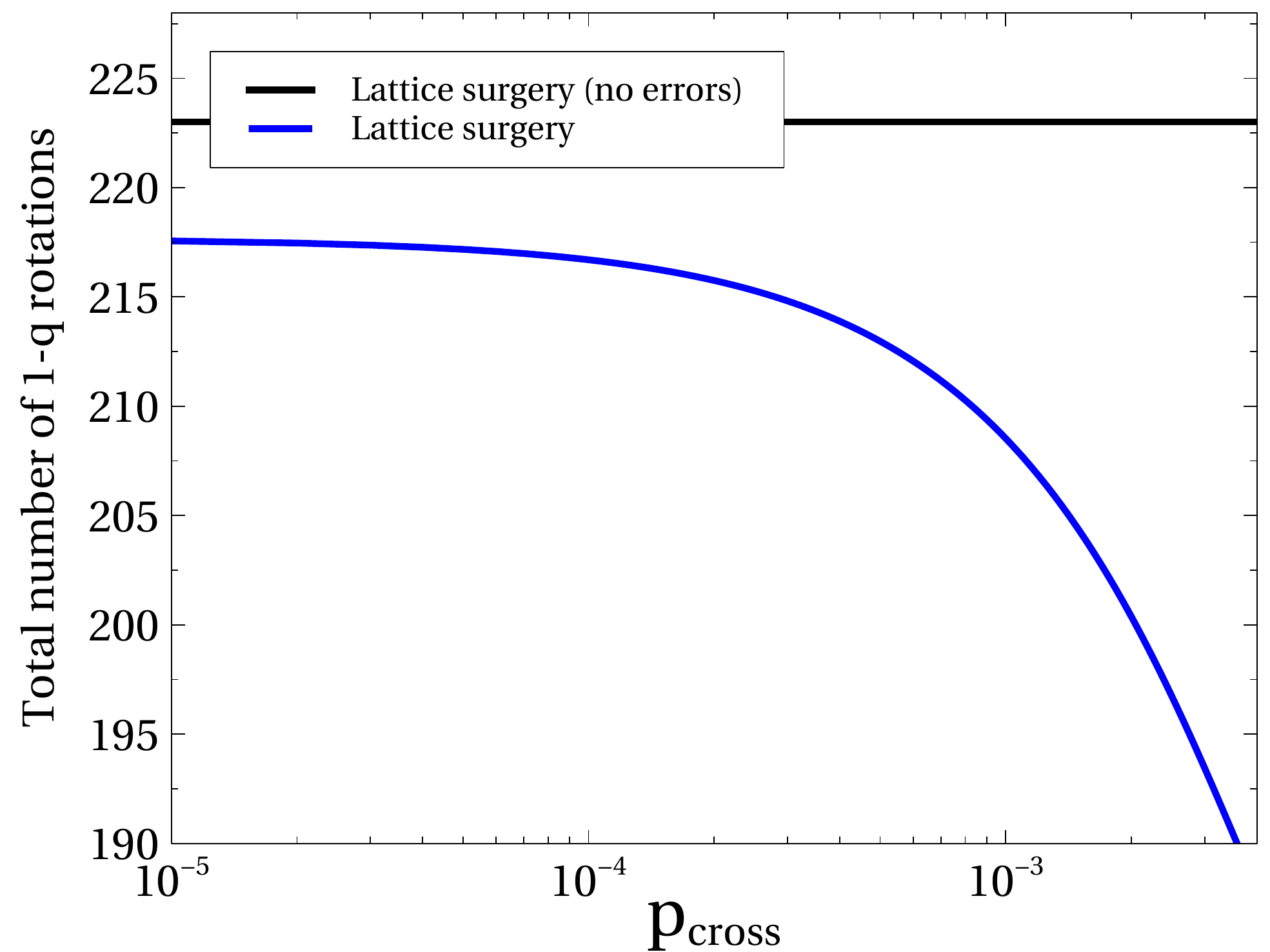}
\includegraphics[scale=0.35]{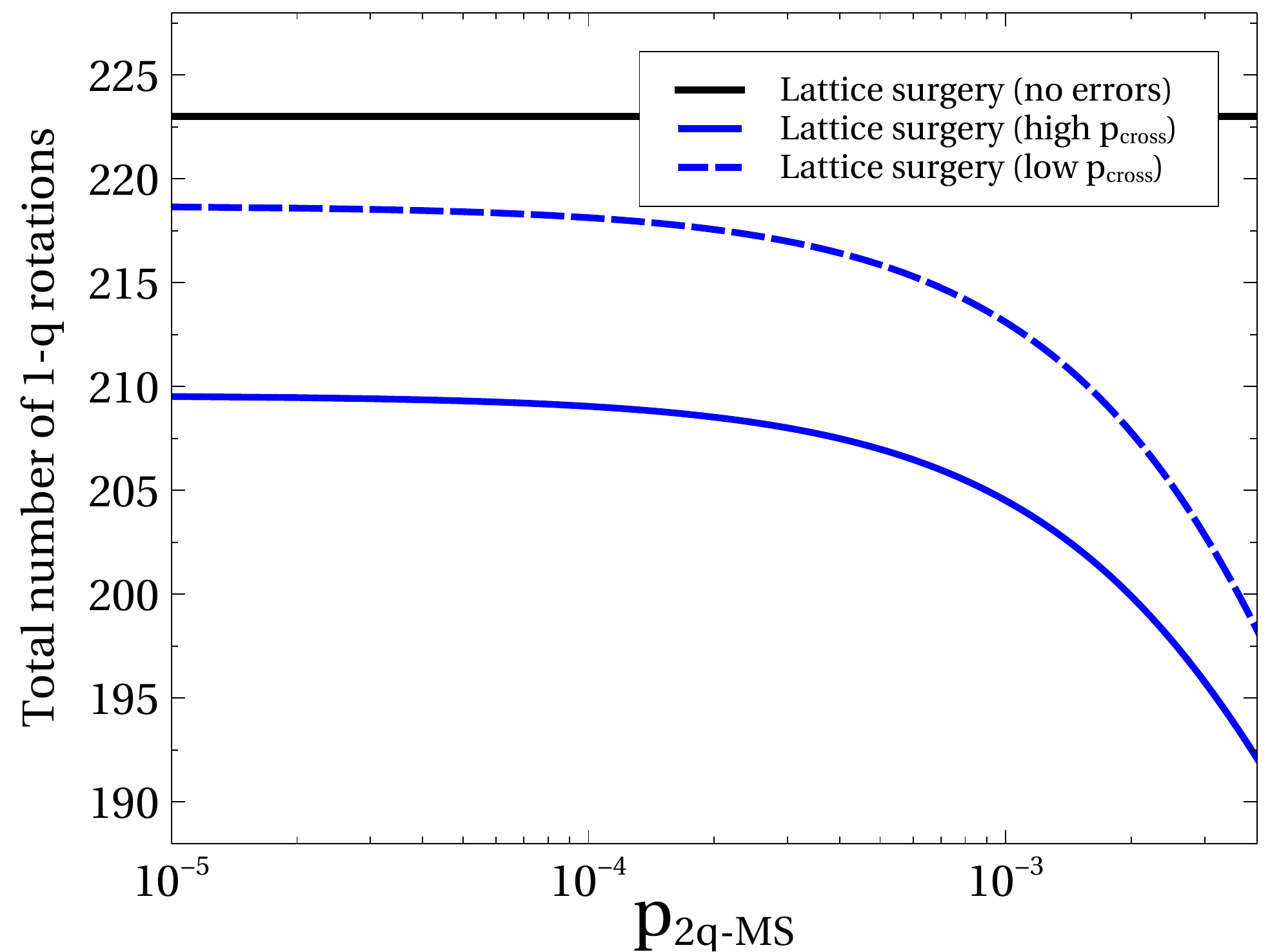}
\includegraphics[scale=0.35]{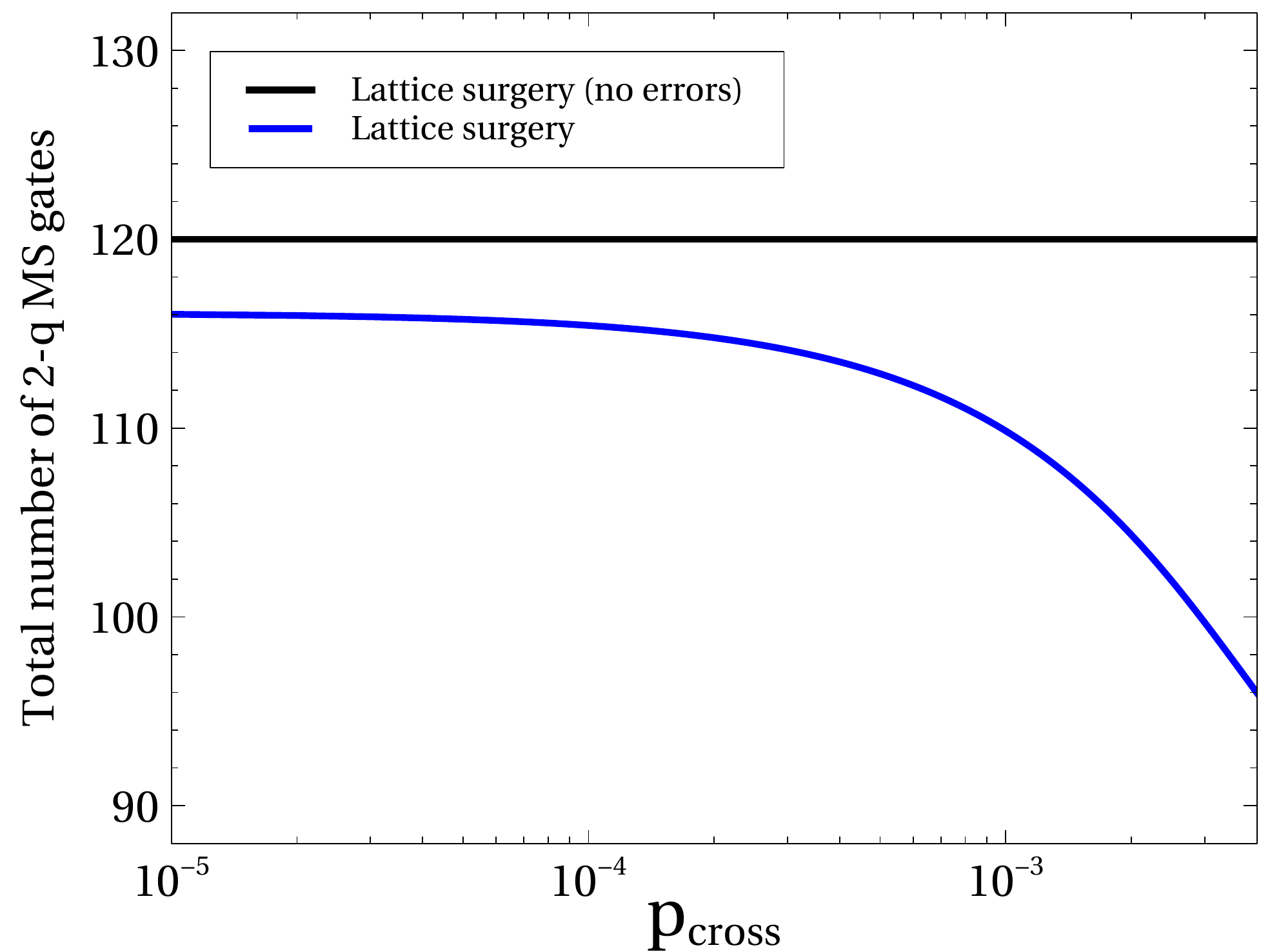}
\includegraphics[scale=0.35]{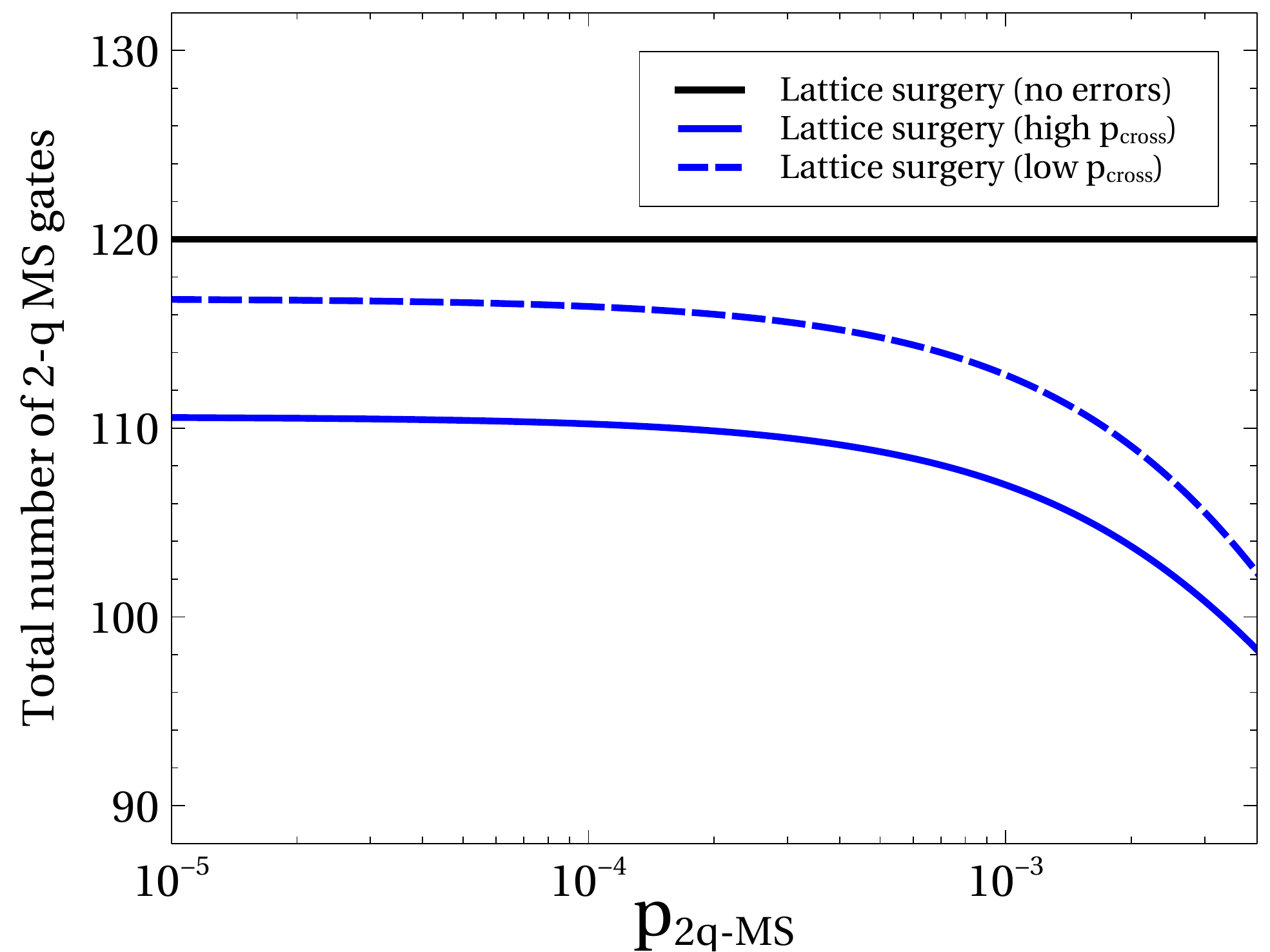}
\includegraphics[scale=0.35]{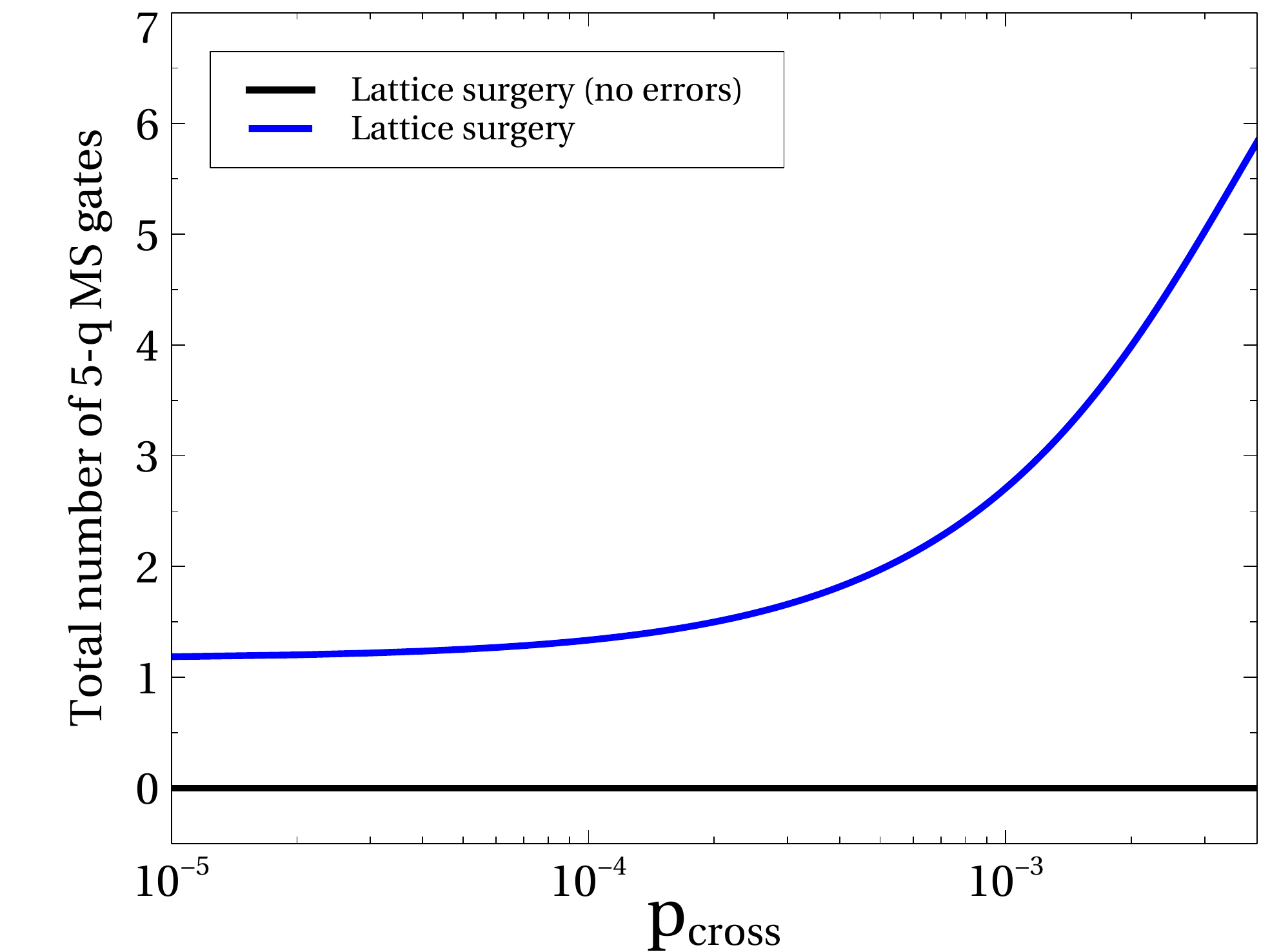}
\includegraphics[scale=0.35]{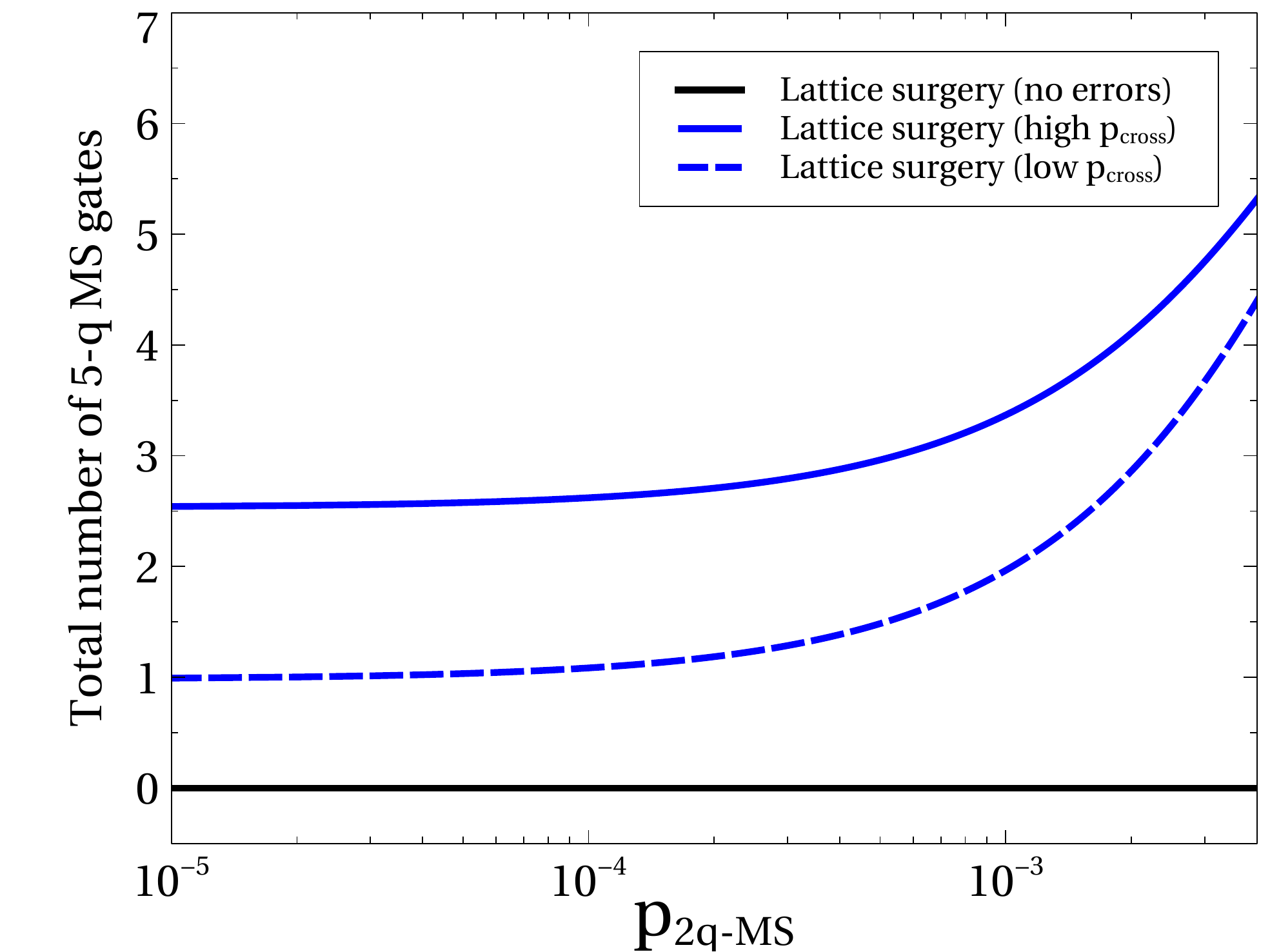}
\includegraphics[scale=0.35]{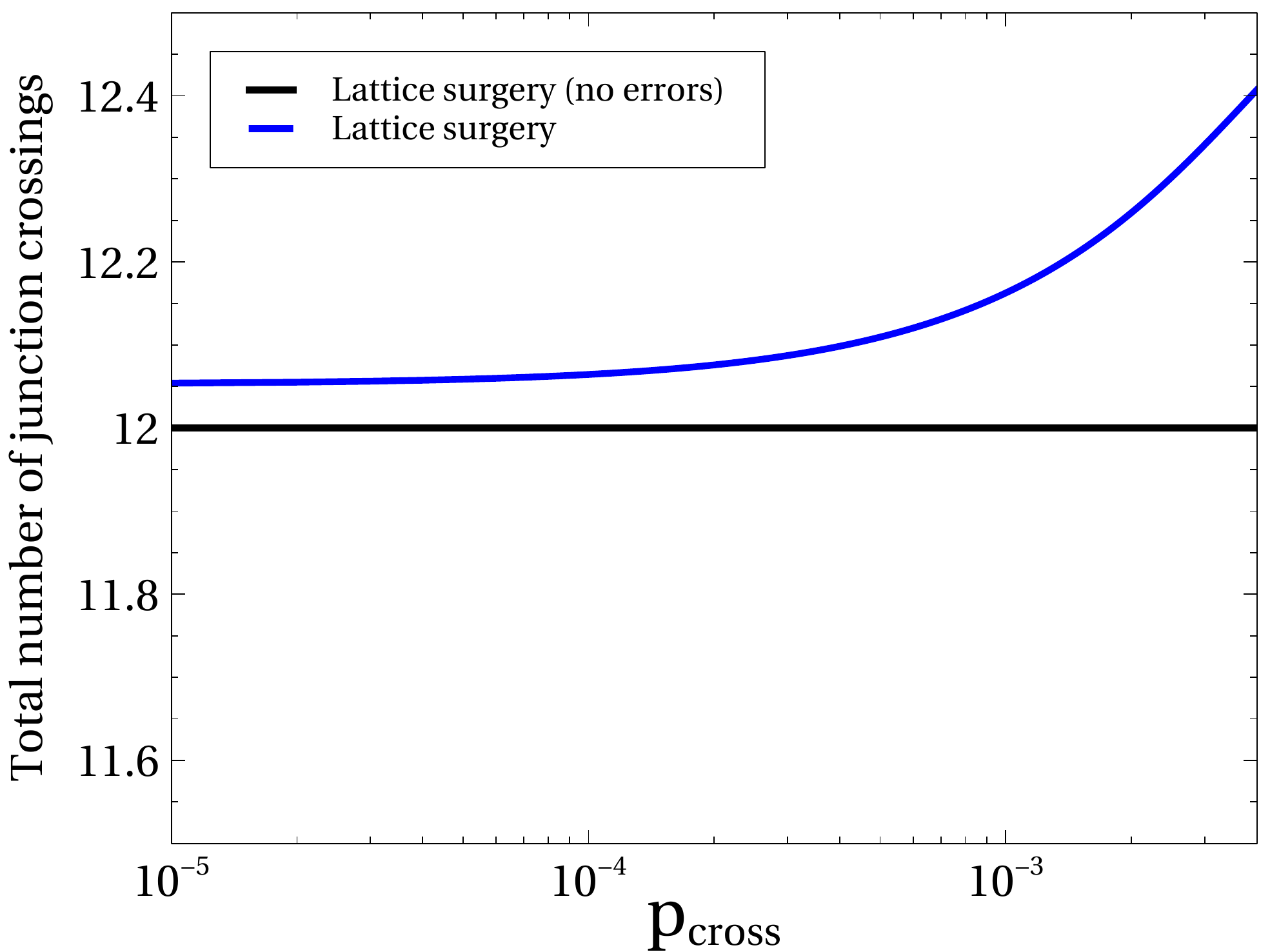}
\includegraphics[scale=0.35]{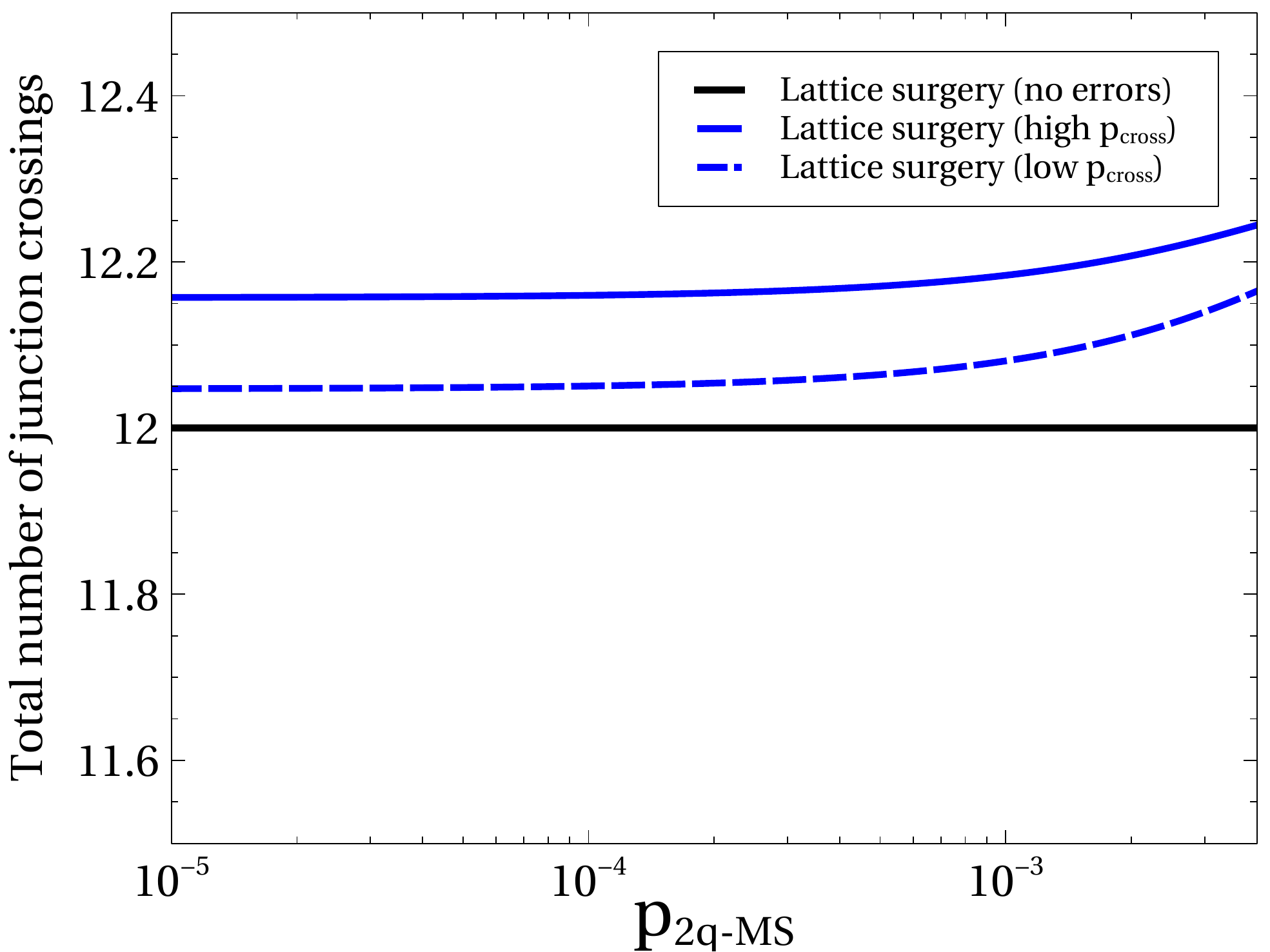}
\end{centering}
\caption{{\bf Gate and junction-crossing resources in the lattice-surgery CNOT: } Average number of $1$-qubit gates, $2$- and $5$-qubit MS gates, and trap-junction crossings performed during the lattice surgery CNOT as a function of two different error strengths, the one associated with the trap-junction crossing $p_{cross}$ (left) and the error rate of the $2$-qubit MS gate (right).   All the other error strengths and durations are set to the anticipated experimental values.  For the figures on the right, the error strength of the $5$-qubit MS gate is set to be $p_{5q} = 5 \, p_{2q}$.  In this case, for high-error crossing we set $p_{cross} = 10^{-3}$, while for the low-error crossing we set $p_{cross} = 10^{-5}$.  The black curves correspond to an error-free regime where every step in the protocol is FT because we never detect an error. In the realistic (faulty) setting, we instead switch to non-FT circuits as soon as an error is detected.  This decreases the total number of $1$-qubit rotations and $2$-qubit MS gates that we have to perform on average.  In contrast, it increases the number of non-FT $5$-qubit MS gates, which we use only after an error has been detected. The total number of trap-junction crossings also increases, but only slightly.}
\label{fig:resources_gates}
\end{figure*}

Once the performance of both strategies has been presented, we discuss in this subsection another important aspect for their comparison, namely a resource analysis quantifying the complexity of the protocols. Let us note that, since the lattice-surgery approach makes use of the active rounds of flag-based QEC,  the number of gates and overall duration shall depend on the instants when the flag triggering occurs and the measurement outcomes, such that the  lattice-surgery resource analysis will be  more involved than the transversal one.   

In Fig.~\ref{fig:resources_gates}, we depict the average number of the various gates and trap-junction crossings that would be required during the lattice surgery CNOT as a function of the strength of two key error parameters, namely the junction crossing error rate and the error rate of the 2-qubit MS gates. In contrast to the transversal strategy, which has a fixed number of gates and reordering operations, the required resources for the lattice surgery CNOT will depend on the error rates. There are two competing effects.  On the one hand, within a given step or substep of the protocol, errors will increase the number of required gates.  This can be exemplified with the measurement of one of the merging operators.  If an error causes two subsequent measurement outcomes to differ, the operator needs to be measured a third time, which increases the total number of gates and reordering operations. On the other hand, because of the low-resource philosophy that we have adopted, for the protocol as a whole, the detection of an error event will decrease the total number of gates and reordering operations, because the remaining steps of the protocol will involve the operationally simpler un-flagged circuits.  

In the error-free regime (limit of all physical error rates approaching zero), where every step in the circuit is FT, the total number of $1$-qubit gates would be $223$.  Interestingly, considering the anticipated experimental rates as displayed in Table~\ref{tab:summary_toolbox}, the number of $1$-qubit gates that would be performed on average becomes lower than in the error-free case  as a consequence of the low-resource philosophy (i.e.~un-flagged circuit in the event of a detected error are less resource-intensive than the flagged ones). Moreover, this number decreases with increasing error strength. This trend  is depicted in the two upper plots in the first row of Fig.~\ref{fig:resources_gates}, where we show the average number of $1$-qubit rotations. However, we note that, even at relatively high error rates, the number of $1$-qubit gates is always considerably larger than the one required by the transversal CNOT approach, which is always limited  to $28$ $1$-qubit gates (resulting from $4$ single-qubit rotations for each physical CNOT). The number of required $2$-qubit MS gates follows  a similar pattern, decreasing from $120$ for the error-free lattice surgery circuit to less than $100$ for high error rates, as depicted in the plots of the second row of Fig.~\ref{fig:resources_gates}. Once again, the transversal CNOT strategy is less resource intensive, employing only $7$ $2$-qubit MS gates. In the plots of the  two lower rows of Fig.~\ref{fig:resources_gates}, we show that the average number of $5$-qubit MS gates and trap-junction crossings display an opposite trend, i.e.~increasing as the microscopic error rate grows.  In the error-free circuit, a $5$-qubit MS gate would never be performed. In the faulty circuit, the average number of $5$-qubit MS gates performed goes from about $1$ in the low-error limit to more than $4$ in the high-error limit. 

\begin{figure*}
\begin{centering}
\includegraphics[scale=0.4]{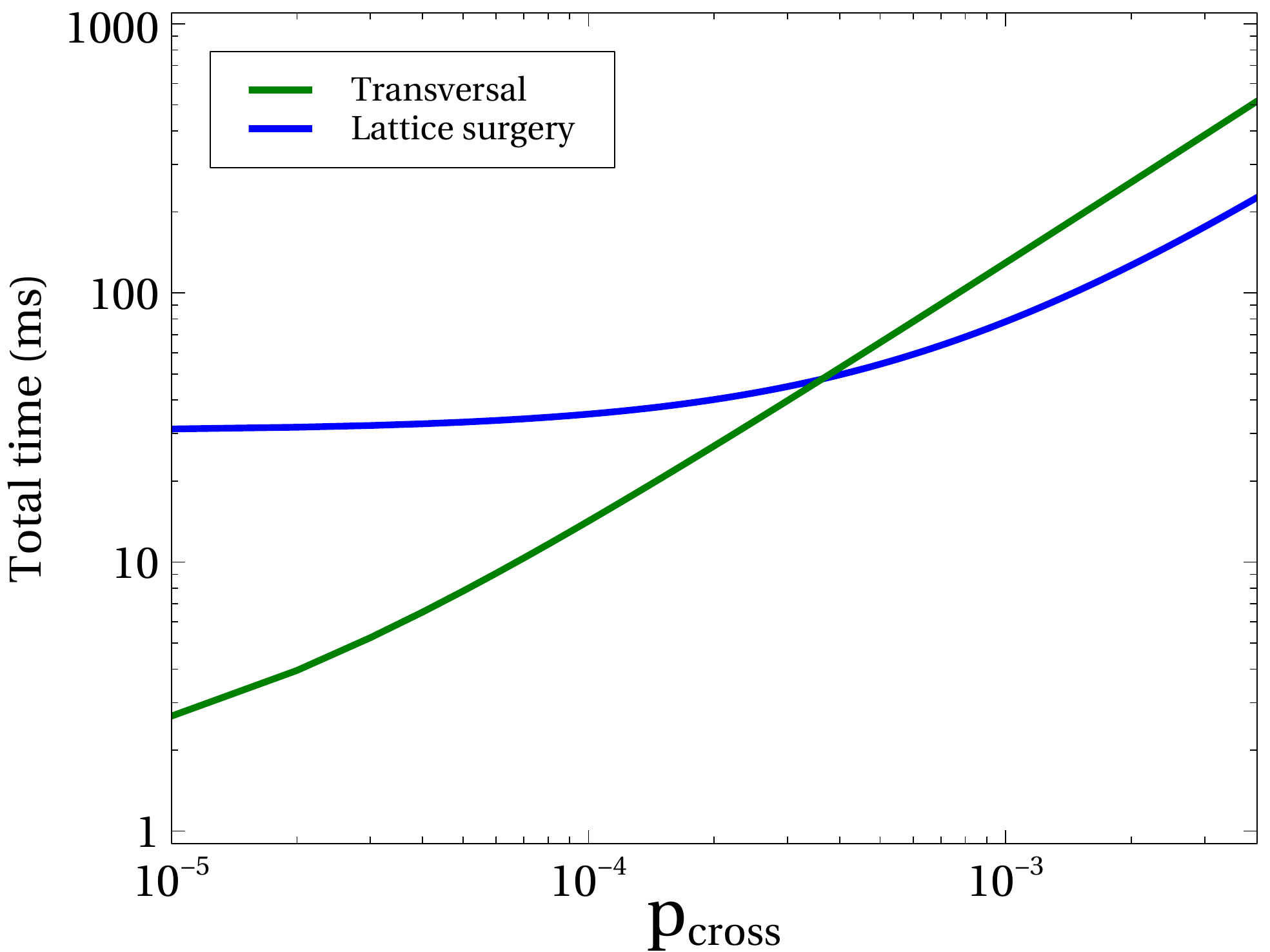}
\includegraphics[scale=0.4]{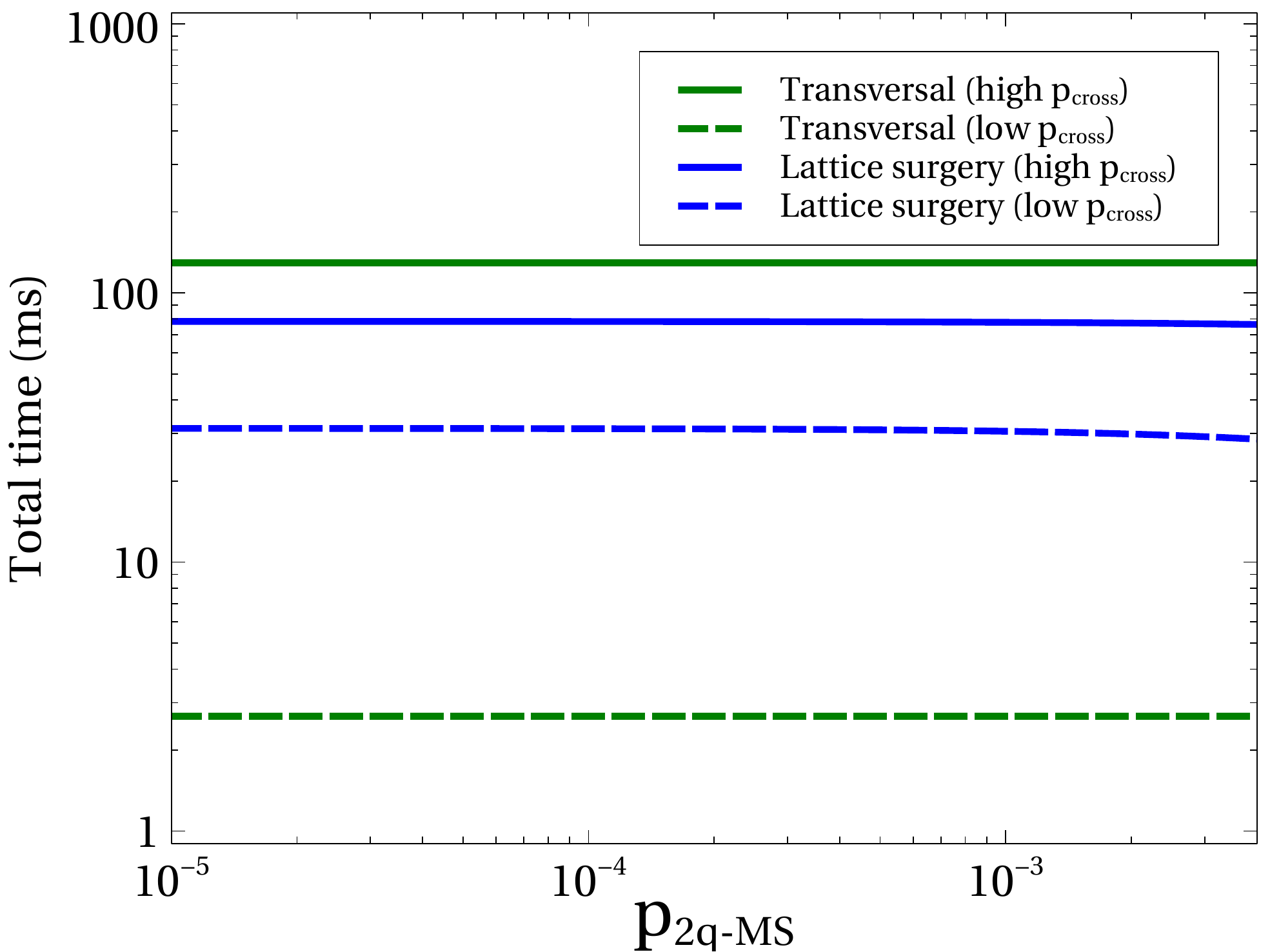}
\end{centering}
\caption{{\bf Comparison of the durations of the transversal and lattice-surgery CNOT gates: } Average duration of the two alternative CNOT strategies as a function of two different error strengths.   the one associated with the trap-junction crossing $p_{\rm cross}$ (left) and the error rate of the $2$-q MS gate (right).   All the other error strengths and durations are set to the anticipated experimental values.  For the figures on the right, the error strength of the $5$-qubit MS gate is set to be $p_{5q} = 5 \, p_{2q}$.  In this case, for high-error crossing we set $p_{cross} = 10^{-3}$, while for the low-error crossing we set $p_{\rm cross} = 10^{-5}$.  The durations of every gate and reordering operation are taken from Table \ref{tab:summary_toolbox}.  The duration of the trap-junction crossing is set to $t_{\rm cross} = -T_2 \ln(1-2p_{cross})$, and therefore increases with $p_{\rm cross}$.}
\label{fig:total_duration_CNOTs}
\end{figure*}

Regarding the junction crossings, the need to measure repeatedly the low-weight operators during the merging and splitting steps in the event of errors, requires some additional crystal reconfigurations which increase the number of junction crossings. In the error-free circuit, the number of trap-junction crossings is exactly $12$: we first shuttle $3$ ions from the ancillary arm to the target arm, and back again to the ancillary arm, in order to measure the $X_L^t X_L^a$ operator, as depicted in Fig~\ref{Fig:schedule_M_XX_operators}.  We then shuttle $3$ ions from the control to the ancillary arm, and back again to the control arm, to measure the $Z_L^c Z_L^a$ operator, as depicted in Fig~\ref{Fig:schedule_M_ZZ_operators}.  In the faulty circuit the number of trap-junction crossings is slightly larger than $12$, because in some cases, after the round-trip shuttling of the $3$ ions from one arm to another, if the QEC detects an error on one of the boundary qubits, either one of the boundary operators will have to be measured again, which implies $2$ ($4$) extra trap-junction crossings for the weight-$2$ (weight-$4$) operator (see Fig.~\ref{fig:FT_MXX} for a reminder of the protocol to measure a joint logical operator).  With our current schedule, the transversal CNOT requires $32$ trap-junction crossings.  Notice that, in principle, one can devise a schedule for the transversal CNOT that only requires $14$ trap-junction crossings, as we only need to shuttle $7$ ions from one arm to another and back.  However, such a schedule would increase considerably the number of necessary intra-arm reconfiguration operations (splitting, merging and, more importantly, rotations). This is the underlying reason for our choice of  a microscopic schedule for the transversal CNOT gate where some zones of the ancillary arm are initially vacated, and  used as a temporary storage zones to simplify the crystal reconfigurations.  Notice that, even with the transversal-CNOT schedule that minimizes the number of junction crossings ($14$), this is still larger than the average number of junction crossings for the lattice surgery CNOT (see lower row of of Fig.~\ref{fig:resources_gates}), which  implies that there will always be a high $p_{\rm cross}$ regime where the lattice surgery method outperforms the transversal CNOT scheme. This enhanced  sensitivity to the junction-crossing error will become higher as the code distance grows, since the transversal scheme requires transporting all data qubits across the junction, and back, in order to couple the equivalent qubits via CNOTs. Moreover, larger codes will certainly require storing the data qubits of a single logical block within various neighboring arms, such that the number of required junction crossings will increase even further. Conversely, the required crossings in the lattice surgery approach increase less rapidly. For instance, a judicious arrangement of the data qubits of the control, target, and ancillary blocks in neighboring trap arms can minimize the number of required crossings for the measurement of the low-weight operators in the merging step. 

Once we have calculated the number of required gates and junction crossings, one understands the qualitative trends of the performance of the two strategies shown in Figs.~\ref{fig:logical_error_pcross} and~\ref{fig:logical_error_MS}, namely the lattice surgery approach is  favorable for   higher junction-crossing errors and lower entangling-gate errors, while the transversal strategy should be adopted in the complementary regime. Moreover, given the average number of gates, we can also calculate 
 the average duration of the transversal and the lattice surgery CNOT strategies as a function of the trap-junction crossing and the $2$-q MS gate error strengths (see Fig.~\ref{fig:total_duration_CNOTs}).  As explained Subsection \ref{sec:error_model}, every reordering operation has an associated dephasing error with a strength given by $p_{\rm idle} = (1-\exp(-t/T_2))/2$, where $t$ is the duration of the operation.  We therefore set the duration of the trap-junction crossing to be $t_{\rm cross} = -T_2 \ln(1-2p_{\rm cross})$, which implies that, even with the low-resource philosophy, the total duration of both protocols will increase with $p_{\rm cross}$.  We observe a similar trend to the logical $Z$ error rate of Fig.~\ref{fig:logical_error_pcross}.  In the low $p_{\rm cross}$ regime, the transversal strategy is advantageous, as its duration is about a tenth of the lattice surgery CNOT duration.  On the other hand, in the high $p_{\rm cross}$ regime, the larger number of trap-junction crossing steps required as compared to the lattice-surgery method becomes a disadvantage and the transversal approach is slower.  Interestingly, the break-even point between the two regimes occurs at $p_{\rm cross} = 4.0 \times 10^{-4}$, very close to the break-even point for the logical $Z$ error rate of Fig.~\ref{fig:logical_error_pcross}.  This reinforces the observation that the main driver behind the logical $Z$ errors is simply the dephasing  during reordering operations, instead of errors associated to gates.

The durations of the protocols do not change significantly with the error strength of the $2$-qubit MS gate, as shown in the right panel of  Fig.~\ref{fig:total_duration_CNOTs}.  For the transversal CNOT, once we fix the duration of the trap-junction crossing, the duration of the entire protocol remains constant. These correspond to $2.68\,$ms and $129\,$ms in the low and high $p_{\rm cross}$ regimes, respectively. The durations of the lattice surgery protocols decrease slightly with increasing $p_{2q}$, due to the low-resource philosophy.  They decrease from $78.4\,$ms to $76.3\,$ms for high $p_{\rm cross}$, and from $31.5\,$ms to $28.5\,$ms for low $p_{\rm cross}$.  

Together with the conclusions from the above section on the performance comparison, this resource analysis shows that the future technological improvements in the junction-crossing capabilities of trapped-ion QCCD architectures, together with further-improving fidelities of the entangling gates, will be crucial achievements by the collaborative community efforts towards a functional fault-tolerant quantum processor.

\section{\bf Conclusions and outlook}
\label{sec:conclusions}

In this work, we have presented a detailed analysis of the prospects of trapped-ion QCCD architectures for the demonstration of a logical CNOT gate  between two logical qubits, here encoded in color-code qubits. Achieving such logical entangling gate is the most demanding ingredient to implement the whole encoded set of logical Clifford unitaries, and a vital ingredient for noise-resilient quantum information processing. The future accomplishment of this challenge with current or next-generation trapped-ion processors, which will still focus on low-distance topological codes, necessarily requires a careful fault-tolerant design of all the required steps in the protocols. In this respect, we leverage from the recently proposed flag-based methods for stabiliser readout, and  adapt them to the particular trapped-ion setup. In this work, we have proven that these flag-based methods  offer a clear advantage over other more resource-demanding FT schemes, and will likely be of key importance for further developments in FT trapped-ion QEC. In addition to the standard FT transversal approach for the CNOT gate using color codes, we have also described a  careful FT design of the lattice-surgery strategy, and discussed in detail the microscopic trapped-ion schedules to realize both protocols using a realistic QEC toolbox with state-of-the-art trapped-ion operations. 

As noted in the main text, a correct assessment of the performance of QEC in general, and these CNOT strategies in particular, requires the use of a realistic and physically-motivated error model for the  architecture at hand, upgrading the over-simplified device-independent models mostly used in the literature. Towards that goal, one of the characteristic features of our study is the use of a microscopic multi-parameter error model with different quantum channels describing the leading effects of noise on the different trapped-ion operations, the parameters of which are set by microscopic calculations and experimental results. Equipped with the above FT designs, microscopic schedules, and realistic error model, we have presented a numerical comparison of both CNOT strategies in view of current and envisioned experimental capabilities.  The development of an efficient multi-parameter subset sampler for the stabiliser formalism  has been crucial  to perform these exhaustive simulations efficiently.  

Our numerical results can be used to identify the experimental regimes where one strategy will be favourable over the other, paying  special attention to the roles of the errors during the transport of ions across the trap junction, which is the most demanding operation on the external degrees of freedom of the ions, and the effect of infidelities in the phonon-mediated entangling gates, which is still the most-demanding operation on the internal degrees of freedom of the ions. We show that depending on the error budget of these two operations, lattice-surgery might be preferred with respect to a transversal approach or vice versa. In general, the lattice surgery approach will be favourable in situations of a higher junction-crossing error and lower entangling-gate errors, while the transversal strategy should be adopted in case that the entangling-gate error is higher but the junction-crossing is more faulty. These results can be understood from our resource analysis, and will be even more pronounced for larger-distance codes, and they will hopefully guide the future experimental trapped-ion progress in the field.     

As an outlook for future work, we note that trapped-ion experiments and other platforms have evaluated the performance of single-qubit and entangling quantum gates at the physical qubit level using randomized benchmarking \cite{knill_RB} and other techniques, such as gate set tomography \cite{robin_GST}. Therefore, to assess the performance of the logical CNOT operation, instead of computing the error rate associated with the preparation of a logical Bell pair, as explored in this work, it would be very interesting and useful to 
extend this to a study of logical randomized benchmarking \cite{logical_RB} of the different strategies hereby described.

Stimulated by the planned experimental developments in the mid-term horizon, it will be very interesting to perform analogous studies for larger-distance codes promising higher resilience against errors. As already noted in the main text, in the QCCD architecture with the HOA-2 segmented trap it will likely not be optimal to store whole logical blocks within a single arm. It will therefore be important to develop microscopic arrangements of the data qubits distributed over neighbouring arms, and work out detailed operational schedules similar to the ones hereby presented that minimize the required resources. With an optimized microscopic design, and depending on how noisy the junction crossings turn out to be in the laboratory, it is plausible that the advantage of the lattice-surgery approach over the transversal CNOT gate will  become apparent  earlier. This trend would constitute a proof of the advantage of schemes with local quantum processing in a practical and realistic scenario that could be experimentally tested in the near future. Let us also mention that, in this work, we have not considered another viable strategy towards the FT implementation of a logical CNOT gate: code deformation \cite{Horsman-njp-2012}. The reason for this omission is that the resource requirements for this strategy are higher for the small-distance codes that will be available in near-term experiments. However, as the trapped-ion technology improves, and larger registers can be handled, it will be very interesting to extend our comparison by including also a realistic and detailed study of a trapped-ion code deformation strategy.

The FT design of the lattice-surgery methods, together with the trapped-ion tools to implement them, can be considered as generic building blocks that can also be useful for other  
applications of lattice surgery, such as switching between different codes e.g.~a surface code for information storage and a color code for information processing by means of transversal gate operations \cite{Bombin2016,Nautrup2017}. It would also be interesting in the future to assess the prospects of this trapped-ion architecture in the realization of such schemes.

\acknowledgements
The research is based upon work supported by the Office of the Director of National Intelligence (ODNI), Intelligence Advanced Research Projects Activity (IARPA), via the U.S. Army Research Office Grant No. W911NF-16-1-0070. The views and conclusions contained herein are those of the authors and should not be interpreted as necessarily representing the official policies or endorsements, either expressed or implied, of the ODNI, IARPA, or the U.S. Government. The U.S. Government is authorized to reproduce and distribute reprints for Governmental purposes notwithstanding any copyright annotation thereon. Any opinions, findings, and conclusions or recommendations expressed in this material are those of the author(s) and do not necessarily reflect the view of the U.S. Army Research Office.  
 
We also acknowledge support by U.S. A.R.O. through Grant No. W911NF-14-1-010. A.B. acknowledges support from Spanish MINECO Projects FIS2015-70856-P, and CAM regional research consortium QUITEMAD+.  We acknowledge the resources and support of High Performance Computing Wales, where most of the simulations were performed. We gratefully thank all members of the eQual consortium (i.e. "Encoded Qubit Alive") for their continuous input and useful discussions during several stages of this work. 

 

\end{document}